\documentclass[aps, prd, twocolumn, showpacs, superscriptaddress, floatfix]{revtex4-2} 

\usepackage{lineno}

\usepackage{graphicx}  
\usepackage{dcolumn}   
\usepackage{bm}        
\usepackage{amssymb}   
\usepackage[caption=false]{subfig}
\usepackage{xcolor}
\usepackage{amsmath}
\makeatletter
\def\maketag@@@#1{\hbox{\m@th\normalfont\normalsize#1}}
\makeatother
\usepackage{placeins}
\usepackage{enumerate}
\usepackage{multirow}
\usepackage{xcolor}
\usepackage{xparse,xcoffins}

\ExplSyntaxOn
\NewCoffin\imagecoffin
\NewCoffin\labelcoffin

\keys_define:nn { miguel/label }
 {
  label   .tl_set:N = \l_miguel_label_tl,
  labelbox .bool_set:N = \l_miguel_label_box_bool,
  labelbox .default:n = true,
  fontsize .tl_set:N = \l_miguel_label_size_tl,
  fontsize .initial:n = \footnotesize,
  pos .choice:,
  pos/nw .code:n = \tl_set:Nn \l_miguel_label_pos_tl { left,up },
  pos/ne .code:n = \tl_set:Nn \l_miguel_label_pos_tl { right,up },
  pos/sw .code:n = \tl_set:Nn \l_miguel_label_pos_tl { left,down },
  pos/se .code:n = \tl_set:Nn \l_miguel_label_pos_tl { right,down },
  pos/n .code:n = \tl_set:Nn \l_miguel_label_pos_tl { hc,up },
  pos/w .code:n = \tl_set:Nn \l_miguel_label_pos_tl { left,vc },
  pos/s .code:n = \tl_set:Nn \l_miguel_label_pos_tl { hc,down },
  pos/e .code:n = \tl_set:Nn \l_miguel_label_pos_tl { right,vc },
  pos .initial:n = nw,
  unknown .code:n   = \clist_put_right:Nx \l_miguel_label_clist
                       { \l_keys_key_tl = \exp_not:n { #1 } }
 }
\clist_new:N \l_miguel_label_clist
\box_new:N \l_miguel_label_box
\box_new:N \l_miguel_label_image_box

\NewDocumentCommand{\zincludegraphics}{O{}m}
 {
  \group_begin:
  \tl_clear:N \l_miguel_label_tl
  \clist_clear:N \l_miguel_label_clist
  \keys_set:nn { miguel/label } { #1 }
  \tl_if_empty:NTF \l_miguel_label_tl
   {
    \miguel_includegraphics:Vn \l_miguel_label_clist { #2 }
   }
   {
    \SetHorizontalCoffin\imagecoffin
     {
      \miguel_includegraphics:Vn \l_miguel_label_clist { #2 }
     }
    \SetHorizontalCoffin\labelcoffin
     {
      \raisebox{\depth}
       {
        \bool_if:NTF \l_miguel_label_box_bool
         { \fcolorbox{white}{white}{\l_miguel_label_size_tl\l_miguel_label_tl} }
         { \l_miguel_label_size_tl\l_miguel_label_tl }
       }
     }
    \SetVerticalPole\imagecoffin{left}{3pt+\CoffinWidth\labelcoffin/2}
    \SetVerticalPole\imagecoffin{right}{\Width-7pt-\CoffinWidth\labelcoffin/2}
    \SetHorizontalPole\imagecoffin{up}{\Height-3pt-\CoffinHeight\labelcoffin/2}
    \SetHorizontalPole\imagecoffin{down}{16pt+\CoffinHeight\labelcoffin/2}
    \use:x{\JoinCoffins\imagecoffin[\l_miguel_label_pos_tl]\labelcoffin[vc,hc]} 
    \TypesetCoffin\imagecoffin
   }
   \group_end:
 }

\NewDocumentCommand{\xincludegraphics}{O{}m}
 {
  \group_begin:
  \tl_clear:N \l_miguel_label_tl
  \clist_clear:N \l_miguel_label_clist
  \keys_set:nn { miguel/label } { #1 }
  \tl_if_empty:NTF \l_miguel_label_tl
   {
    \miguel_includegraphics:Vn \l_miguel_label_clist { #2 }
   }
   {
    \SetHorizontalCoffin\imagecoffin
     {
      \miguel_includegraphics:Vn \l_miguel_label_clist { #2 }
     }
    \SetHorizontalCoffin\labelcoffin
     {
      \raisebox{\depth}
       {
        \bool_if:NTF \l_miguel_label_box_bool
         { \fcolorbox{white}{white}{\l_miguel_label_size_tl\l_miguel_label_tl} }
         { \l_miguel_label_size_tl\l_miguel_label_tl }
       }
     }
    \SetVerticalPole\imagecoffin{left}{3pt+\CoffinWidth\labelcoffin/2}
    \SetVerticalPole\imagecoffin{right}{\Width-15pt-\CoffinWidth\labelcoffin/2}
    \SetHorizontalPole\imagecoffin{up}{\Height-3pt-\CoffinHeight\labelcoffin/2}
    \SetHorizontalPole\imagecoffin{down}{30pt+\CoffinHeight\labelcoffin/2}
    \use:x{\JoinCoffins\imagecoffin[\l_miguel_label_pos_tl]\labelcoffin[vc,hc]} 
    \TypesetCoffin\imagecoffin
   }
   \group_end:
 }

\NewDocumentCommand{\setlabel}{m}
 {
  \keys_set:nn { miguel/label } { #1 }
 }

\cs_new_protected:Nn \miguel_includegraphics:nn
 {
  \includegraphics[#1]{#2}
 }
\cs_generate_variant:Nn \miguel_includegraphics:nn { V }

\ExplSyntaxOff

\begin{document}


\title{Measurement of charged-current $\nu_\mu$ and $\bar{\nu}_\mu$ cross sections on hydrocarbon in a shallow inelastic scattering region} 

\newcommand{\Rutgers}{Rutgers, The State University of New Jersey, Piscataway, New Jersey 08854, USA}
\newcommand{\Hampton}{Hampton University, Dept. of Physics, Hampton, VA 23668, USA}
\newcommand{\Dortmund}{Institute of Physics, Dortmund University, 44221, Germany }
\newcommand{\Otterbein}{Department of Physics, Otterbein University, 1 South Grove Street, Westerville, OH, 43081 USA}
\newcommand{\JMU}{James Madison University, Harrisonburg, Virginia 22807, USA}
\newcommand{\Florida}{University of Florida, Department of Physics, Gainesville, FL 32611}
\newcommand{\UCIrvine}{Department of Physics and Astronomy, University of California, Irvine, Irvine, California 92697-4575, USA}
\newcommand{\CBPF}{Centro Brasileiro de Pesquisas F\'{i}sicas, Rua Dr. Xavier Sigaud 150, Urca, Rio de Janeiro, Rio de Janeiro, 22290-180, Brazil}
\newcommand{\PUCP}{Secci\'{o}n F\'{i}sica, Departamento de Ciencias, Pontificia Universidad Cat\'{o}lica del Per\'{u}, Apartado 1761, Lima, Per\'{u}}
\newcommand{\INRM}{Institute for Nuclear Research of the Russian Academy of Sciences, 117312 Moscow, Russia}
\newcommand{\Jlab}{Jefferson Lab, 12000 Jefferson Avenue, Newport News, VA 23606, USA}
\newcommand{\Pittsburgh}{Department of Physics and Astronomy, University of Pittsburgh, Pittsburgh, Pennsylvania 15260, USA}
\newcommand{\Guanajuato}{Campus Le\'{o}n y Campus Guanajuato, Universidad de Guanajuato, Lascurain de Retana No. 5, Colonia Centro, Guanajuato 36000, Guanajuato M\'{e}xico.}
\newcommand{\Athens}{Department of Physics, University of Athens, GR-15771 Athens, Greece}
\newcommand{\Tufts}{Physics Department, Tufts University, Medford, Massachusetts 02155, USA}
\newcommand{\WM}{Department of Physics, William \& Mary, Williamsburg, Virginia 23187, USA}
\newcommand{\FNAL}{Fermi National Accelerator Laboratory, Batavia, Illinois 60510, USA}
\newcommand{\Purdue}{Department of Chemistry and Physics, Purdue University Calumet, Hammond, Indiana 46323, USA}
\newcommand{\MCLA}{Massachusetts College of Liberal Arts, 375 Church Street, North Adams, MA 01247}
\newcommand{\UMD}{Department of Physics, University of Minnesota -- Duluth, Duluth, Minnesota 55812, USA}
\newcommand{\Northwestern}{Northwestern University, Evanston, Illinois 60208}
\newcommand{\UNI}{Facultad de Ciencias F\'{i}sicas, Universidad Nacional Mayor de San Marcos, CP 15081, Lima, Per\'{u}}
\newcommand{\Rochester}{Department of Physics and Astronomy, University of Rochester, Rochester, New York 14627 USA}
\newcommand{\Austin}{Department of Physics, University of Texas, 1 University Station, Austin, Texas 78712, USA}
\newcommand{\USM}{Departamento de F\'{i}sica, Universidad T\'{e}cnica Federico Santa Mar\'{i}a, Avenida Espa\~{n}a 1680 Casilla 110-V, Valpara\'{i}so, Chile}
\newcommand{\Geneva}{University of Geneva, 1211 Geneva 4, Switzerland}
\newcommand{\Chicago}{Enrico Fermi Institute, University of Chicago, Chicago, IL 60637 USA}
\newcommand{\hired}{}
\newcommand{\OregonState}{Department of Physics, Oregon State University, Corvallis, Oregon 97331, USA}
\newcommand{\oxford}{Oxford University, Department of Physics, Oxford, OX1 3PJ United Kingdom}
\newcommand{\umiss}{University of Mississippi, Oxford, Mississippi 38677, USA}
\newcommand{\upenn}{Department of Physics and Astronomy, University of Pennsylvania, Philadelphia, PA 19104}
\newcommand{\AMU}{Department of Physics, Aligarh Muslim University, Aligarh, Uttar Pradesh 202002, India}
\newcommand{\wroclaw}{University of Wroclaw, plac Uniwersytecki 1, 50-137 Wroa\l{}aw, Poland}
\newcommand{\Mohali}{Department of Physical Sciences, IISER Mohali, Knowledge City, SAS Nagar, Mohali - 140306, Punjab, India}
\newcommand{\CINVESTAV}{Departamento de Fisica Col. San Pedro Zacatenco, 07360 Mexico, DF, Av. Instituto PolitÃ©cnico Nacional, Mexico}
\newcommand{\york}{York University, Department of Physics and Astronomy, Toronto, Ontario, M3J 1P3 Canada}
\newcommand{\ND}{Department of Physics and Astronomy, University of Notre Dame, Notre Dame, Indiana 46556, USA}
\newcommand{\ICL}{The Blackett Laboratory,  Imperial College London,  London SW7 2BW, United Kingdom}
\newcommand{\warwick}{Department of Physics, University of Warwick, Coventry, CV4 7AL, UK}
\newcommand{\qmul}{G O Jones Building, Queen Mary University of London, 327 Mile End Road, London E1 4NS, UK}

\newcommand{\adrianThanks}{Now at Department of Physics, Drexel University, Philadelphia, Pennsylvania 19104, USA}
\newcommand{\ricfregianThanks}{now at Department of Physics and Astronomy, University of California at Davis, Davis, CA 95616, USA}
\newcommand{\mascencioThanks}{Now at Iowa State University, Ames, IA 50011, USA}
\newcommand{\finerThanks}{Now at Los Alamos National Laboratory, Los Alamos, New Mexico 87545, USA}
\newcommand{\kleykampThanks}{now at Department of Physics and Astronomy, University of Mississippi, Oxford, MS 38677}
\newcommand{\bamThanks}{Now at University of Minnesota, Minneapolis, Minnesota 55455, USA}
\newcommand{\byaeggyThanks}{Now at Department of Physics, University of Cincinnati,  Cincinnati, Ohio 45221, USA}

\author{A.~Lozano}\thanks{\adrianThanks}  \affiliation{\CBPF}
\author{G.~Silva}                         \affiliation{\CBPF}
\author{G.~Caceres}\thanks{\ricfregianThanks}  \affiliation{\CBPF}
\author{S.~Akhter}                        \affiliation{\AMU}
\author{Z.~~Ahmad~Dar}                    \affiliation{\WM}  \affiliation{\AMU}
\author{V.~Ansari}                        \affiliation{\AMU}
\author{M.~V.~Ascencio}\thanks{\mascencioThanks}  \affiliation{\PUCP}
\author{M.~Sajjad~Athar}                  \affiliation{\AMU}
\author{J.~L.~Bonilla}                    \affiliation{\Guanajuato}
\author{A.~Bravar}                        \affiliation{\Geneva}
\author{G.A.~D\'{i}az~}                   \affiliation{\FNAL}  \affiliation{\Rochester}
\author{H.~da~Motta}                      \affiliation{\CBPF}
\author{J.~Felix}                         \affiliation{\Guanajuato}
\author{L.~Fields}                        \affiliation{\ND}
\author{R.~Fine}\thanks{\finerThanks}     \affiliation{\Rochester}
\author{A.M.~Gago}                        \affiliation{\PUCP}
\author{H.~Gallagher}                     \affiliation{\Tufts}
\author{P.K.Gaur}                         \affiliation{\AMU}
\author{R.~Gran}                          \affiliation{\UMD}
\author{E.Granados}                       \affiliation{\Guanajuato}  \affiliation{\Guanajuato}
\author{D.A.~Harris}                      \affiliation{\york}  \affiliation{\FNAL}
\author{J.~Kleykamp}\thanks{\kleykampThanks}  \affiliation{\Rochester}
\author{A.~Klustov\'{a}}                  \affiliation{\ICL}
\author{M.~Kordosky}                      \affiliation{\WM}
\author{D.~Last}                          \affiliation{\Rochester}  \affiliation{\upenn}
\author{X.-G.~Lu}                         \affiliation{\warwick}  \affiliation{\oxford}
\author{S.~Manly}                         \affiliation{\Rochester}
\author{W.A.~Mann}                        \affiliation{\Tufts}
\author{C.~Mauger}                        \affiliation{\upenn}
\author{K.S.~McFarland}                   \affiliation{\Rochester}
\author{B.~Messerly}\thanks{\bamThanks}   \affiliation{\Pittsburgh}
\author{O.~Moreno}                        \affiliation{\WM}  \affiliation{\Guanajuato}
\author{J.G.~Morf\'{i}n}                  \affiliation{\FNAL}
\author{J.K.~Nelson}                      \affiliation{\WM}
\author{C.~Nguyen}                        \affiliation{\Florida}
\author{A.~Olivier}                       \affiliation{\ND}  \affiliation{\Rochester}
\author{V.~Paolone}                       \affiliation{\Pittsburgh}
\author{G.N.~Perdue}                      \affiliation{\FNAL}  \affiliation{\Rochester}
\author{C.~Pernas}                        \affiliation{\WM}
\author{K.-J.~Plows}                      \affiliation{\oxford}
\author{M.A.~Ram\'{i}rez}                 \affiliation{\upenn}  \affiliation{\Guanajuato}
\author{D.~Ruterbories}                   \affiliation{\Rochester}
\author{H.~Schellman}                     \affiliation{\OregonState}
\author{C.~J.~Solano~Salinas}             \affiliation{\UNI}
\author{D.~S.~Correia}                    \affiliation{\CBPF}
\author{M.~Sultana}                       \affiliation{\Rochester}
\author{V.S.~Syrotenko}                   \affiliation{\Tufts}
\author{A.V.~Waldron}                     \affiliation{\qmul}  \affiliation{\ICL}
\author{B.~Yaeggy}\thanks{\byaeggyThanks}  \affiliation{\USM}

\collaboration{The MINERvA Collaboration}
\noaffiliation
\date{\today}

\begin{abstract}
This MINERvA analysis is the first neutrino and antineutrino study of a shallow inelastic scattering region, which is the transition region between resonant production and deep inelastic scattering processes. This transition is explicitly included in this study by expanding the scope of shallow inelastic scattering to include not only the mainly lower-$Q^2$ nonresonant pion production but also the kinematic region where pion production off quarks within the nucleon becomes significant with $Q^2$ below the onset of the deep-inelastic scattering region defined in this analysis. To reduce the resonance background the kinematic region 1.5 $<$ $W$ $<$ 2 GeV/$c^2$ was chosen. In addition to the inclusive differential cross section measurements, to emphasize SIS interactions off quarks within the nucleon a sample with $Q^2$ $\geq$ 1 GeV/$c^2$ was also analyzed. The measurements of one-dimensional cross sections at $\left\langle E_\nu \right\rangle \sim 6$ GeV on hydrocarbon of $Q^2$, Bjorken x and muon momentum variables are compared with modified predictions from the GENIE 2 neutrino generator as well as predictions of other neutrino simulators GiBUU, NEUT, NuWro and an alternative GENIE 3 version. Significant discrepancies both in shape and magnitude between measurements and neutrino simulator predictions of all variables have been observed.
\end{abstract}

\maketitle

\section{Introduction}
In accelerator-based experimental studies of neutrino scattering, the interactions have been broadly classified with increasing hadronic mass ($W$) such as the quasielastic (QE) interactions \cite{Sobczyk:2011bi} and initial lower-$W$ resonance production dominated by the $\Delta$ (1232) resonance \cite{Paschos:2012va}. Although there has been rather extensive detailed experimental and theoretical \cite{AlvarezRuso2018} analyses of QE and $\Delta$(1232) production, the contribution of nonresonant meson (mainly pion) production and higher $W$ resonances have received minimal attention \cite{Sajjad2020}. 
  
With $W$ increasing beyond the $\Delta$(1232), the importance of including $Q^2$, the square of the four-momentum transferred from the (anti)neutrino to the nucleon, in the classification of interactions becomes important. Although resonance production across $Q^2$ remains important as a possible background, the present study emphasizes meson production in the kinematic region with sufficient $Q^2$ where interactions on quarks within the nucleon become significant. Consideration of the types of $Q^2$–dependent quark interactions is essential to carefully define Deep-Inelastic Scattering (DIS) as well as to then introduce the expanded definition of Shallow Inelastic Scattering (SIS) quark interactions.
  
The accurate understanding and subsequent modeling of this expanded SIS production is also important in order to reduce systematic uncertainties in the determination of oscillation probabilities for experiments such as the Deep Underground Neutrino Experiment (DUNE) \cite{DUNE2015} that will have around 50\% of events with $W$ greater than the $\Delta$(1232) resonance. The MINERvA experiment is the only neutrino experiment that has concentrated on this SIS kinematic region, presenting here the experimental analyses of neutrino and antineutrino interactions in a shallow inelastic scattering region on a hydrocarbon target.

\subsection{Main SIS kinematic variables}
The fundamental variables measured by the MINERvA experiment are the energy and angle of the outgoing (anti)muon relative to the (anti)neutrino direction, $E_\mu$ and $\theta_\mu$ as well as $E_{vis}$, the available (detectable) energy of the final state hadrons. The initial hadronic energy, $E_{had}$, is determined from $E_{vis}$ by correcting for missing energy. Using these fundamental variables the reconstructed energy of the incoming neutrino, $E_\nu$ $\equiv$ ($E_{had}$ + $E_\mu$) and the momentum of the outgoing muon, $p_\mu$, are obtained. These variables along with the nucleon mass, $M_N$, and the muon mass, $M_\mu$ yield the main measured variables of this study:
\begin{enumerate}[(i)]
    \item $Q^2$ - the square of the four-momentum transferred from the (anti)neutrino to the nucleon, 
        \begin{eqnarray}
            Q^2 &=& \frac{2E_{\nu}}{c} \left(\frac{E_{\mu}}{c} - p_\mu\cos\theta_\mu \right) - M^2_{\mu}c^2\,,\label{exp_vars1}
        \end{eqnarray}
    \item $W$ - the experimental invariant mass of the hadronic system assuming  the target nucleon is at rest in the lab frame,
        \begin{eqnarray}
            W &=& \frac{1}{c} \sqrt{M_N^2c^2+2E_{had}M_N-Q^{2}}\,,\label{exp_vars2}
        \end{eqnarray}
     \item $x$ - the Bjorken scaling variable representing the distribution of quarks within the nucleon. Although $x$ is normally a higher $Q^2$ variable, the predicted $x$ distributions in this analysis already contain consideration of low $Q^2$ effects (see section \ref{Sec:SISregion}),
        \begin{eqnarray}
            x_{Bj} &=& \frac{Q^2}{2M_N E_{had}}\ \equiv{x},\label{exp_vars5}
        \end{eqnarray}
    \item ${p_{||}}_\mu$ and ${p_\mathrm{T}}_\mu$ - the parallel and transverse components of the outgoing (anti)muon momentum with respect to the (anti)neutrino direction, 
        \begin{eqnarray}
            {{p_{||}}_\mu} &=& p_\mu \cos \theta_\mu\,,\label{exp_vars3}\\ 
            {{p_\mathrm{T}}_\mu} &=& p_\mu \sin \theta_\mu\,.\label{exp_vars4}
        \end{eqnarray}
\end{enumerate}   

In addition, the analysis requires two versions of the kinematic variables \ref{exp_vars1}-\ref{exp_vars4}: the reconstructed and the true versions extracted from event generator simulations. The reconstructed version uses the measured values of these fundamental variables. The true version, on the other hand, is computed using simulated values of the fundamental variables without considering detector and measurement effects. 

\subsection{The Shallow Inelastic Scattering Region}
\label{Sec:SISregion}
The definition of Shallow Inelastic Scattering depends primarily on $Q^2$ with SIS interactions occurring across a wide range of $W$. In addition to nonresonant pion production, the mainly lower $Q^2$ component of SIS, there is a higher $Q^2$ component of SIS involving interactions off quarks within the nucleon. This higher $Q^2$ SIS region begins with sufficient $Q^2$ for initial interactions off quarks within the nucleon consisting of, for example, multi-quark or multidimensional (i.e. including transverse component) non-perturbative Quantum Chromodynamic (QCD) quark interactions. This SIS region proceeds with increasing $Q^2$ until it reaches the rigorous DIS requirement roughly defined by the global QCD parton distribution fitting community as sufficient $Q^2$ for perturbative QCD interaction within the nucleon, which is, for example, $Q^2$ $\geq$ 4 GeV$^2/c^2$ according to the CTEQ global QCD fitting group \cite{Muzakka:2022wey}.

Mostly due to the challenge of accumulating statistics, the neutrino experimental community has often defined ``DIS" as the point where interactions could begin to occur within the nucleon at roughly $Q^2$ $\geq$ 1 GeV$^2/c^2$. This is well below the perturbative QCD defined DIS interaction region.  It is rather the start of the expanded SIS multi-quark region. Analysis of this SIS non-perturbative QCD region, that has received minimal detailed experimental and theoretical study, includes effects such as Target Mass Corrections (TMC) \cite{Ruiz:2023ozv} and higher twist (HT) \cite{Braun:2022gzl} and will be referred to as the ``multi-quark" SIS region in this study. That this lower restriction on $Q^2$ to be greater than 1 GeV$^2/c^2$ necessitates the inclusion of such multi-quark SIS effects in global analyses is demonstrated by the nCTEQ global analysis \cite{Segarra:2020gtj} to lower $Q^2$/lower $W$ data.

With SIS interactions being primarily a function of $Q^2$, the inclusion of $W$ dependence in the study of SIS enters mainly to minimize resonance contribution to the studied region. Although resonance production decreases with increasing $Q^2$ and $W$, there is still a background of nucleon resonances above the $\Delta$(1232) and their observed decay chains in the $W$ region (1.5 $< W <$ 2.0 GeV/$c^2$) selected for this higher $Q^2$ SIS multi-quark study.  

Since there are no experimental measurements of resonance production on hydrocarbon in the $W$ - $Q^2$ region of this study, neutrino event simulator predictions that give model-dependent approximations of the resonance background in the chosen kinematic region will be summarized.

Experimentally, it has not been possible to distinguish individual resonant from either nonresonant or SIS multi-quark pion production events. Consequently, the SIS region and transition to DIS can best be experimentally studied in terms of inclusive production within kinematic cuts that emphasize SIS and minimize the resonance and DIS backgrounds. Considering the kinematics of SIS events, to better define the contribution of the SIS signal, we restrict $W$ to 1.5 $< W < $ 2 GeV/$c^2$. The $W$ greater than 1.5 GeV/$c^2$ was chosen to decrease the probability that the major contributor to resonance pions, $\Delta$(1232) \cite{MINERvA:2016sfc} is included, while the $W$ less than 2 GeV/$c^2$ helps isolate this study from ongoing MINERvA and community studies in the $W > $ 2 GeV/$c^2$ region. It should be emphasized that SIS events can and do occur across the $W$ spectrum and not only in the W range chosen for this study. We further divide this $W$ restricted region into:
\begin{enumerate}[(i)]
    \item SIS definition I - events with $Q^2$ $\geq$ 0 GeV$^2$/$c^2$ that include all nonresonant and multi-quark SIS production as well as resonant background.
    \item SIS definition II - events with $Q^2$ $\geq$ 1 GeV$^2$/$c^2$ that eliminate the low $Q^2$ resonant meson background while emphasizing the multi-quark SIS contribution. 
\end{enumerate}

\begin{figure}[htbp]
    \centering
    \includegraphics[scale=0.45]{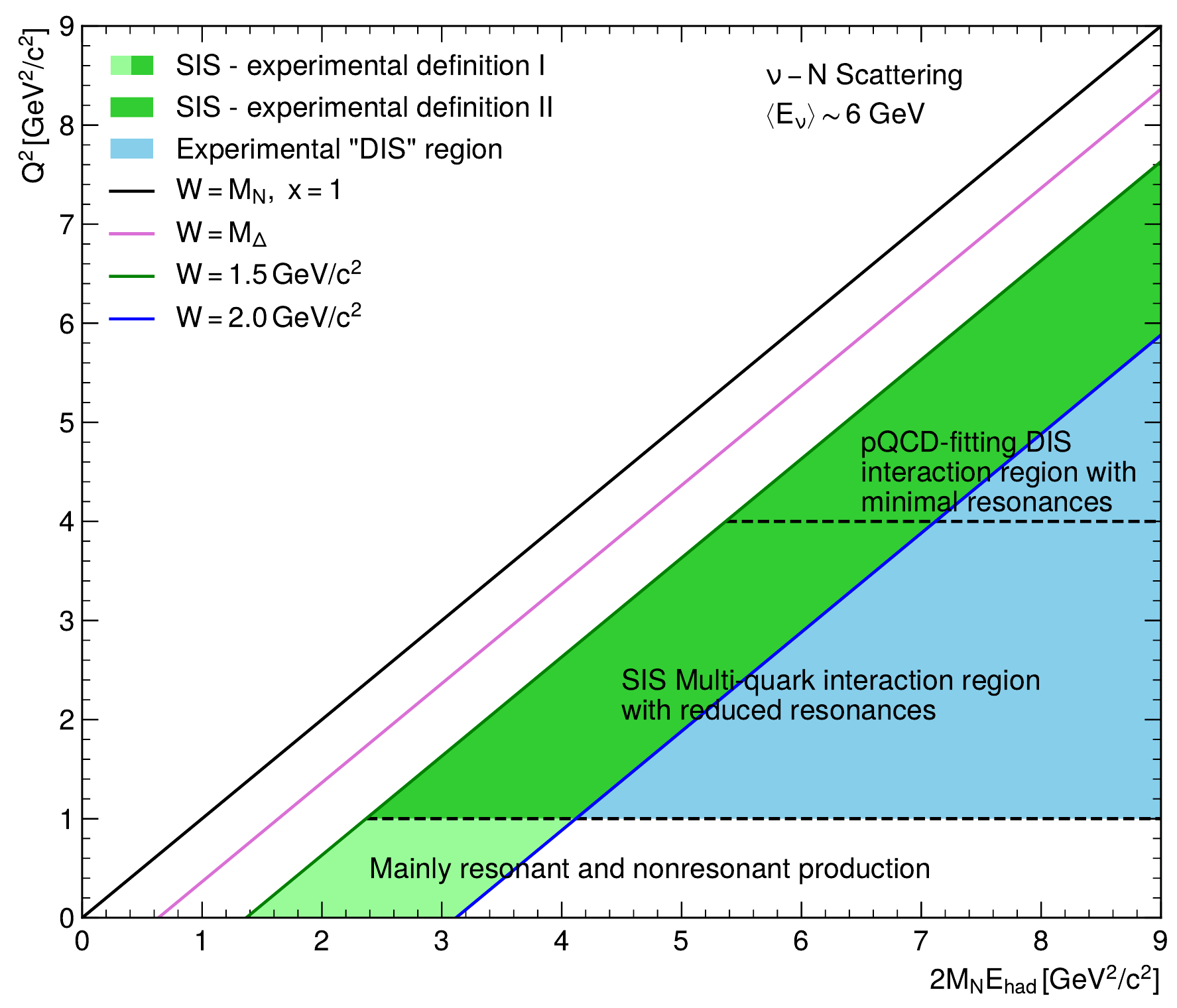}
    \caption{SIS experimental regions I (light plus dark green) and II (dark green only) defined for this study and their relation with other community kinematic regions. The dominant types of interaction for different ranges of $Q^2$ are also indicated. The $Q^2$ values of events in this study are mainly below 4 GeV$^2$/c$^2$.}
\label{sis_definition_6}
\end{figure}

Figure \ref{sis_definition_6} plots the numerator and denominator of Bjorken $x$ (Eq. 3) on separate axes and illustrates how these SIS regions relate with other kinematic regions defined by the neutrino-scattering community. The neutrino flux and accumulated statistics of this restricted $W$ study limit the $Q^2$ values of events to mainly below 4 GeV$^2$/c$^2$.

The SIS definition I and II measured experimental results are compared with the neutrino event simulators the community employs. Currently, several MC generators have been developed within the experimental community, such as GENIE \cite{GENIE:2022qrc}, NEUT \cite{Hayato:2021heg}, NuWro \cite{Golan:2012rfa}, and the GiBUU nuclear transport model \cite{Soplin:2023nxg}. These simulation programs each have variations of a nuclear interaction model that contains the initial state nuclear medium effects, the knowledge of the initial $\nu_l/\bar\nu_l$ - nucleon cross sections, and the final state interactions of the produced hadrons within the nucleus. Although they have mostly different kinematic ranges and methods to define the components of the interaction, one common method for the considered MC generators, such as GENIE, is the use of the Bodek-Yang (B-Y) \cite{Bodek:2010km} model to extend the predicted total cross sections to low $Q^2$ and low $W$. For comparisons with measured data it is important to note that this B-Y extension already includes corrections for low-$Q^2$ target mass and higher-twist effects in predicted kinematic variables.

In this analysis, the SIS signal is considered as (anti)neutrino charged-current interactions on a hydrocarbon target, satisfying the SIS definitions I or II, and having $\theta_\mu \leq 20^\circ$ and $2 < E_\mu < 20$ GeV. The (anti)muon kinematic requirements included in the definition are due to detector and flux restrictions as described in section \ref{Sec:SRdefinition}.

\section{The MINERvA Experiment}
MINERvA is a dedicated (anti)neutrino scattering experiment whose main goal is to understand the nature of (anti)neutrino-nucleus interactions. To this end, it studies the nuclear effects and tests the models involved in the simulation of exclusive and inclusive final states. The MINERvA detector sat on the axis of the main injector neutrino beam (NuMI) produced at Fermilab, just upstream of the MINOS (Main Injector Neutrino Oscillation Search) \cite{MINOS:2008hdf} near detector, which was used as muon spectrometer for particles leaving the MINERvA detector that enter MINOS.

\begin{figure}[htbp]
   \centering
   \includegraphics[width=0.23\textwidth]{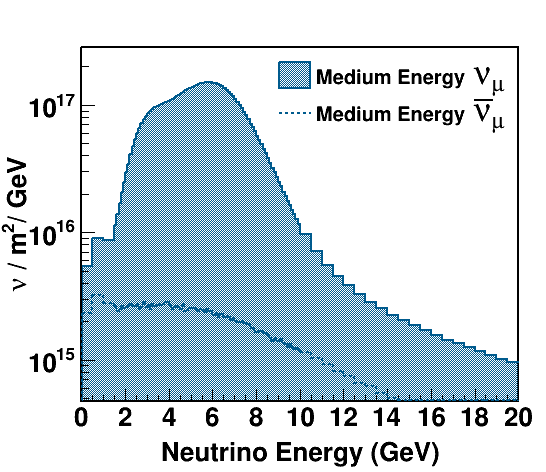}
   \includegraphics[width=0.23\textwidth]{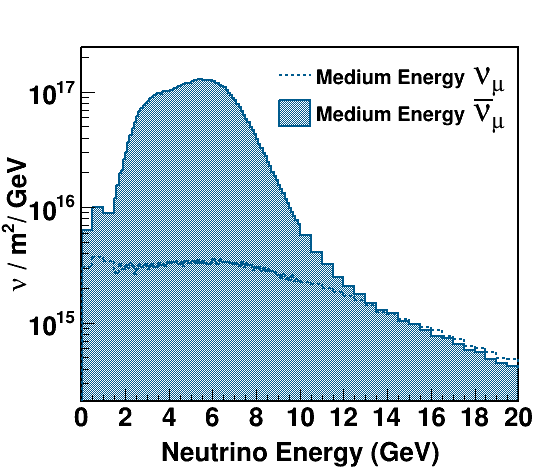}
   \caption{Medium Energy fluxes in the neutrino (left) and antineutrino (right) focused mode at MINERvA.}
   \label{nu_energy}
\end{figure}

The MINERvA experiment used a fine-grained tracking detector to record interactions of neutrinos coming from the high intensity NuMI beamline at Fermilab \cite{Adamson2016}. The neutrino beam results from the decays of pions and kaons produced by collisions of 120 GeV/$c$ protons from the Main Injector onto a graphite target. Two magnetic ``horns" focus the positive (negative) mesons that produce a $\nu_\mu$ ($\bar{\nu}_\mu$) beam \cite{Anderson1998, Zwaska2005} upon decay. 
This analysis uses data from the Medium Energy (ME) NuMI beam, which has a $\nu_\mu$ or $\overline{\nu}_\mu$ flux peaked near 6 GeV as can be seen in Figure \ref{nu_energy}. The protons on target (POT) used for this analysis add up to a total of $1.06 \times 10^{\text{21}}$ with the polarity of the NuMI horns set to produce a beam almost exclusively of muon neutrinos, and exposure of $1.12 \times 10^{\text{21}}$ POT to produce a beam of mostly muon antineutrinos.

The MINERvA detector consisted of an upstream region of nuclear targets intermixed with scintillator planes followed by a region of purely polystyrene scintillator planes called the active tracker region, surrounded by calorimeters. This analysis includes only those interactions in this active tracking region with a fiducial mass of 5.48 tons, consisting of 62 tracking modules with two scintillator planes per module and each scintillator plane is comprised of 127 scintillating plane strips. The scintillator plane strips are oriented in three different directions, $0^\circ$ and $\pm 60^\circ$ relative to the vertical axis of the detector to provide a three-dimensional reconstruction. Wavelength-shifting fibers embedded in the scintillator strips are read out by optical cables that are connected to photomultiplier tubes. The photomultiplier tubes read out the scintillation light and the associated electronics achieve 3 ns timing resolution per hit. The downstream electromagnetic calorimeter is made up of layers of 2 mm thick lead planes with scintillator planes. The downstream and side hadronic calorimeter is made up of alternating 2.54 cm thick steel planes and scintillator planes.

The MINOS near detector, situated two meters downstream of the MINERvA detector, served as a magnetized muon spectrometer. It was essential for the detection and reconstruction of charged-current interactions in the MINERvA experiment. Muon tracks which exit the downstream end of MINERvA are matched to tracks in the MINOS near detector. The events used in this analysis are required to be within the MINOS acceptance, a condition that restricts the sample to events with $\theta_\mu \leq 20^\circ$.

\section{Event Simulation}
\label{Simulation Experiment}
The simulation of events in the MINERvA experiment is divided into three stages: neutrino flux, neutrino interactions with the nucleon in the nuclear environment plus final state interactions (FSI), and interactions with the components of the MINERvA detector.

\subsection{Neutrino Flux Simulation}
The neutrino flux entering the MINERvA detector is predicted by a simulation of the NuMI beamline based on GEANT4 \cite{Agostinelli2003}. In the interest of improving the GEANT4 prediction, the hadron production model is constrained \cite{AliagaThesis2016, Aliaga2016} using external data from relevant hadron production measurements such as the NA49 experiment \cite{NA49exp2007} at CERN that uses a thin carbon target with an incident proton momentum of 158 GeV/c. In addition to these hadron model corrections, the neutrino flux prediction has been constrained by measuring processes with precisely known cross sections such as neutrino-electron elastic scattering \cite{Valencia2019} and inverse muon decay reactions \cite{RuterboriesIMD2021}. The combined constraints reduce the uncertainty of the flux for neutrinos of energy between 2 and 20 GeV from 7.62\% (7.76\%) to 3.27\% (4.66\%) during the $\nu_\mu$ ($\bar{\nu}_\mu$) mode operation \cite{Zazueta2023}.

\subsection{Neutrino Interaction Simulation and MINERvA Tunes}
Neutrino interactions are simulated using the GENIE neutrino event generator \cite{Andreopoulos2010} version 2.12.6 with additions and tunes. The physics models used in the simulation can be broadly categorized into nuclear physics models, cross section models, and hadronization models. As a nuclear physics model, GENIE uses the Bodek and Ritchie version of the relativistic Fermi gas (RFG) nuclear model \cite{Bodek:1980ar}, which incorporates a high momentum tail to the initial nucleon momentum distribution to represent short-range nucleon-nucleon correlations (SRC) \cite{LlewellynSmith1972}. Nuclear reinteractions, or final state interactions, of the hadrons produced within the nucleus are modeled using the INTRANUKE-hA package \cite{Dytman:2007zz}. This option selects a single kinetic-energy dependent rescattering fate (including no rescattering) for each nucleon and pion as an approximation of a full cascade simulation.

The cross section model provides the calculation of the differential and total cross sections of (anti)neutrino-nucleon interactions. Quasielastic (1 particle 1 hole - 1p1h) interactions are modeled using an implementation of the Llewellyn-Smith model \cite{LLEWELLYNSMITH1972261} with the vector form factors modeled using the BBBA05 model parameters \cite{BRADFORD2006127}. The dipole form is assumed for the axial vector form factor with an axial mass of $M_A$ = 0.99 GeV/c$^2$. Meson exchange current (MEC) reactions (2 particles 2 holes - 2p2h) are implemented using the Valencia model with a kinematic cutoff at 1.2 GeV/c of momentum transfer \cite{schwehr2017genie}. The model version ported to GENIE v2 is equivalent to the one used in the latest version of GENIE.

Baryon resonance production in neutral and charged current channels is simulated using the Rein-Sehgal (R-S) model \cite{Rein1981}. This version uses the original dipole form factors and an axial mass $M_A^{RES}$ of 1.12 GeV/c$^2$ tuned to bubble chamber data. A total of 16 baryon resonances that mainly decay to a single pion are included in the R-S model. However, resonance and nonresonance production above $W$ = 1.7 GeV/$c^2$ and most resonance multi-pion decays as well as the interference between neighboring resonances are not considered by the R-S simulation in this version of GENIE.

Nonresonance production, multi-pion resonance decays, interference between resonances, shallow and deep inelastic scattering are simulated using an effective leading order model based on a modified B-Y prescription \cite{Bodek2003}, where HT and TMC are accounted for through the use of a new scaling variable and modifications are made to the low $Q^2$ parton distributions. This B-Y based model is referred to by GENIE as ``DIS". This ``GENIE DIS" definition is not the experimental community definition of DIS that depends on higher $Q^2$ and $W$ but rather uses the same abbreviation ``DIS" to refer to the inclusive production across all $Q^2$ and $W$ $>$ $M_N + m_\pi$, with $m_\pi$ being the pion mass. Which ``DIS" is being considered in this study will be clearly designated.

Within the GENIE simulation, the charged current neutrino nucleon inelastic differential cross section of the resonance and GENIE DIS region can be expressed as:
\begin{equation}
    \frac{d^2 \sigma^{inel}}{d Q^2 dW} = \sum_k \left(\frac{d^2 \sigma^{R-S}}{d Q^2 dW}\right)_k +  \frac{d^2 \sigma^{DIS}}{d Q^2 dW}\,.
    \label{eq.RESRS}
\end{equation}

The first term represents the contribution of the inelastic channels that proceed via resonance production. They are mainly single-pion decays, but higher mass resonances have multi-pion and kaons, etas, and hyperons in their decay chain following estimates from the particle data group. Some resonances decay to a photon and a nucleon. The k index in the resonance term runs over all 16 baryon resonances up to 1.7 GeV/$c^2$ in GENIE. The second term represents the contribution from the GENIE DIS interactions, in which the part of the total differential cross section with $W$ $<$ 1.7 GeV/c$^2$ is modulated to agree with inclusive, 1-pion, and 2-pion cross section data to avoid double counting the resonances.

\begin{figure}[!htbp]
    \centering
    \includegraphics[width=0.23\textwidth]{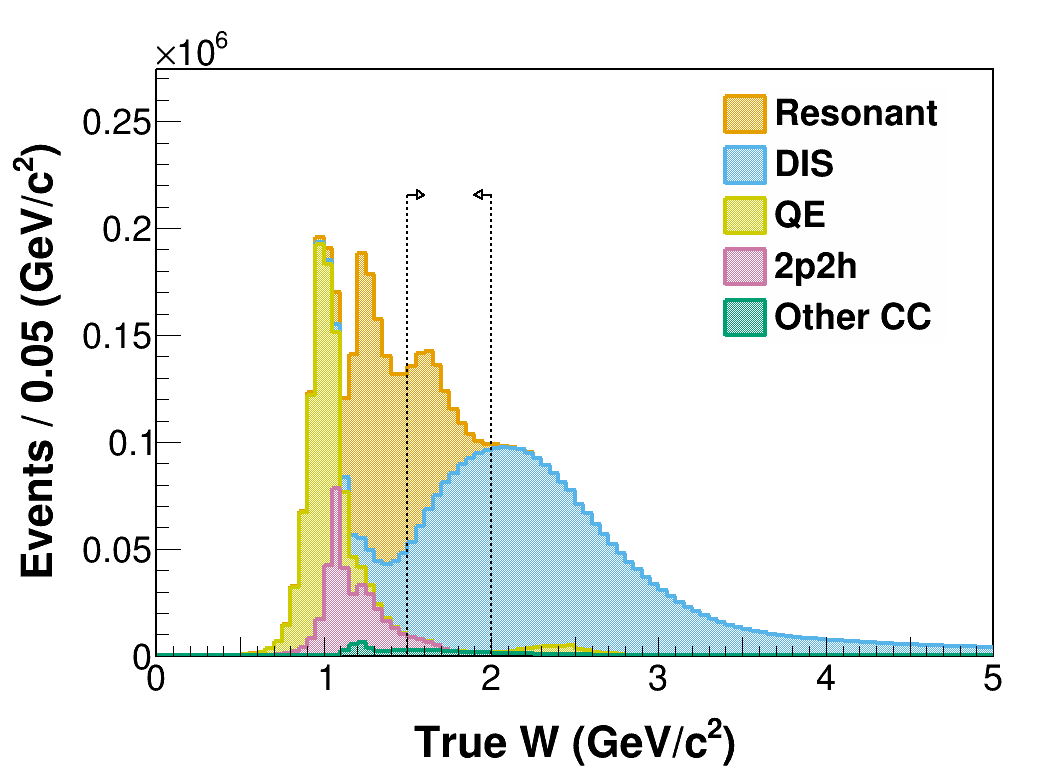}
    \includegraphics[width=0.23\textwidth]{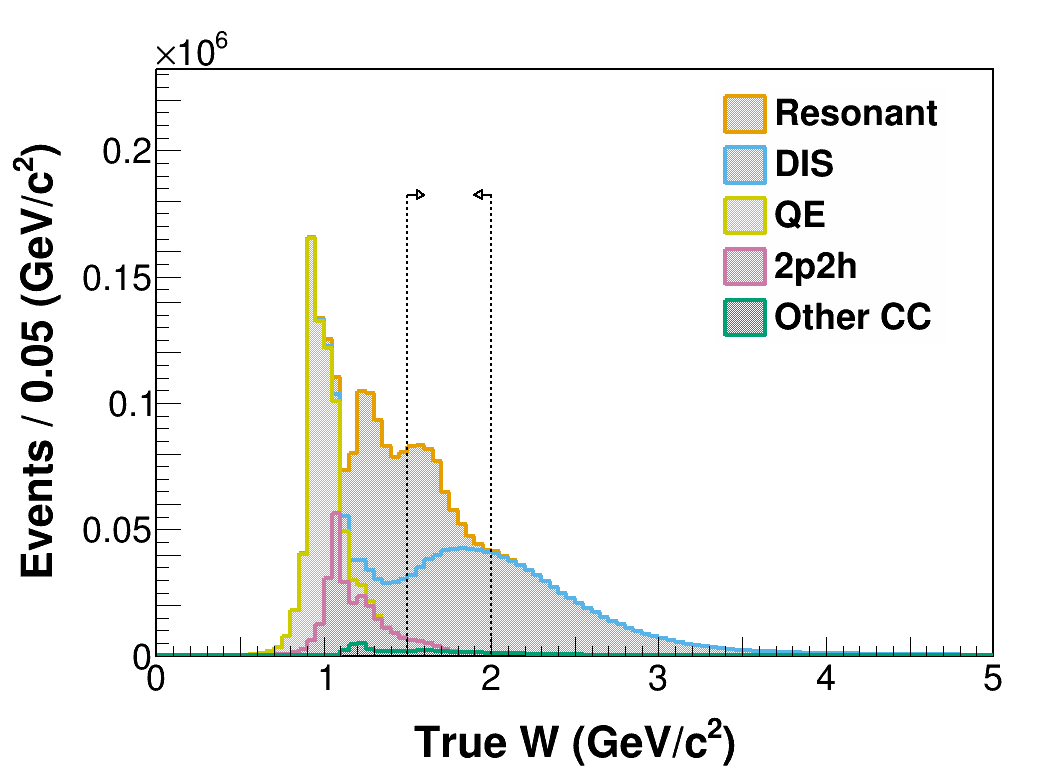}

    \includegraphics[width=0.23\textwidth]{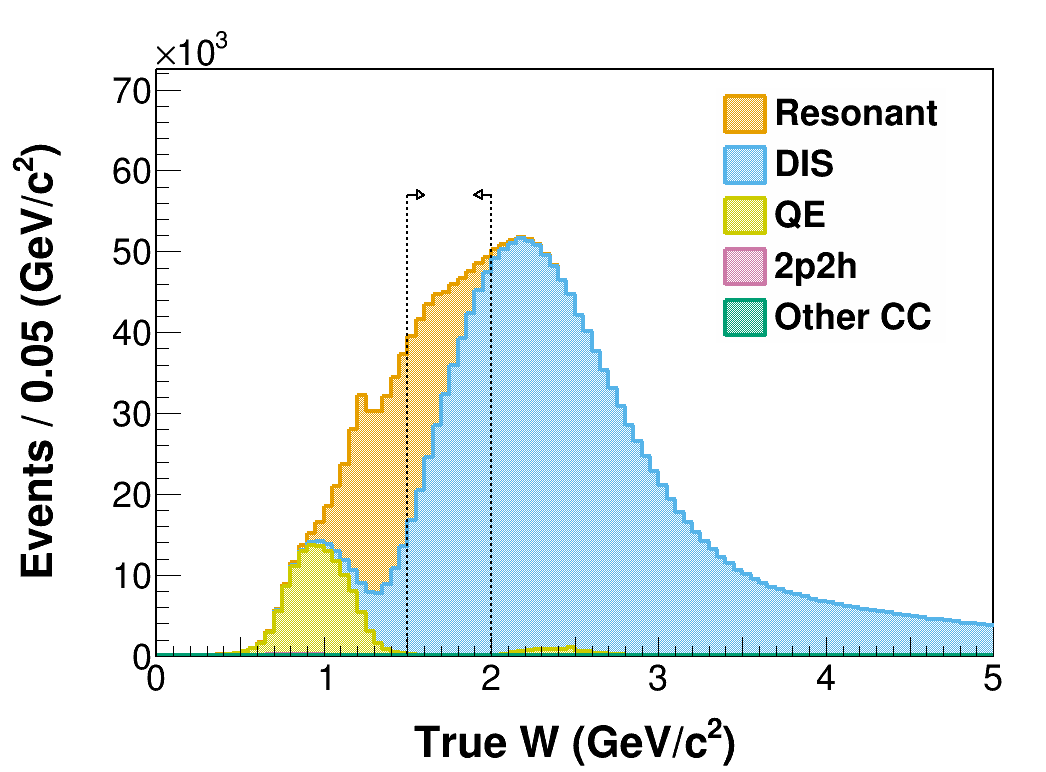}
    \includegraphics[width=0.23\textwidth]{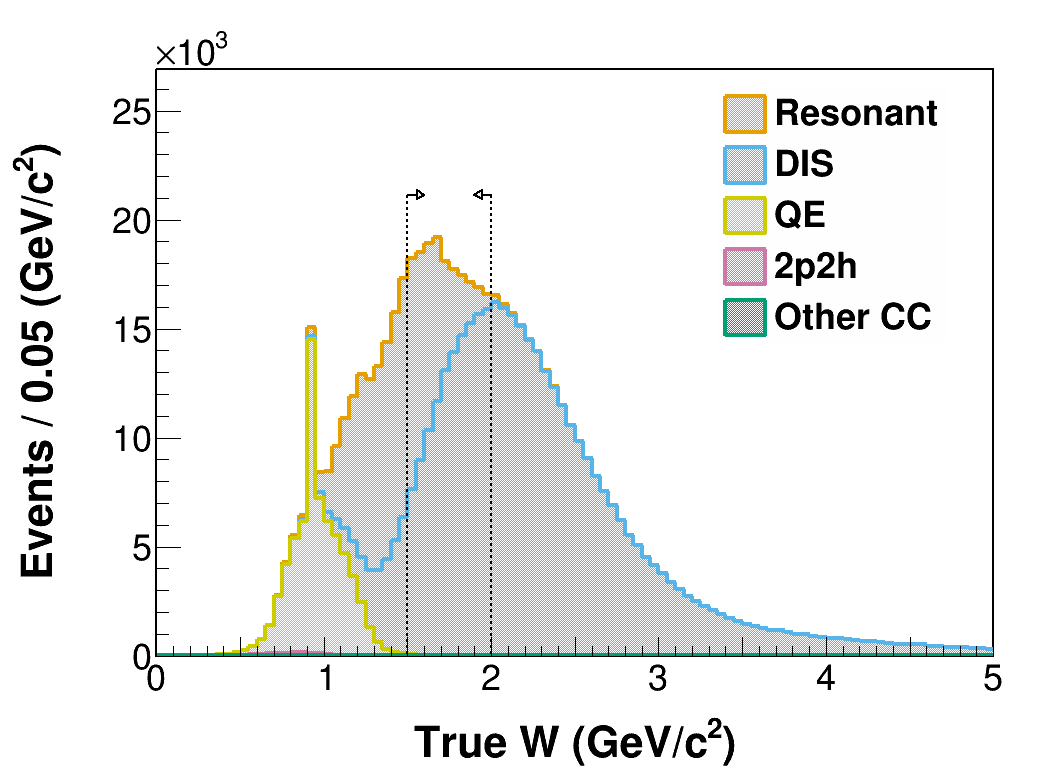}

    \caption{The true $W$ distribution for the inclusive sample, divided into GENIE components. The region between the arrows corresponds to SIS signal events. Top: SIS definition I. Bottom: SIS definition II. Left: Neutrino. Right: Antineutrino.}
    \label{genie_components}
\end{figure}

The GENIE 2.12.6 cross section models are ``tuned" by the MINERvA collaboration incorporating new deuterium analyses and the results from their measurements using a lower energy neutrino beam.
The simulation of QE events was improved by including the random phase approximation (RPA) effect described by the Valencia model \cite{PhysRevC.70.055503, gran2017model} for a Fermi gas \cite{PhysRevC.94.015501, NIEVES2017455}, while the simulation of 2p2h events is improved by fitting the simulation with the data in the low three-momentum transfer analyses \cite{PhysRevLett.116.071802, PhysRevLett.120.221805}. This enhances the rate in the region between the QE and $\Delta$(1232) peaks. The nonresonant pion production simulated by the GENIE DIS component with $W$ less than 2 GeV/$c^2$ was reduced by 43\% based on the fits made in \cite{Rodrigues2016}. However, the fit was made to neutrino data with lower $W$ and $Q^2$ and extrapolated to SIS kinematics. To reduce the tensions between the simulation of baryon resonance production and data found in MINERvA and other experiments at low $Q^2$, an analysis was performed to tune the GENIE pion production model with data using an empirical function to describe the shape of this low $Q^2$ resonant suppression \cite{Stowell2019}. The version of the simulation including all modifications except the low $Q^2$ resonant suppression is referred to as MINERvA Tune v1. The version including the low $Q^2$ resonant suppression is referred to as MINERvA Tune v2. The resulting distributions of events in true $W$ after applying MINERvA Tune v2 are shown in Figure \ref{genie_components}. The main physical categories of events are explicitly depicted in these distributions based on simulation. This information is important for the background estimate described in section \ref{Sec:Bkg}. Recall that the true $W$ is computed using the true values of the fundamental variables of simulated events without considering the detector and measurement resolution effects. The lower-$W$ spike in the definition II antineutrino distribution of Figure \ref{genie_components} is due to the QE interaction of antineutrinos with the H component of hydrocarbon (CH).

The AGKY model \cite{Yang:2009zx} is the default hadronization model in GENIE. At high hadronic invariant masses, it consists of the PYTHIA/JETSET \cite{Torbjorn_Sjostrand_2006} model, which is based on the Lund string fragmentation framework \cite{ANDERSSON198331}, to simulate the neutrino-induced hadronic showers. Since it is known that the validity of the PYTHIA/JETSET model decreases as $W$ decreases to the experimental DIS threshold, the low $W$ invariant mass hadronization is carried out by a phenomenological description based on Koba-Nielsen-Olesen (KNO) scaling \cite{KOBA1972317}. The AGKY effort then further tunes the pion multiplicity to bubble chamber data. The transition from the KNO-based model to the PYTHIA/JETSET model is done gradually by using an invariant mass transition window $[W_\mathrm{min}, W_\mathrm{max}]$ over which the fraction of neutrino events for which the hadronization is performed by the PYTHIA/JETSET model increase linearly from $0\%$ at $W_\mathrm{min}$ to $100\%$ at $W_\mathrm{max}$. In this way, the continuity of all simulated observables is ensured, before additional tunings are performed. The default values used in the AGKY model are $W_\mathrm{min} = 2.3\ \mathrm{GeV/c^2}$ and $W_\mathrm{max} = 3.0\ \mathrm{GeV/c^2}$. 

\subsection{Detector Simulation}
The last step consists of simulating the interaction of the particles with the detector. These were produced with the neutrino interacting with a nucleon in the nuclear environment and following the interaction products passing through the various materials of the detector using GEANT4 \cite{Agostinelli2003} version 4.9.3p6 with the QGSP\_BERT physics list, where the Bertini cascade is the most relevant for MINERvA hadron systems. The simulation accounts for the detector's optical and electronic response. The absolute energy scale of minimum ionizing energy depositions was set using through-going muons, the procedure used to match the simulation to the data can be found in Reference \cite{ALIAGA2014130}. The absolute energy response of the detector to charged hadrons was estimated by measurements using a charged particle test beam \cite{Aliaga:2015aqe} with a scaled-down version of the MINERvA detector. With the higher energy neutrino beam of this analysis, the probability of multiple interactions in a single gate increased and this pile-up effect was added to the simulation by adding simulated events to the 16 $\mu$s windows of activity taken from actual data.

\section{Event Reconstruction and Selection}
A general reconstruction algorithm is run on all MINERvA's data and simulation to classify events and quantify the kinematics of the interactions. The reconstructed information is then used to select the events of interest for analysis.

\subsection{Reconstruction in MINERvA}
The NuMI beam is delivered to the MINERvA detector in 10 $\mu$s bursts every 1.3 s. The MINERvA detector records data over a 16 $\mu$s gate. Multiple neutrino interactions can occur in one spill, but the timing resolution of the electronics means that individual neutrino interactions can be isolated in time. The energy deposits in the scintillator within one gate are grouped into time windows called time slicing. After grouping the hits in time slices, it is necessary to group together contiguous hits within a plane. A set of hits in a plane is called a cluster, even if only one hit is registered in the plane.

The next step in the reconstruction is to group the clusters and build tracks. The longest track in the detector usually belongs to a muon and its starting point is used as an anchor. Then, the other track candidates are merged and attached to the vertex of the muon candidate. The muon tracks leaving the MINERvA detector must match tracks in the MINOS detector. The track matching is done when the two tracks are within a 200 ns time window. The MINERvA tracks must also end within the last five modules of MINERvA, and the MINOS track must start within the first four modules of MINOS. Finally, a track is considered matched to MINOS if the projected distance at the front face of MINOS between the two tracks is less than 40 cm. If the previous method does not find any matches, the MINOS track is projected toward MINERvA, and the MINERvA track is projected toward MINOS, and the point of closest approach of the two tracks is found. The magnetic field in the MINOS detector can be used to measure the muon momentum and its charge. The muon momentum is measured by range for the lower energy subset of these data, and by curvature for high energy muons.

The final hadronic energy ($E_{had}$) is reconstructed calorimetrically and includes corrections to compensate for the loss of detectable energy due to final state interactions, neutral particles, and the binding energy of the ejected nucleons. In this way, the calorimetry algorithm produces an estimator for the energy transfer. Visible energy, defined as available energy from the hadronic shower of the neutrino interaction ($E_{vis}$) is reconstructed from the visible energy of all clusters involved in each event that are not part of the muon track within a time window of [-25, +35] ns around the interaction vertex time. Two corrections to $E_{vis}$ are applied at a hit-by-hit level to all calorimetric clusters to account for the energy loss due to the passive materials of the medium. The first is for the passive components of the scintillator planes, like the epoxy glue and translucent coating. The second correction, of a few percent, is for the light attenuation of the optical fibers that connect the strips with the PMTs. The light attenuation was measured using a radioactive source after the scintillator planes were constructed. Test beam data at the Fermilab Test Beam Facility was used to validate the MINERvA detector simulation and evaluate systematic uncertainties \cite{Aliaga:2015aqe}.

\subsection{Signal Region Definition and Event Selection}
\label{Sec:SRdefinition}

The signal region (SR) definition consists of requirements using reconstructed information to define a region with high purity and efficiency of SIS events. The SR is characterized by charged-current $\nu_\mu$ ($\bar{\nu}_\mu$) events in the reconstructed sample with a $\mu^-$ ($\mu^+$) and a hadronic shower with an invariant mass 1.5 $<$ $W_\mathrm{reco}$ $<$ 2 GeV/$c^2$. For the SIS definition II, the $Q_\mathrm{reco}^2 \geq 1\,\mathrm{GeV}^2$/$c^2$ cut is added to the SR definition. 

The tags ``reco" in the variables used for the signal region definition explicitly indicate that they were obtained by reconstructed information of observed quantities computed from the particles emerging from the nucleus after the final state interactions. When necessary, the ``reco" and ``true" tags will be used to distinguish reconstructed from true GENIE or GEANT4 information present in the simulated samples. 

To further define the signal region sample that will be used in the estimation of a cross section, the cuts are made using reconstructed information as follows:
\begin{enumerate}[(i)]
    \item The interaction vertex must be in the active tracker region within a hexagonal fiducial area with apothem 850 mm in the xy plane between the 27 and 80 modules of the z-axis. 
    \item To ensure a good acceptance in the MINOS detector, two cuts are applied. The muon (antimuon) energy is required to be greater than 2 GeV and the muon (antimuon) angle less than 20$^\circ$ with respect to the neutrino beamline direction.
    \item The muon (antimuon) energy has to be less than 20 GeV to ensure an accurate flux prediction. 
    \item A muon (antimuon) whose track matches in both MINERvA and MINOS detectors must be a negative (positive) curvature in the MINOS magnetic field, which ensures that the event is produced by a muon neutrino (antineutrino).
    \item The muon (antimuon) reconstructed by curvature in the MINOS detector must have a curvature with a significance level of at least five times the relative error on charge over momentum for the MINOS track.
    \item The endpoints of muon (antimuon) tracks in the MINOS detector must be at a distance R from the MINOS magnetic coil between 210 $< R < 2500$ mm.
\end{enumerate}

The analysis samples after applying all the selection cuts and the restricted $W$ region contain $617,953$ ($364,082$) events in data for the $\nu_\mu$ ($\bar{\nu}_\mu$) samples that correspond to the SIS definition I, while there are $174,089$ ($74,380$) events for the SIS definition II. The sample purity, defined as the fraction of selected events that are signal events and estimated using the simulation, is 65.8\% (68.9\%) for SIS definition I and 52.9\% (55.3\%) for SIS definition II.

\section{Background}
\label{Sec:Bkg}
The SIS signal events are defined in Sec.~\ref{Sec:SISregion} using true information from the MC events. On the other hand, the experimental signal region, used in calculating the differential cross sections, is defined in Sec.~\ref{Sec:SRdefinition} using reconstructed (measured) information. There are two signal region definitions, with and without the $Q^2$ cut, depending on the SIS definition being considered, as can be seen in Figure \ref{Regions_Definition}. 

For each SIS definition, the background events are the ones that are not SIS events but are reconstructed inside the signal region. For SIS definition I, the background events have similar contributions from GENIE DIS and resonances representing about 30-35\% of the events in the SR. In turn, for SIS definition II, the background represents about 50-55\% of the events in the SR being dominated by GENIE DIS, as expected due to the $Q^2$ $\geq$ 1 GeV$^2/c^2$ cut. To subtract the correct amount of background events from the initial total data distribution, the simulated background sample within the signal region is corrected by data-driven factors estimated using the measured data in regions called sidebands that are kinematically adjacent to the signal region.

\subsection{Templates and Sidebands Definition}
\begin{figure}[htbp]
	\centering
    \includegraphics[scale=0.3]{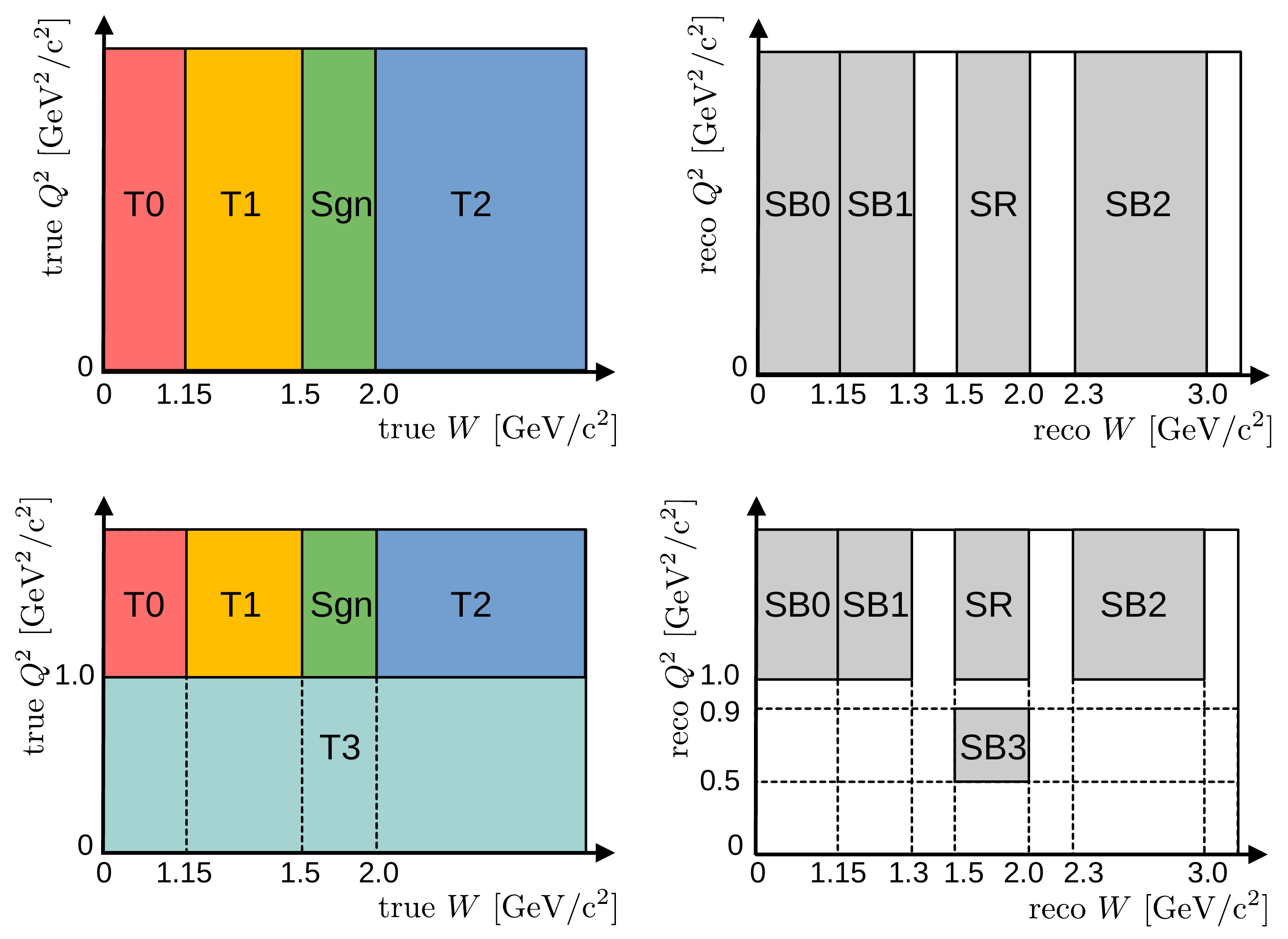}
	\caption{Charged-current $\nu_\mu$ ($\bar{\nu}_\mu$) event categories (left) and sideband and signal region definitions (right). Top: Setup for the SIS definition I. Bottom: Setup for SIS definition II.}
	\label{Regions_Definition}
\end{figure}

For the SIS definition I, without the $Q^2$ cut, the background events are divided into three categories called templates. Templates 0, 1, and 2 (or T0, T1, and T2) are CC $\nu_\mu$ ($\bar{\nu}_\mu$) events mostly populated by QE, resonant, and higher $W$ DIS events, respectively, and their definition can be seen in the upper plots of Figure \ref{Regions_Definition}. The non-CC $\nu_\mu$ ($\bar{\nu}_\mu$) events yield a minimal contribution in the inclusive sample (0.1\% and 0.3\% respectively) and are estimated directly from the simulation. The sidebands 0, 1, and 2 (SB0, SB1, and SB2), where each of the CC $\nu_\mu$ ($\bar{\nu}_\mu$) templates T0, T1, and T2 are respectively corrected to the data, are defined with selection cuts in reconstructed information resulting in regions dominated by the associated template and small signal contamination. The sideband definitions are shown as well in the upper plots of Figure \ref{Regions_Definition}. 

For the SIS definition II, the background events are divided into four categories. The templates T0, T1, T2, and T3 are CC $\nu_\mu$ ($\bar{\nu}_\mu$) events and their definitions are shown in the lower plots of Figure \ref{Regions_Definition}. The additional CC $\nu_\mu$ ($\bar{\nu}_\mu$) template, T3, is the source of background events with low $Q^2$. The non-CC $\nu_\mu$ ($\bar{\nu}_\mu$) events, since minimal, are estimated directly from the simulation. The sidebands SB0, SB1, SB2, and SB3, where the CC $\nu_\mu$ ($\bar{\nu}_\mu$) templates T0, T1, T2, and T3 are respectively corrected to the data, are shown as well in the lower plots of Figure \ref{Regions_Definition}.

\subsection{Tuning Procedure and Background Subtraction}
To estimate the background of non-SIS events in the designated SIS signal region, the magnitude and shape of the total background simulation are compared to the measured data magnitude and shape in each sideband. For this inclusive analysis, these backgrounds primarily appear in the signal region because of $W$ resolution effects. By tuning the contribution of each background channel, the total background is tuned to the measured data. This sideband tuning is then applied to the simulated background in the signal region to get a final estimate of the background to be subtracted from the data distributions in the signal region.

The tuning of the sidebands that corrects the simulated prediction to better describe the data, is done using a $\chi^2$ minimization technique, with $\chi^2$ defined as:
{\footnotesize
\begin{eqnarray}
    \displaystyle{
	\chi^2 = \sum_{s}\sum_{i}\frac{\left ( N^\mathrm{MC}_{s,i} - N^\mathrm{data}_{s,i} \right )^2}{N^\mathrm{data}_{s,i}}\,,\ \ \ N^\mathrm{MC}_{s,i} = \sum_{c} f_{c,i} N_{c,s,i},
    }
    \label{Chi2_minimization}
\end{eqnarray}
}
where $N^\mathrm{MC}_{s,i}$ and $N^\mathrm{data}_{s,i}$ are, respectively, the number of simulated and data events in the bin $i$ of the distribution of a selected variable inside the sideband $s$, $N_{c,s,i}$ is the number of simulated events of the event category $c$ in the bin $i$ of the distribution inside the sideband $s$, and $f_{c,i}$ is the factor correcting the contribution of the event category $c$ in the bin $i$ of the distribution. The event categories comprise events from all templates plus the non-CC and minimal signal events. In the SIS definition I, the templates T0, T1, and T2 are corrected by factors optimized in the fit. In turn, the templates T0, T1, T2, and T3 are the event categories corrected in the fit for the SIS definition II. The background events after this correction are referred to as ``tuned".  

For the SIS definition I, the selected variable used in the fit was $Q^2_{reco}$ because it was observed that correcting the $Q^2_{reco}$ distribution resulted in the correction of all the main variables of interest for the analysis. The template T0 was corrected by a scale factor, $f_{0,i} = c_0$, the template T1 was corrected by a combination of a linear and constant function of $Q^{2}_\mathrm{true}$, $f_{1,i} = f_{1}(Q^{2}_\mathrm{true};a_1,b_1)$, and the template T2 was corrected by an exponential function of $Q^{2}_\mathrm{true}$, $f_{2,i} = f_{2}(Q^{2}_\mathrm{true};a_2,b_2,c_2)$, where
{\footnotesize
\begin{eqnarray}
    f_{1}(Q^{2}_\mathrm{true};a_1,b_1) &\equiv& \left\{ \begin{array}{ll}
    a_1{\cdot}(0.3 - Q^{2}_\mathrm{true}) + b_1  & \mbox{if } Q^{2}_\mathrm{true} < 0.3 \\
    b_1 & \mbox{if } Q^{2}_\mathrm{true} \geq 0.3
    \end{array} \right. \\
    f_{2}(Q^{2}_\mathrm{true};a_2,b_2,c_2) &\equiv& a_2{\cdot}e^{-b_2{\cdot}Q^{2}_\mathrm{true}} + c_2.
\end{eqnarray}
}
The functions were chosen through observation of the data to MC ratio of the $Q^2_{reco}$ distribution in the sidebands, and the use of $Q^{2}_\mathrm{true}$ in the functions is allowed since $Q^2$ is closely related to the transverse muon momentum and the muon momentum and direction are well-measured at MINERvA with a nearly diagonal migration matrix between true and reconstruction values.

For the SIS definition II with the $Q^2$ cut, $Q^2$ was not the appropriate variable to use in the tuning. Rather ${\theta_\mu}_{reco}$ was used in the fit since it is little affected by the kinematic cuts in $Q^2$ and $W$. A different tuning approach was used since no obvious functional behavior of data to MC ratio of the ${\theta_\mu}_{reco}$ distribution in the sidebands was observed. A bin-by-bin scale factor approach was used and the correction of the ${\theta_\mu}_{reco}$ distribution also corrected the differences in the other kinetic variables.

\begin{figure}[htbp]
    \centering
    \includegraphics[width=0.23\textwidth]{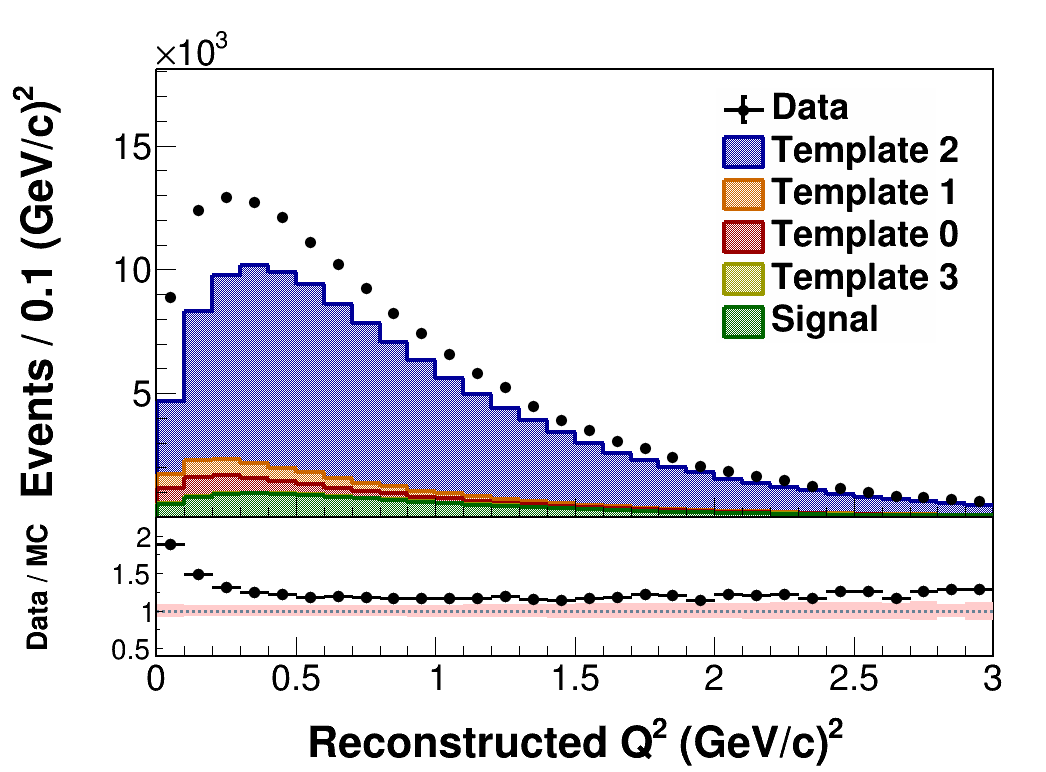}
     \includegraphics[width=0.23\textwidth]{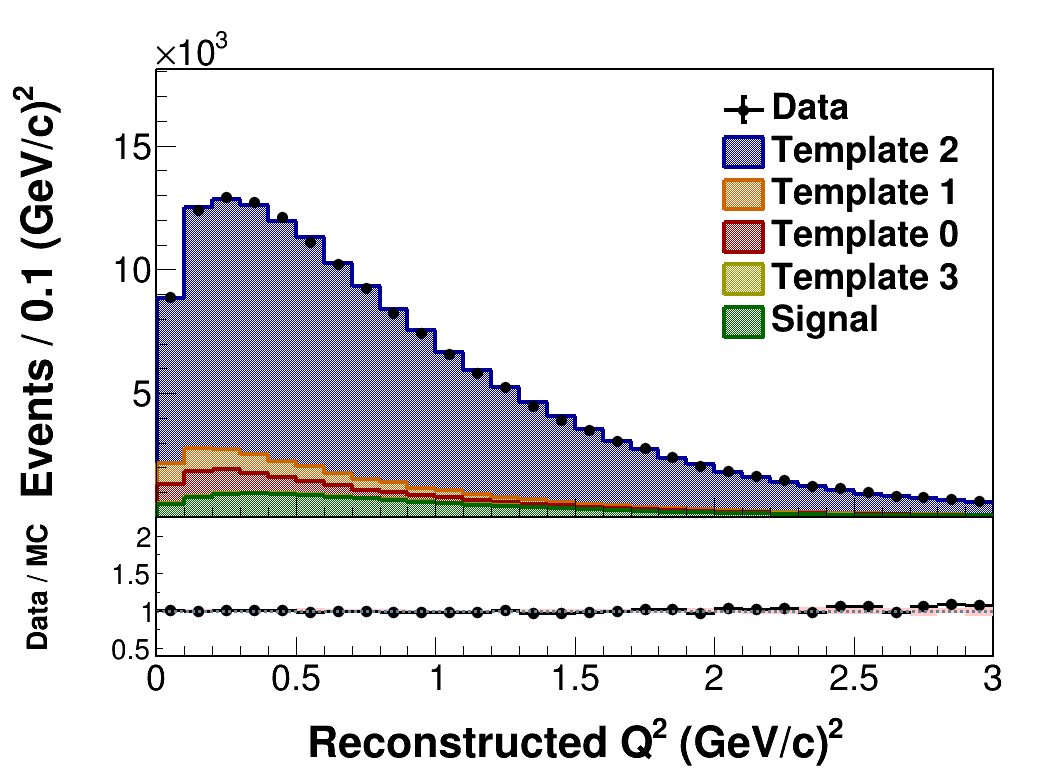}

    \includegraphics[width=0.23\textwidth]{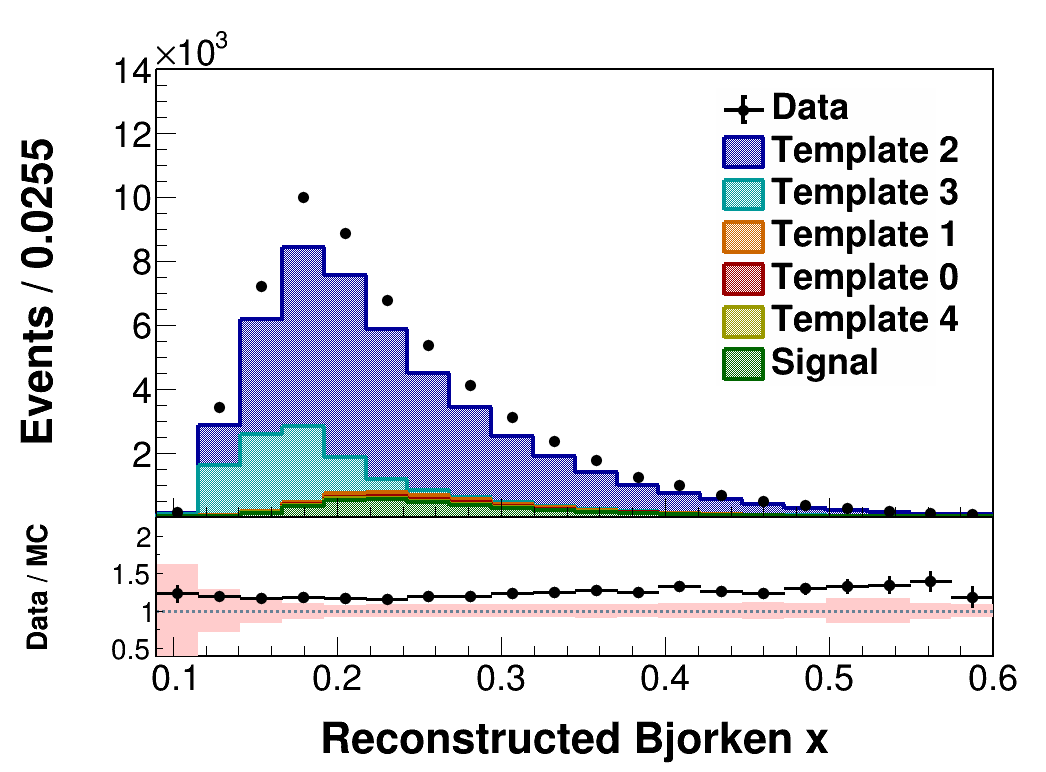}
    \includegraphics[width=0.23\textwidth]{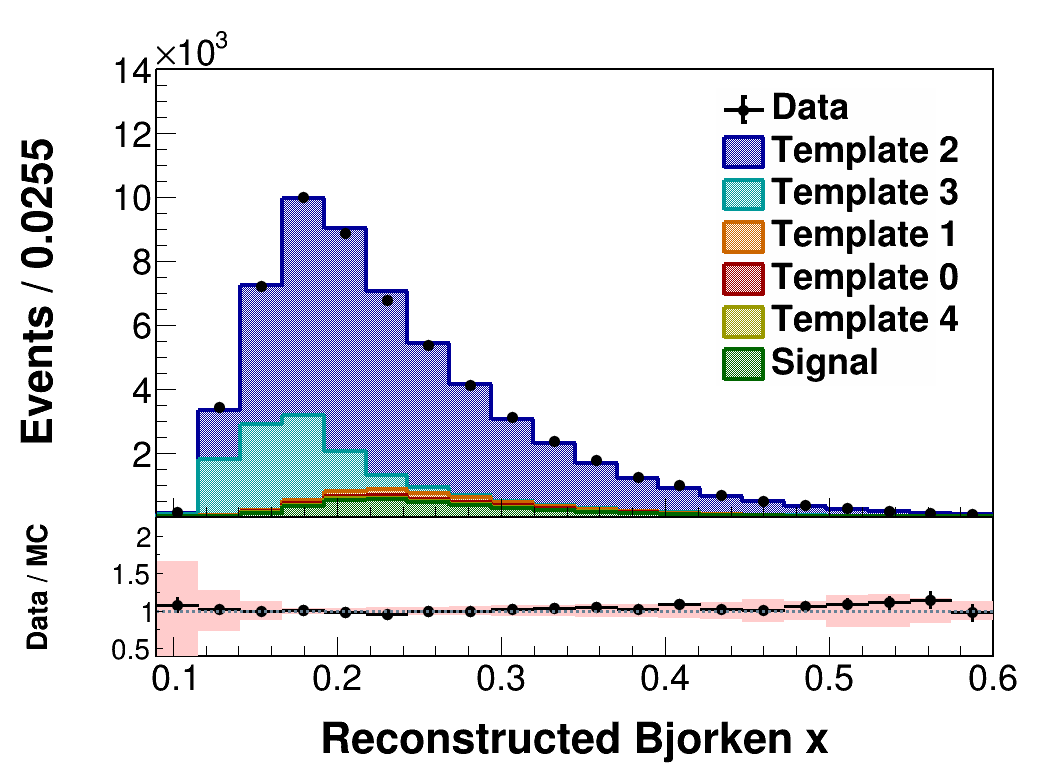}
    \caption{Antineutrino event distribution and data MC ratio in the SB2 as a function of $Q^2$ for the SIS definition I (top) and $x$ for the SIS definition II (bottom). Left: Before the tuning procedure. Right: After the tuning procedure.}
    \label{fit_Q2_defI_sb2}
\end{figure}

An example of the definition I $Q^2$ and definition II $x$ data and MC distributions in SB2 before and after the tuning procedure is shown in Figure \ref{fit_Q2_defI_sb2}.

\begin{figure}[htbp]
    \centering
    \includegraphics[width=0.23\textwidth]{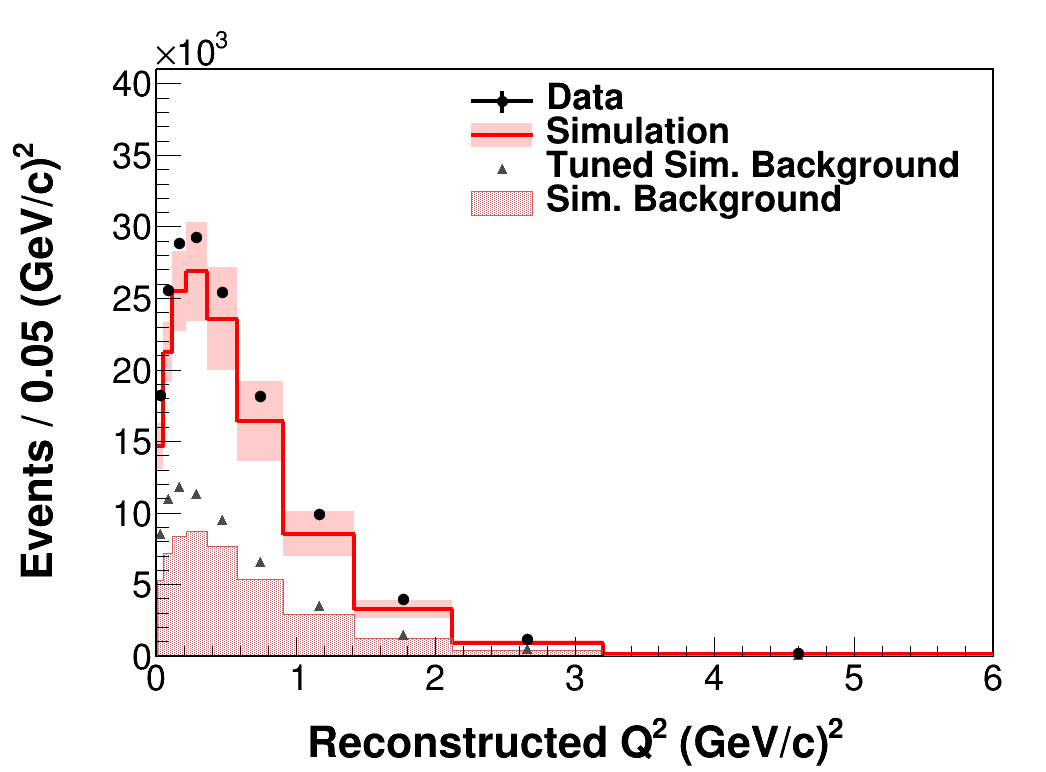}
    \includegraphics[width=0.23\textwidth]{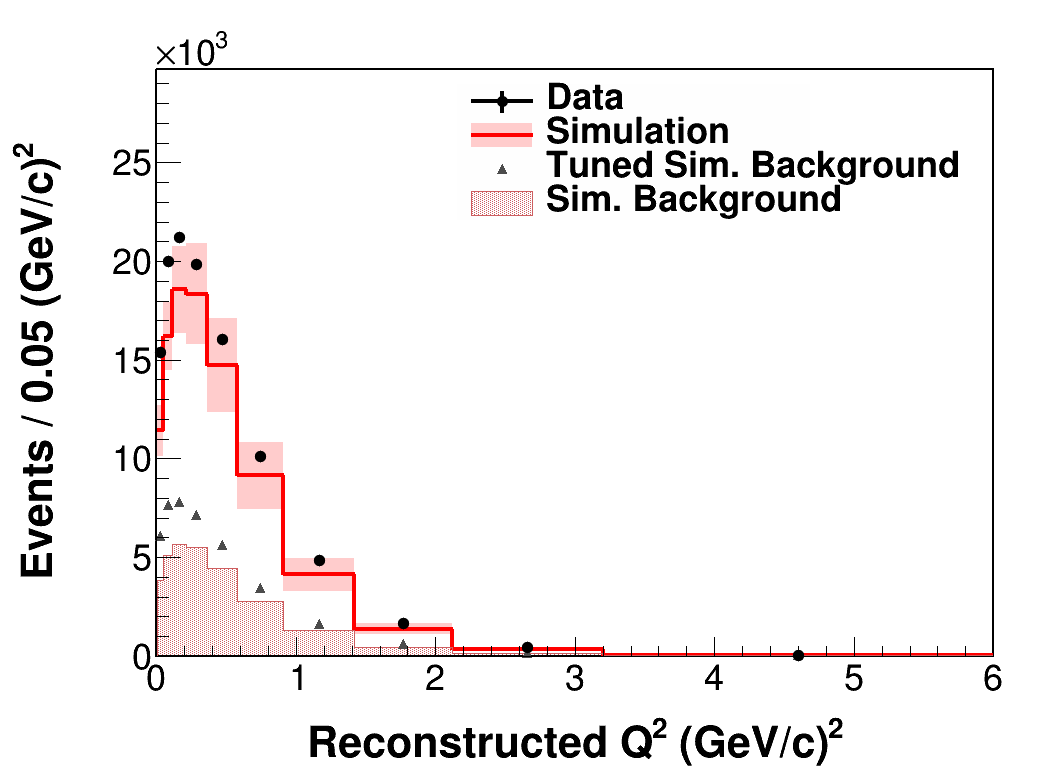}

    \includegraphics[width=0.23\textwidth]{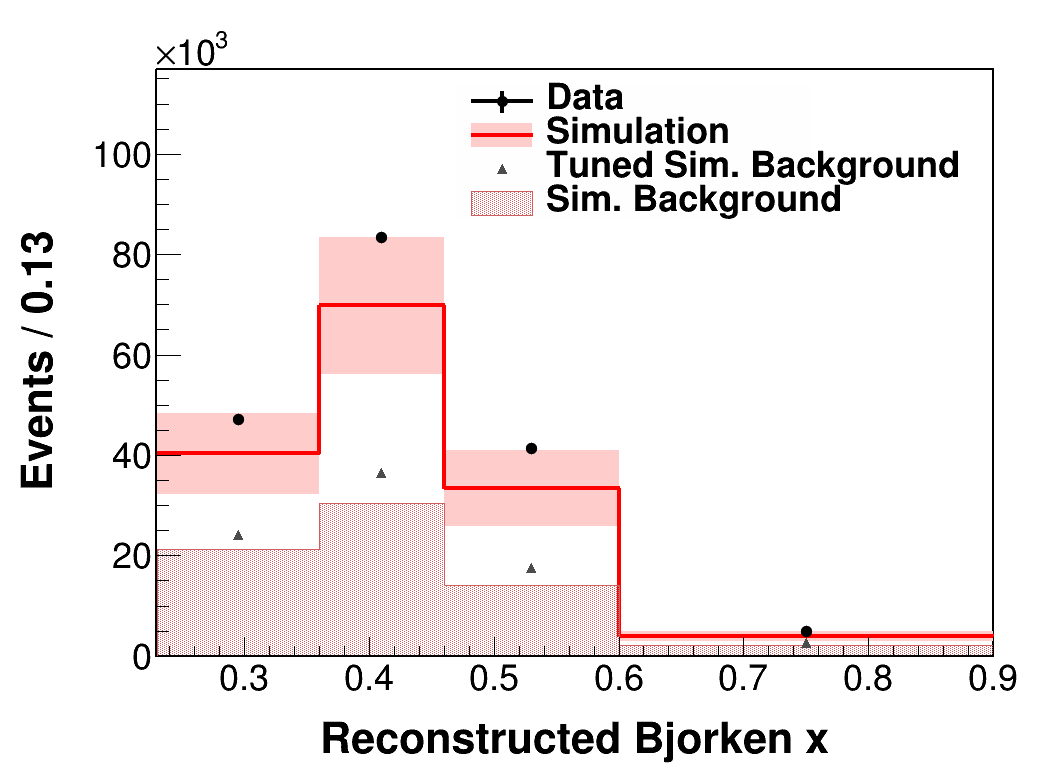}
    \includegraphics[width=0.23\textwidth]{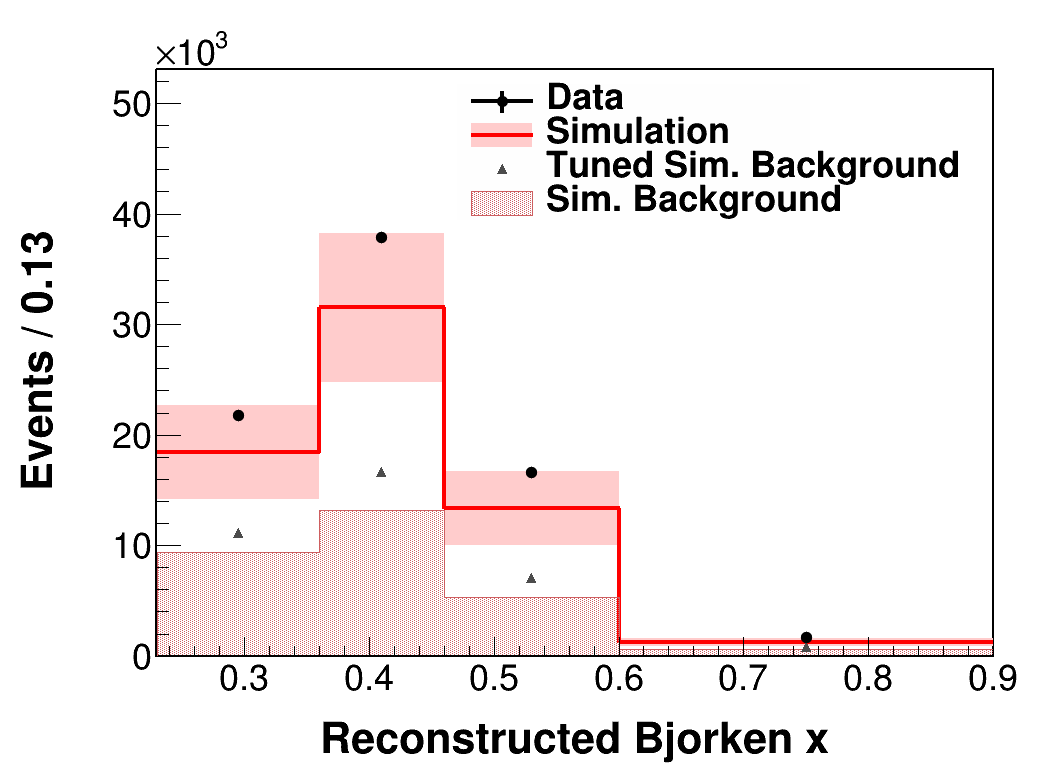}
    
    \caption{Data and MC distributions in the signal region as a function of $Q^2$ for the SIS definition I (top) and $x$ for the SIS definition II (bottom) showing explicitly the tuned background contamination that is used in the background subtraction. Left: Neutrino. Right: Antineutrino.}
    \label{SignalRegion_Q2ge0}
\end{figure}

The tuning determined in the sidebands is applied to the simulated background contribution inside the signal region to yield the corrected (tuned) background. 
As an example, Figure \ref{SignalRegion_Q2ge0} displays the distributions in the signal region with the tuned background contribution that will be subtracted from the data distribution for $Q^2$ in SIS definition I and $x$ in SIS definition II. In this figure, the simulation prediction is reported with statistical uncertainty while the uncertainties associated with the data and background predictions are omitted.
Following the trend of observed data exceeding the MC predictions in the sidebands, exposed in the examples of Figure \ref{fit_Q2_defI_sb2}, there is an excess of tuned background over MC prediction in the SR for neutrino and antineutrino, all variables, and both SIS definitions as illustrated in the examples of Figure \ref{SignalRegion_Q2ge0}.

The tuned background is subtracted bin-by-bin from the data signal distributions. As a result, the SIS background-subtracted distribution for each variable is obtained. The systematic uncertainties are propagated to the data.
The background-subtracted data samples contain $383,494$ ($235,017$) events for the SIS definition I and $93,410$ ($40,048$) events for the SIS definition II suggesting that 75\% of events have $Q^2 <$ 1 GeV$^2/c^2$.

\section{Cross Section Extraction}
\subsection{Unfolding}
The reconstructed kinematic distributions differ from the true kinematic distribution due to detector resolution and reconstruction imperfections. This smearing effect is characterized by the migration matrix, also known as the smearing matrix or the response matrix. Simulation is used for the construction of the matrix, which corrects distortions and recovers the true kinematic distribution. For this purpose, it is necessary to find the inverse of the migration matrix. In many cases, it is not invertible using traditional algebra and other methods should be employed. The MINERvA collaboration uses the iterative D’Agostini method \cite{DAgostini1995} based on Bayes’ theorem, which reduces the variance at the price of increasing the bias, is regularized by the number of iterations, and is implemented in the RooUnfold framework \cite{Adye2011}. The RooUnfold framework needs the reconstructed distribution, the migration matrix, and the number of unfolding iterations as input. The first two are known, and the number of iterations must be determined from an additional study called the warping study. This study allows us to test the unfolding procedure and see how well it does with the distributions of interest.

In order to determine the number of iterations that recover the original distribution, fake data were used. These fake data were based on model predictions and functions that mimic the data / MC ratio in each bin and re-reweight the MC. To understand the bias in this procedure and eliminate the effect of statistical fluctuations, 500 variations of the fake data are made using Poisson throws. The variations or universes, as it is also known, are unfolded over a range of iterations using the migration matrix. The statistical uncertainty of the unfolded distribution and the unfolding covariance matrix are modified with a factor that accounts for the finite Monte Carlo statistics in the migration matrix. Finally, the unfolded fake data were compared to their true distributions via a $\chi^2$ test taking full consideration of correlations. The optimal number of iterations with the minimum modeling bias is determined when the $\chi^2$ per degree of freedom approaches a stable value that is often around one.

These studies indicated that the $Q^2$, $p_{T\mu}$, and $p_{\parallel\mu}$ background subtracted data distributions could be unfolded using three iterations for neutrinos and antineutrinos in the SIS definition I analysis.  For the definition II analysis with the $Q^2 \geq 1$ GeV$^2/c^2$ cut that emphasizes the multi-quark contribution to SIS, $x$ was unfolded as well as $Q^2$, $p_{T\mu}$, and $p_{\parallel\mu}$. Unfolding $x$ required 6 iterations for the neutrino analysis and 4 iterations for the antineutrino analysis.

\subsection{Efficiency and Normalization}
An efficiency correction is applied to recover missing events in the final sample due to limitations of the detector, the reconstruction algorithm, and the selection cuts. The average selection efficiency is 29.8\% (34.7\%) for neutrinos (antineutrinos) for the analysis without the $Q^2$ cut and 16.6\% (18.7\%) for neutrinos (antineutrinos) for the analysis with the $Q^2$ cut. The signal definition that restricts the sample to $\theta_\mu < 20^\circ$ avoids a region of unreliable efficiency. In general, the $\theta_\mu$ requirement drives the loss of efficiency with the efficiency decreasing with increasing $\theta_\mu$ due to MINOS acceptance. The unfolded background subtracted distribution is divided by the efficiency bin by bin in each variable for the particular cross section reported. 

The last step to obtain the cross section is to take the efficiency corrected, unfolded and background subtracted distribution and normalize by the integrated flux, the number of targets, and the bin widths. The flux distribution is integrated over $0 \leq E_{\nu} \leq 120$ GeV and scaled to the data exposure, which is given as the number of protons on target. The integrated flux is $6.24 \times 10^{-8}$ $\nu_{\mu}$/cm$^2$/POT for neutrino and $4.77 \times 10^{-8}$ $\bar{\nu}_{\mu}$/cm$^2$/POT for antineutrino. The number of targets in the fiducial volume is expressed as the number of nucleons, resulting in $3.23 \times 10^{30}$. 

\subsection{Systematic Uncertainties}
The simulation models and event reconstruction insert systematic uncertainties in the cross section measurement. They arise during the background subtraction, efficiency correction, flux, and unfolding steps. Each systematic uncertainty is evaluated by creating a systematic ``universe" at the beginning of the analysis process and evaluating the shifted result at the end. 

\begin{figure}[htbp]
    \centering
    \zincludegraphics[width=0.23\textwidth,label=(a),pos=se,labelbox=false,fontsize=\small]{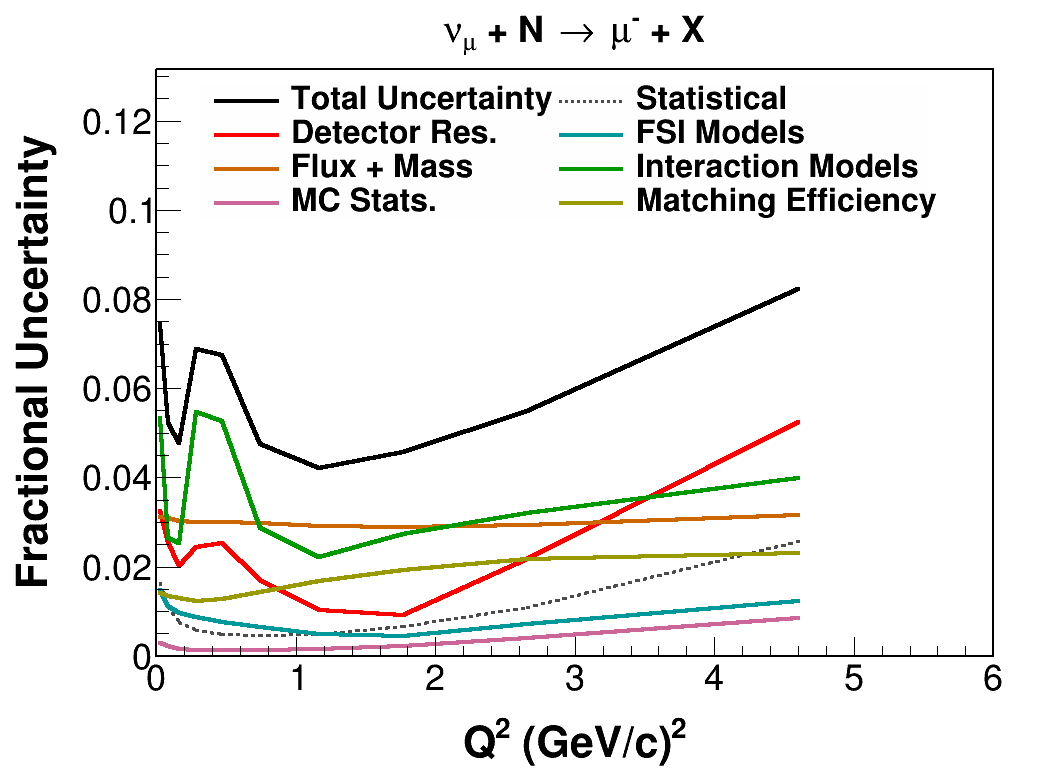}
    \zincludegraphics[width=0.23\textwidth,label=(b),pos=se,labelbox=false,fontsize=\small]{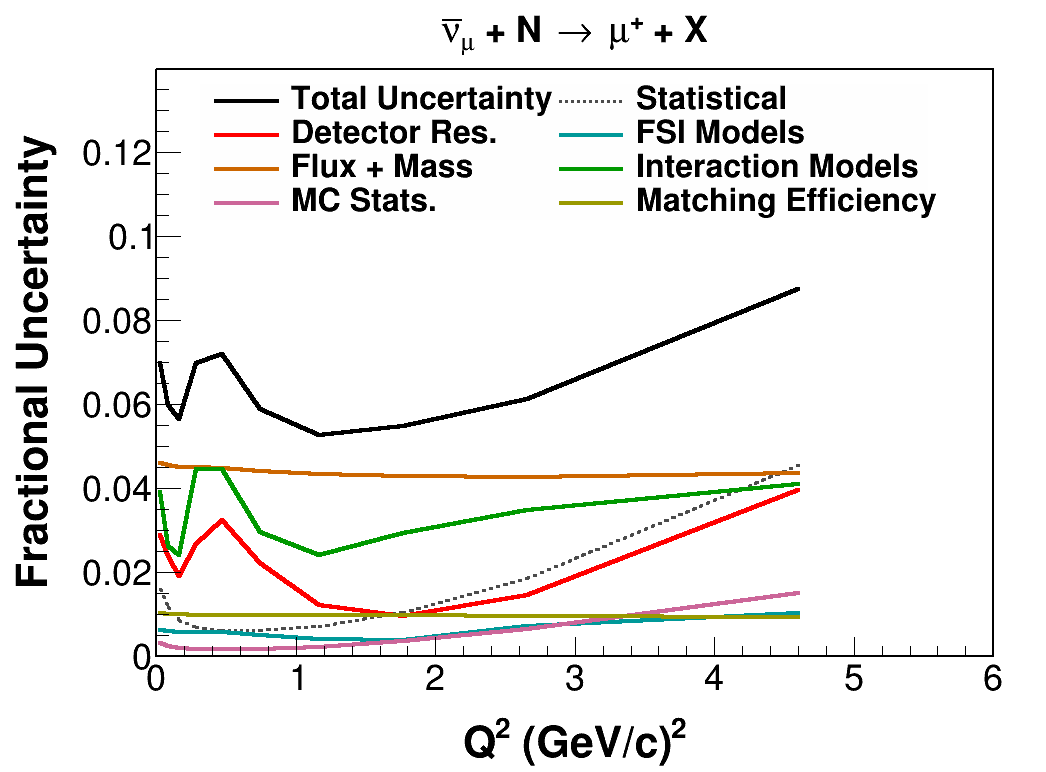}
    
    \zincludegraphics[width=0.23\textwidth,label=(c),pos=se,labelbox=false,fontsize=\small]{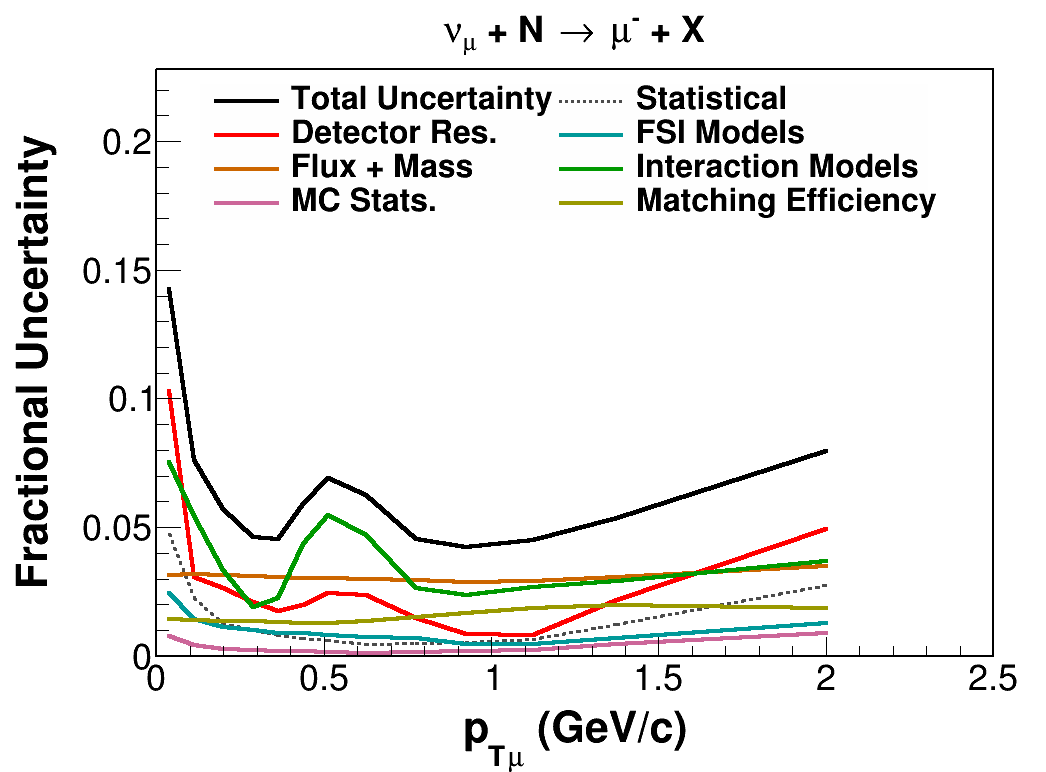}
    \zincludegraphics[width=0.23\textwidth,label=(d),pos=se,labelbox=false,fontsize=\small]{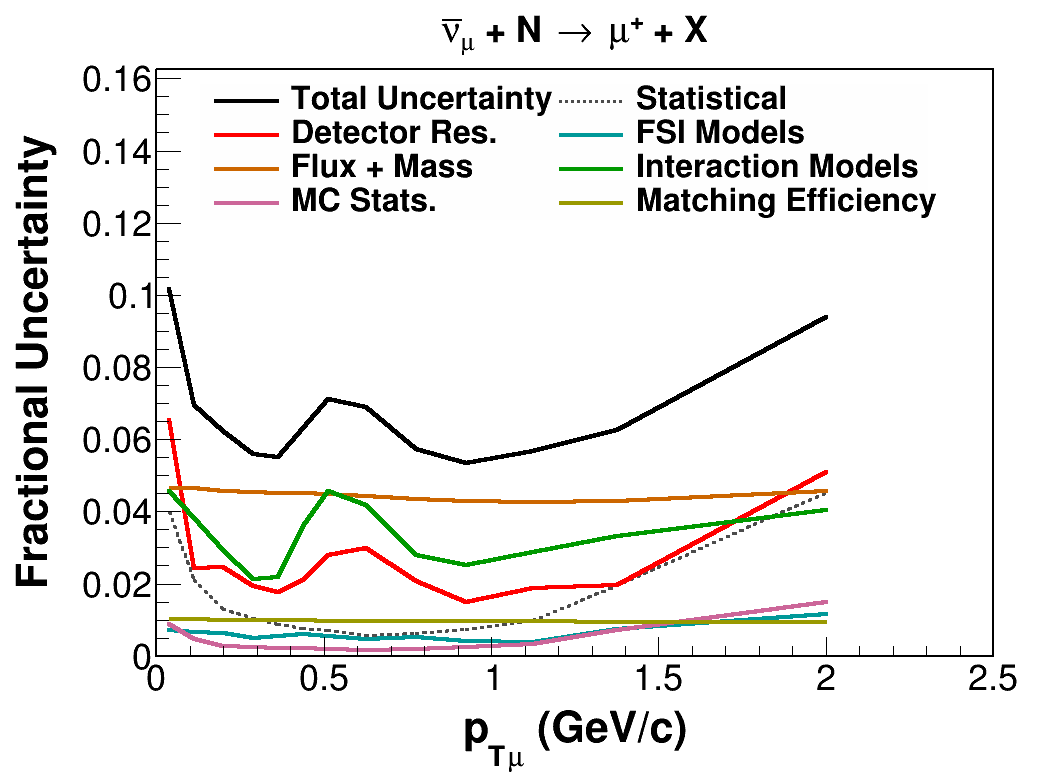}
    
    \zincludegraphics[width=0.23\textwidth,label=(e),pos=se,labelbox=false,fontsize=\small]{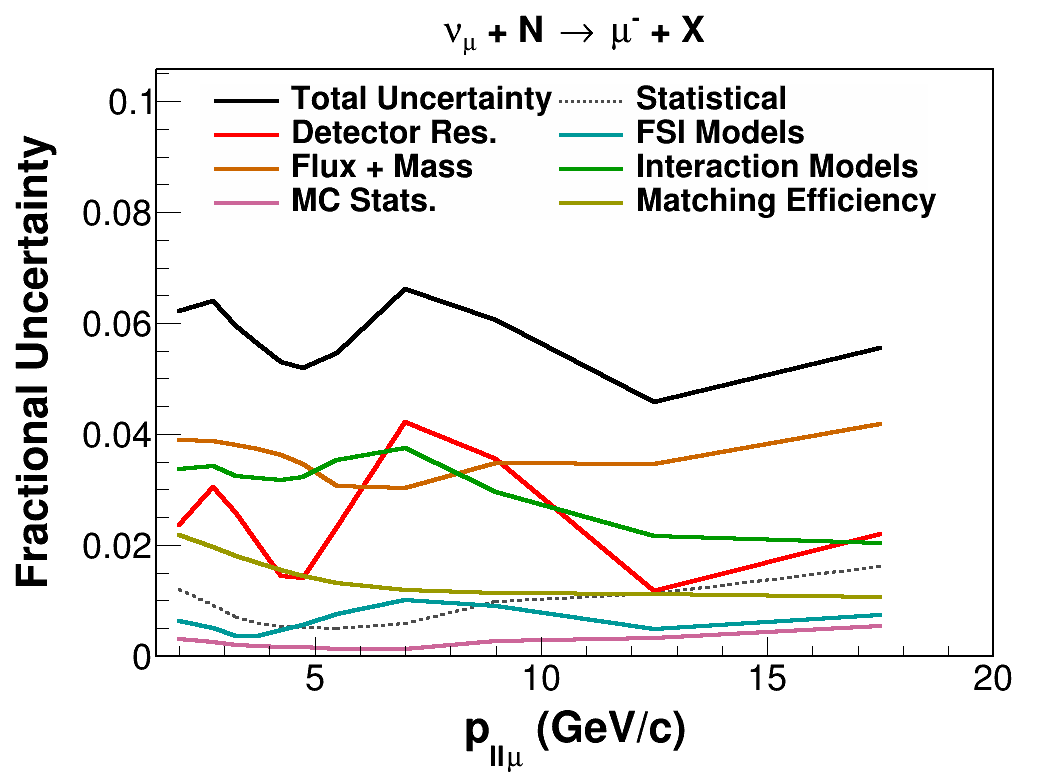}
    \zincludegraphics[width=0.23\textwidth,label=(f),pos=se,labelbox=false,fontsize=\small]{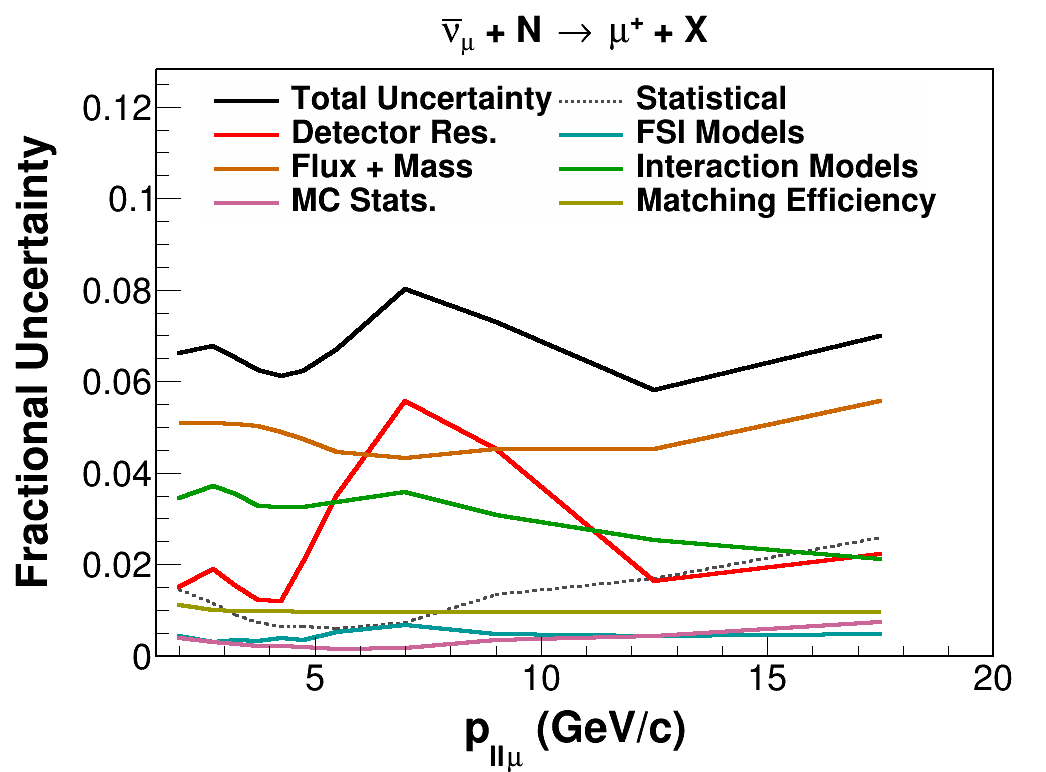}

    \caption{Definition I component and total fractional uncertainties on the measured cross sections as a function of $Q^2$ (a-b), $p_{T\mu}$ (c-d), and $p_{\parallel\mu}$ (e-f). Left: Neutrino. Right: Antineutrino.}
    \label{CrossSectionErrors_Q2ge0}
\end{figure}

\begin{figure}[htbp]
    \centering
    \zincludegraphics[width=0.23\textwidth,label=(a),pos=se,labelbox=false,fontsize=\small]{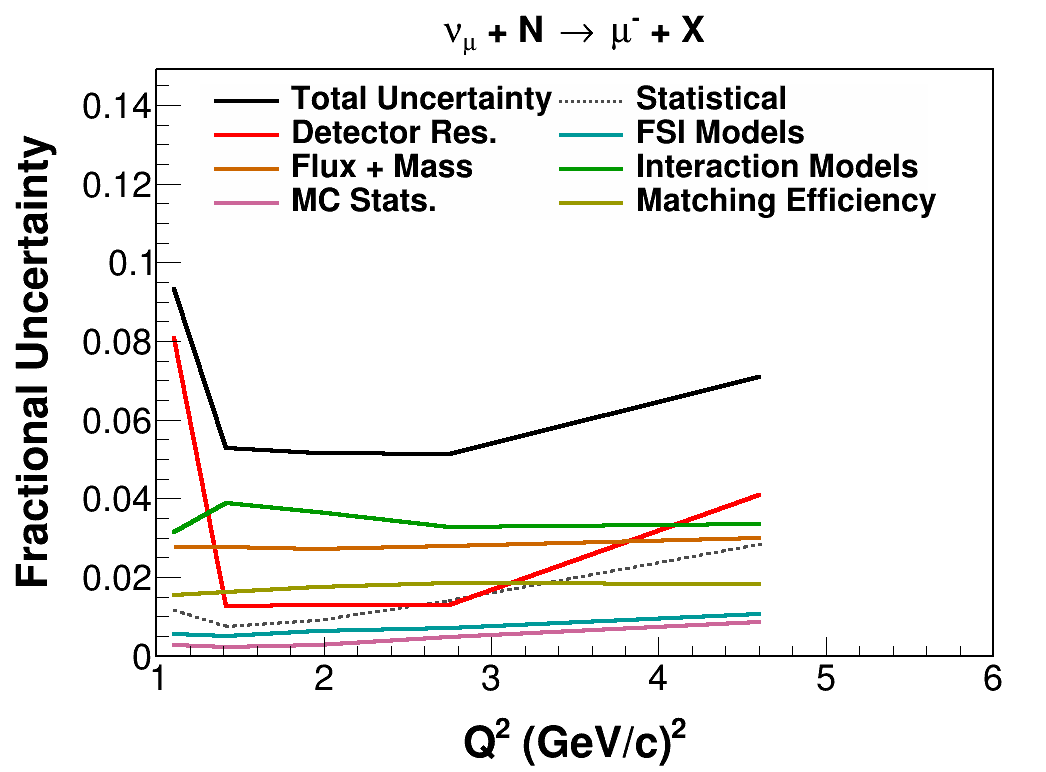}
    \zincludegraphics[width=0.23\textwidth,label=(b),pos=se,labelbox=false,fontsize=\small]{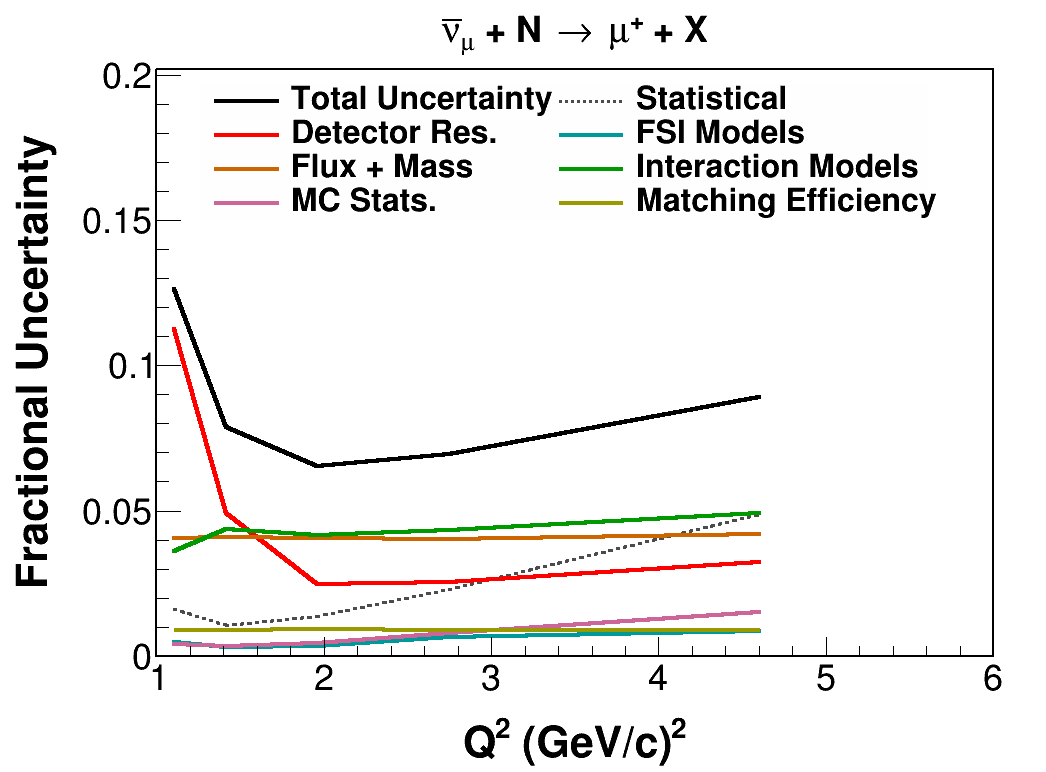}
    
    \zincludegraphics[width=0.23\textwidth,label=(c),pos=se,labelbox=false,fontsize=\small]{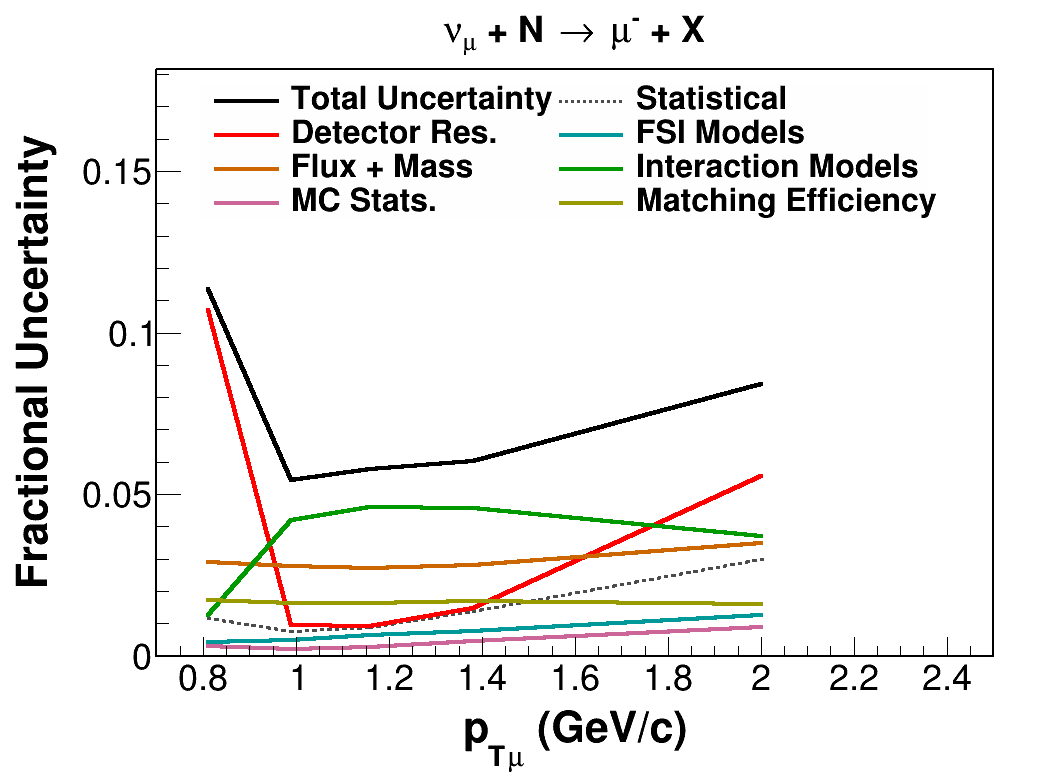}
    \zincludegraphics[width=0.23\textwidth,label=(d),pos=se,labelbox=false,fontsize=\small]{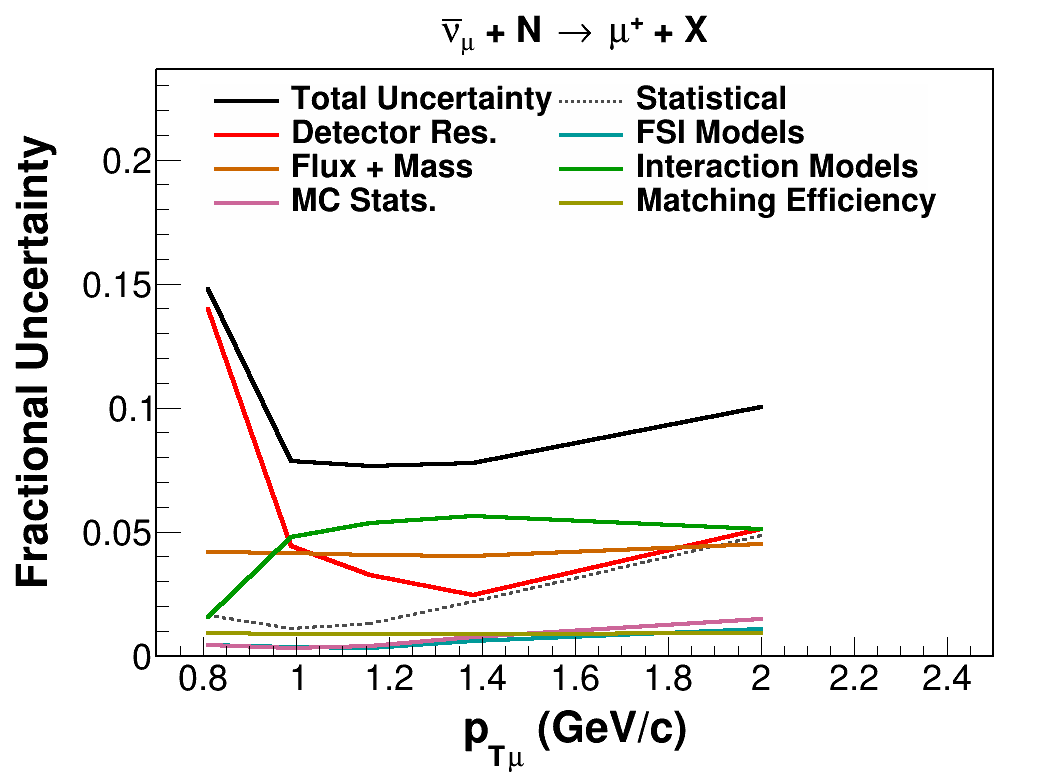}
    
    \zincludegraphics[width=0.23\textwidth,label=(e),pos=se,labelbox=false,fontsize=\small]{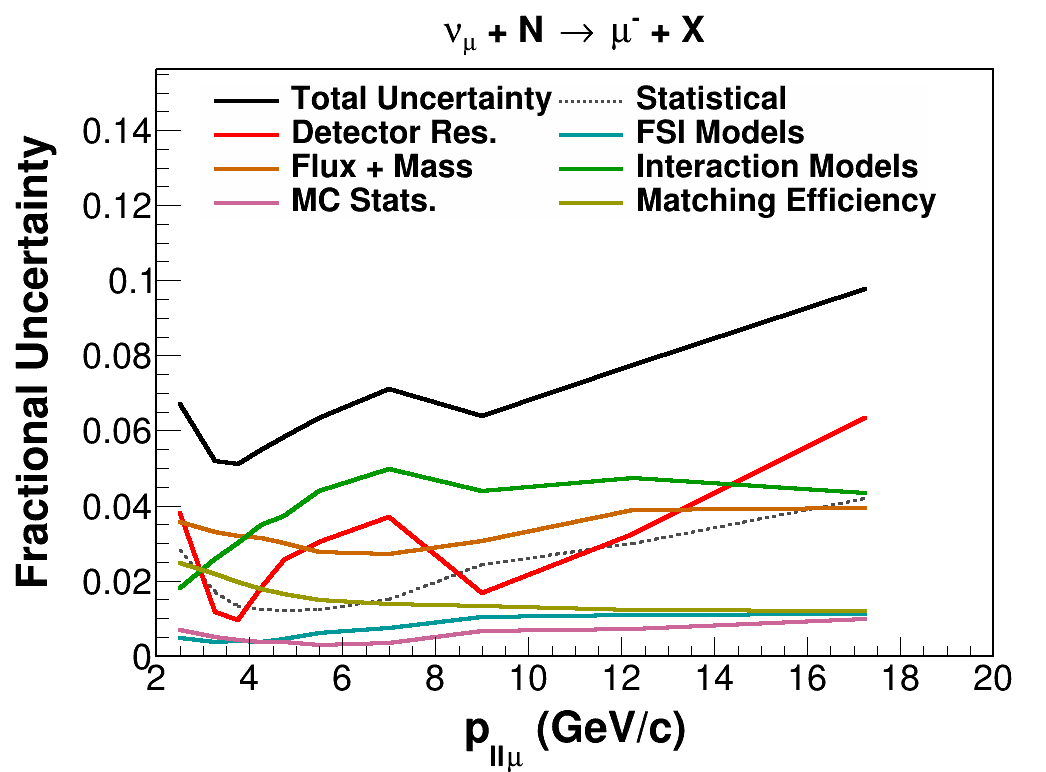}
    \zincludegraphics[width=0.23\textwidth,label=(f),pos=se,labelbox=false,fontsize=\small]{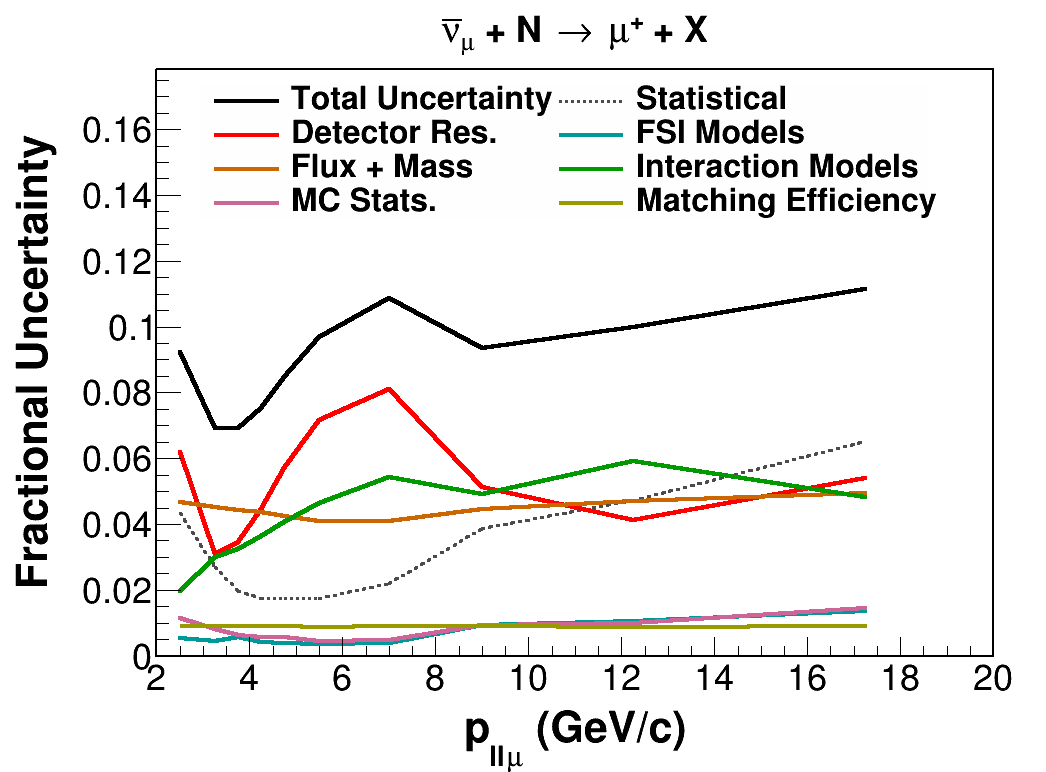}

    \zincludegraphics[width=0.23\textwidth,label=(g),pos=se,labelbox=false,fontsize=\small]{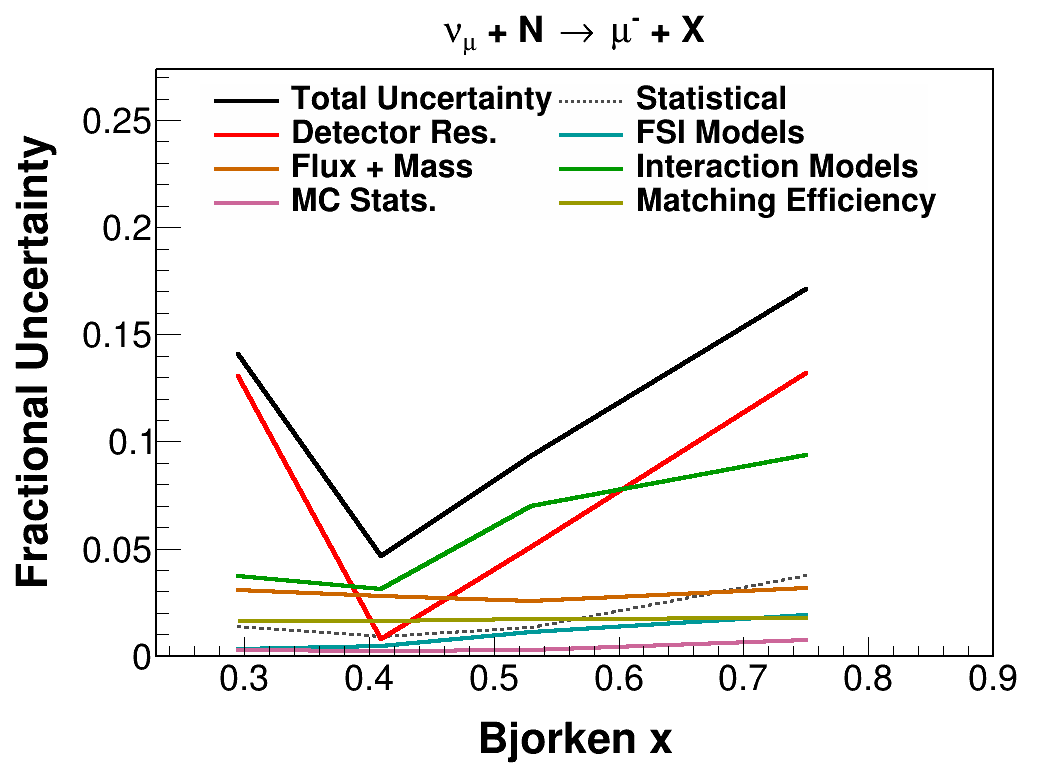}
    \zincludegraphics[width=0.23\textwidth,label=(h),pos=se,labelbox=false,fontsize=\small]{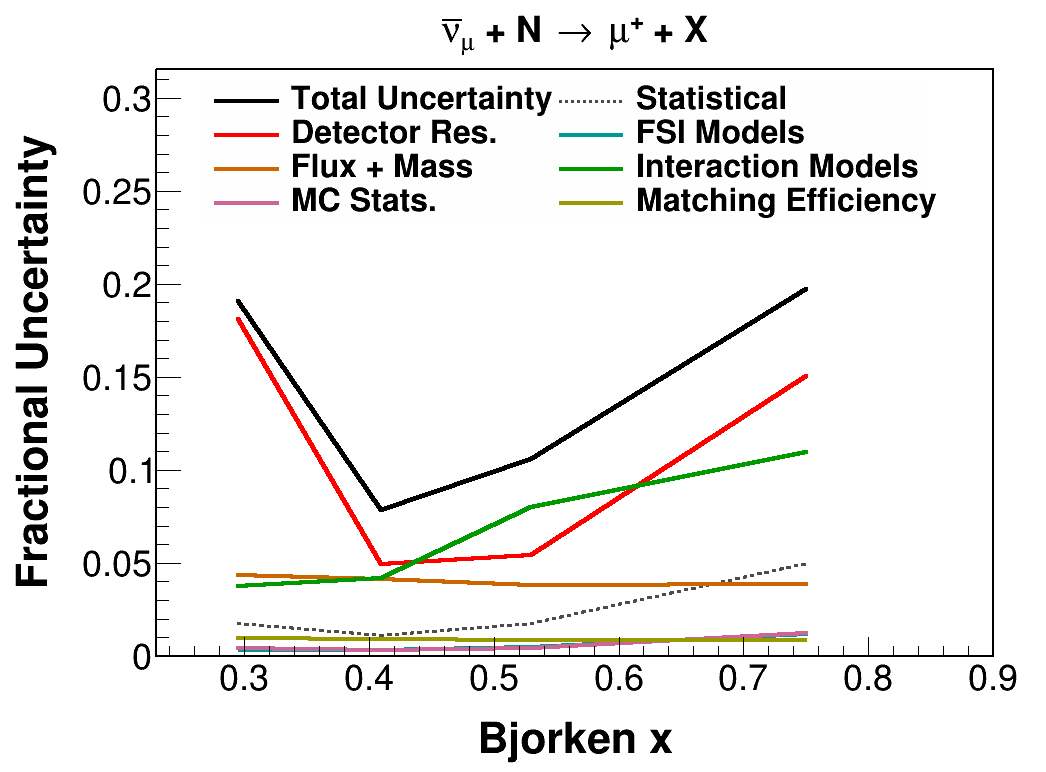}

    \caption{Definition II component and total fractional uncertainties on the measured cross sections as a function of $Q^2$ (a-b), $p_{T\mu}$ (c-d), $p_{\parallel\mu}$ (e-f), and $x$ (g-h). Left: Neutrino. Right: Antineutrino.}
    \label{CrossSectionErrors_Q2ge1}
\end{figure}

The sources of systematic uncertainties are the detector resolution, the flux and the detector mass, the general interaction models, the final state interactions within the nucleus models, the MINOS $\mu$ matching efficiency, and MC statistics. Each source is briefly described in this section and their fractional contribution to the total systematic uncertainty for all variables can be seen in Figures \ref{CrossSectionErrors_Q2ge0} for SIS definition I and \ref{CrossSectionErrors_Q2ge1} for SIS definition II. The figures show that for neutrino and antineutrino, and both SIS definitions, the total fractional uncertainty is usually dominated by interaction models, flux + mass, and detector resolution uncertainties.

Detector resolution uncertainties or experimental biases are associated with quantities measured in the MINERvA and MINOS detectors, such as the tracking and energy estimation for muon energy, muon angle, and hadronic energy. This uncertainty introduces smearing in the measured $E_\mu$, $\theta_\mu$ and $E_{had}$ that moves events in and out of the selected SIS region.

The flux uncertainty is derived from hadron production and beam focusing parameters. The hadron production uncertainties are related to the uncertainty introduced by the collision of the proton beam with the graphite target to produce hadrons that decay to neutrinos such as pions and kaons. The hadron production cross sections are constrained with proton-carbon scattering data from CERN's NA49 experiment and data from other hadron production experiments \cite{AliagaThesis2016, Adamson2016} extrapolated to the MINERvA's range of energy. Beam-focusing parameter uncertainties come from the position and magnetic fields of the magnetic horns that comprise the NuMI beamline. In addition, the flux uncertainty is reduced by applying the neutrino scattering from atomic electrons and the inverse muon decay constraints to their universes, which reduces the uncertainty on the $\nu_\mu$ ($\bar{\nu}_\mu$) flux from 7.62\% (7.76\%) to 3.27\% (4.66\%) at the focusing peak \cite{Zazueta2023}. The target mass uncertainty corresponds to the uncertainty in the measurement of the number of nucleons, which is 1.4\% for the scintillator target. 

The final state interactions models uncertainties correspond to the uncertainties in the intranuclear hadron model that GENIE employs to simulate the final state interactions within the nucleus. The uncertainties of the weights applied to GENIE are also included in this group, for instance, the 2p2h model, RPA model, and low $Q^2$ resonance suppression. The contribution of this group to the cross section is less than 1\%. 

The interaction model uncertainties are characterized by the uncertainties in the cross section models that GENIE uses to simulate the neutrino interactions. GENIE includes several parameters that the experiments can measure and constrain. The uncertainties have been studied extensively by MINOS and T2K collaborations. $M^{RES}_A$ and $M^{RES}_V$ in the R-S dipole form factors are the default GENIE uncertainties 1.12 $\pm$ 0.20 GeV and 0.84 $\pm$ 0.10 GeV (uncorrelated) and are especially significant. The $M^{CCQE}_A$ form factor uncertainty is important at low $Q^2$, as is the GENIE parameter Rvn2pi which scales two-pion production from GENIE DIS events on a quark in a neutron for $W <$ 2 GeV/$c^2$. For this analysis, both are effectively scale factors on the QE and GENIE DIS cross sections. The particular shape at $Q^2$ $\approx$ 0.15 GeV$^2$/$c^2$ is mainly determined by $M^{RES}_A$.

The MINOS $\mu$ matching efficiency uncertainty is associated with the correction applied to account for the difference in the data and MC efficiency of the muon tracks reconstructed in the MINOS detector. The uncertainty of the efficiency correction for neutrinos depends on the muon angle. In the case of antineutrino, it is flat and set at 1\%. 

There is also a systematic uncertainty due to the finite-sized MC samples used in the analysis. The MC samples are four times the size of the data samples.

\section{Results}
The measured differential cross sections of the variables $Q^2$, ${p_{||}}_\mu$, ${p_\mathrm{T}}_\mu$ and $x$ test a wide range of predictions and also provide connections to other on-going MINERvA and community analyses. The muon variables ${p_{||}}_\mu$ and ${p_\mathrm{T}}_\mu$ connect these SIS measurement with inclusive (all $W$) and lower $W$ resonance results while the variables $Q^2$ and $x$ connect these SIS measurements to higher $W$ MINERvA and community experimental results that can be useful in testing quark–hadron duality in neutrino interactions.  

The initial set of predictions for comparison is a series of tuned modifications of the GENIE 2.12.6 production models that were produced by modifying, among other things, the resonant and nonresonant pion production within the default GENIE 2.12.6 generator. The tunes are described as follows:
\begin{enumerate}[(i)]
    \item MINERvA Tune v2, which is our central value, includes a large reduction of nonresonant single pion production applied to GENIE DIS for $W$ $\leq$ 2 GeV/$c^2$ and an additional low $Q^2$ ($\leq$ 0.7 GeV$^2$/$c^2$) suppression of pion production applied to the R-S model. The large reduction in nonresonant single pion production was based on fits to bubble chamber data with $W$ $<$ 1.4 GeV and $Q^2$ $<$ 1 GeV$^2/c^2$ and was simply applied to the higher $W$ and $Q^2$ of this SIS analysis (see section~\ref{Simulation Experiment} for details).
    \item MINERvA Tune v1, which is MINERvA Tune v2 without the reduction in pion production at low $Q^2$.
    \item v2 - Nonres Pion, which is the central value without the reduction in the nonresonant single pion production.
    \item GENIE 2.12.6, which is the version without applied tunes.
\end{enumerate}

To better appreciate how best to model these SIS measurements, it is informative to also determine if other neutrino event simulators with alternative models are consistent with these MINERvA SIS measurements. It is important to emphasize that each considered alternative neutrino/antineutrino simulator employs different treatments of nuclear effects as well as different resonance models. The generators have similar, but not identical, DIS models that depend on the total inelastic cross section based on the B-Y low $Q^2$, low $W$ modifications to the GRV parton distributions \cite{Bodek:2010km}.

A brief summary of the considered simulation programs includes the following:
\begin{enumerate}[(i)]
  \item GENIE 3.0.6: (\cite{GENIE:2022qrc}) This uses the upgraded GENIE generator (G18-02a-02-11a and G18-10a-02-11a)) with modified form factors. These configurations also include a tuning fitting scale factors for resonances, and 1 and 2-$\pi$ nonresonant pion production process. It also includes resonances up to $W$ of 1.93 GeV/c$^2$ mixed with DIS model interactions, after which all events are from the DIS model.
  \item NEUT 5.4.1: (\cite{Hayato:2021heg}) Updated R-S single pion production model, a custom $\geq$ 2 pion model and the B-Y modified GRV DIS model that includes nonresonant pion production. A mix of resonance and DIS in $W$ from 1.3 to 2 GeV/$c^2$ with significant importance of the multiplicity model employed. Above $W$ = 2 GeV/$c^2$ the simulation is pure DIS.
  \item NuWro 19.02: (\cite{Golan:2012rfa} for neutrino only) Their own, not R-S, $\Delta$(1232) resonance model and B-Y modified GRV DIS model including nonresonant pion production. There is a transition mix of resonance and DIS for $W$ between 1.3 and 1.6 GeV/$c^2$, DIS controls predictions above $W$ = 1.6 GeV/$c^2$, which leads to a lower fraction of events from the resonance model.
  \item GiBUU 2021: (\cite{Soplin:2023nxg}) This is a quite different simulation of neutrino interactions compared to the previous generators. The resonance model and the form factors are from Lalakulich \cite{Lalakulich:2010ss} rather than Rein-Sehgal. In addition, it incorporates nuclear transport theory and mean-field dynamics concepts to describe the evolution of hadronic systems in a nuclear potential by considering their interactions through scattering, decay, and production processes. 
\end{enumerate}

Particular versions of the generators were used to predict cross sections. While the version that contains RFG employs the Relativistic Fermi Gas model, the LFG versions employ the Local Fermi Gas model as the nuclear ground state. To describe the final-state interactions, hA is an effective intranuclear transport model that approximates the full shower with a single interaction.

While the figures displaying the measured data points and simulator predictions give a first impression of the comparison, for a more quantitative estimate of the agreement of the SIS measurements with both MINERvA tunes and alternative generators we use the standard $\chi^2$. This is calculated using Eq.~\ref{Chi2_Models}, which takes into account correlations between bins from the measured covariance matrix $V$:
\begin{multline}
    \chi^2 =\sum_{i,j} \left( x_{i,\mathrm{measured}} - x_{i,\mathrm{model}} \right) \times \\ V^{-1}_{ij} \times \left( x_{j,\mathrm{measured}} - x_{j,\mathrm{model}} \right),
    \label{Chi2_Models}	
\end{multline}
where $x_\mathrm{measured}$ is the measured cross section and $x_\mathrm{model}$ is the predicted cross section. The results for both SIS definition I in table \ref{Chi2_ModelsData} and SIS definition II in table \ref{Chi2_ModelsData_SISdefinitionII} contain the comparison of measured data to predictions of tunes of GENIE 2.12.6 as well as the other presented community simulation generator tunes.

\begin{table}[htbp]
	\centering
	\begin{tabular}{lrrrrrr}
        \hline\hline
        \multirow{2}{*}{Process Variant} & \multicolumn{2}{c}{$Q^2$} & \multicolumn{2}{c}{$p_{T\mu}$} & \multicolumn{2}{c}{$p_{\parallel{\mu}}$}    \\ \cline{2-7} 
		\multicolumn{1}{c}{}  &	$\nu_\mu$	&	$\bar{\nu}_\mu$	& $\nu_\mu$   &  $\bar{\nu}_\mu$	&	$\nu_\mu$	&	$\bar{\nu}_\mu$    \\	\hline
        MINERvA Tune v2     & 4.8		& 4.9      & 4.0     & 4.2		& 14.1    &	5.7    \\
        MINERvA Tune v1     & 19.2		& 23.6     & 15.2    & 17.9		& 17.1    &	8.1    \\
		v2 - Non-Res Pion   & 4.7		& 5.1	   & 4.3	 & 4.7	    & 14.3    &	5.5    \\ 
		GENIE 2.12.6        & 18.7		& 19.2     & 14.7	 & 14.6	    & 18.2    &	8.4    \\ \hline
		GENIE 3.0.6 RFG hA  & 6.7		& 5.9      & 6.0	 & 5.9		& 14.0    &	5.5    \\
		GENIE 3.0.6 LFG hA  & 9.8		& 8.3      & 8.3	 & 7.2		& 15.0    &	6.3    \\
		NEUT 5.4.1 LFG      & 18.4		& 13.1     & 14.9    & 8.8		& 16.9    & 4.3    \\
		NuWro 19.02 LFG     & 4.4		& ---      & 4.7     & ---      & 7.0     & ---    \\
        GiBUU 2021          & 18.1		& 14.4     & 16.5    & 11.8     & 17.7    & 7.4    \\  \hline\hline
	\end{tabular}
 	\caption{$\chi^2/ndf$ of various model variants compared to data for the SIS definition I. The number of degrees of freedom for $Q^2$, $p_{T\mu}$, and $p_{\parallel\mu}$ is 10, 13, and 12, respectively.}
 	\label{Chi2_ModelsData}	
\end{table}

\begin{table}[htbp]
	\centering
	\begin{tabular}{lrrrrrrrr}
        \hline\hline
		\multirow{2}{*}{Process Variant}  &	\multicolumn{2}{c}{\textbf{$Q^2$}} & \multicolumn{2}{c}{\textbf{$p_{T\mu}$}} & \multicolumn{2}{c}{\textbf{$p_{\parallel}$}} & \multicolumn{2}{c}{\textbf{$x$}}  \\  \cline{2-9}
		\multicolumn{1}{c}{}  &	$\nu_\mu$	&	$\bar{\nu}_\mu$	& $\nu_\mu$   &  $\bar{\nu}_\mu$	&	$\nu_\mu$	&	$\bar{\nu}_\mu$   & $\nu_\mu$  &  $\bar{\nu}_\mu$     \\	\hline
        MINERvA Tune v2     & 4.8		& 3.7      & 4.5     & 3.5		& 7.3     &	3.0    & 3.6     &	2.0   \\
        MINERvA Tune v1     & 4.8		& 3.7      & 4.5     & 3.5		& 7.3     &	3.0    & 3.6     &	2.0   \\
		v2 - Non-Res Pion   & 1.9		& 2.1	   & 2.5	 & 2.3	    & 7.3     &	2.0    & 0.5     &	0.8   \\ 
		GENIE 2.12.6        & 1.9		& 2.1      & 2.5	 & 2.3	    & 7.3     &	2.0    & 0.5     &  0.8   \\ \hline
		GENIE 3.0.6 RFG hA  & 1.1		& 1.2      & 1.2	 & 1.1		& 8.0     &	1.9    & 1.7     &	1.1   \\
		GENIE 3.0.6 LFG hA  & 1.5		& 1.4      & 1.5	 & 1.2		& 8.5     &	1.9    & 2.3     &	1.2   \\
		NEUT 5.4.1 LFG      & 0.2		& 0.2      & 0.3     & 0.7		& 7.3     & 2.1    & 0.2     &	0.3   \\
		NuWro 19.02 LFG     & 2.4		& ---      & 6.3     & ---      & 8.9     & ---    & 1.2     &	---   \\
        GiBUU 2021          & 6.3		& 4.3      & 6.5     & 4.2      & 7.5     & 2.9    & 6.1     &	4.3   \\  \hline\hline
	\end{tabular}
 	\caption{$\chi^2/ndf$ of various model variants compared to data for the SIS definition II. The number of degrees of freedom for $Q^2$, $p_{T\mu}$, $p_{\parallel\mu}$, and $x$ is 5, 5, 10, and 4 respectively.}
 	\label{Chi2_ModelsData_SISdefinitionII}	
\end{table}

\subsection{SIS Measurements compared to MINERvA Tunes}
\begin{figure*}[htbp]
    \centering
    \xincludegraphics[width=0.45\textwidth,label=(a),pos=se,labelbox=false,fontsize=\large]{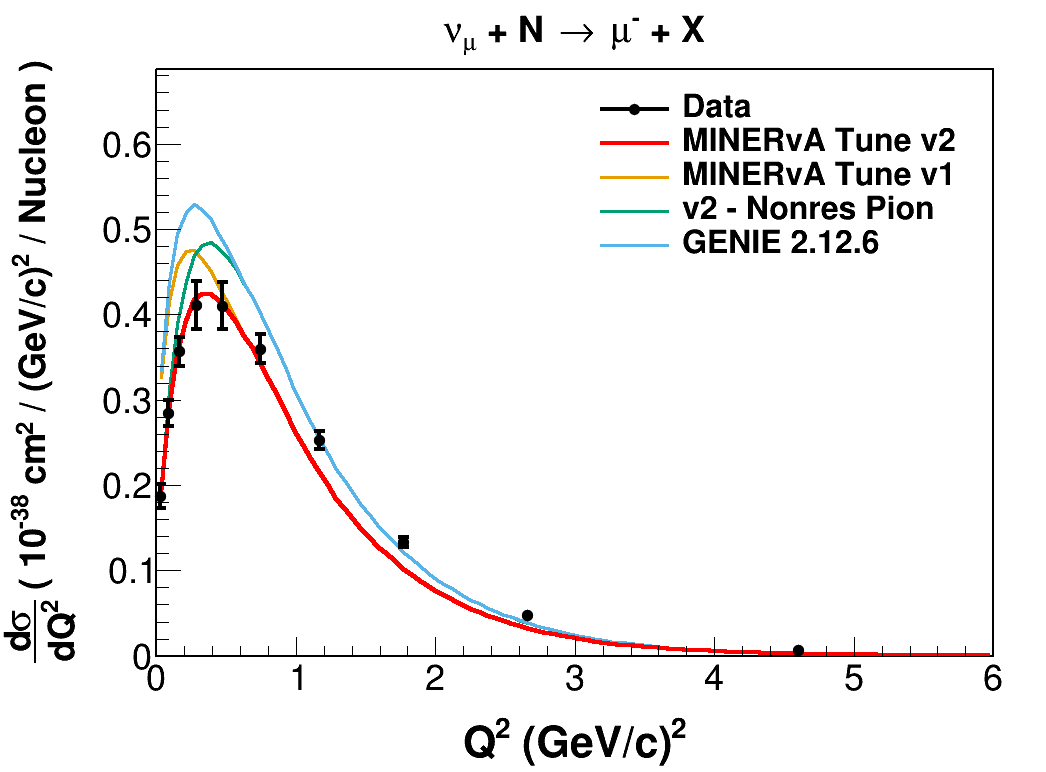}
    \xincludegraphics[width=0.45\textwidth,label=(b),pos=se,labelbox=false,fontsize=\large]{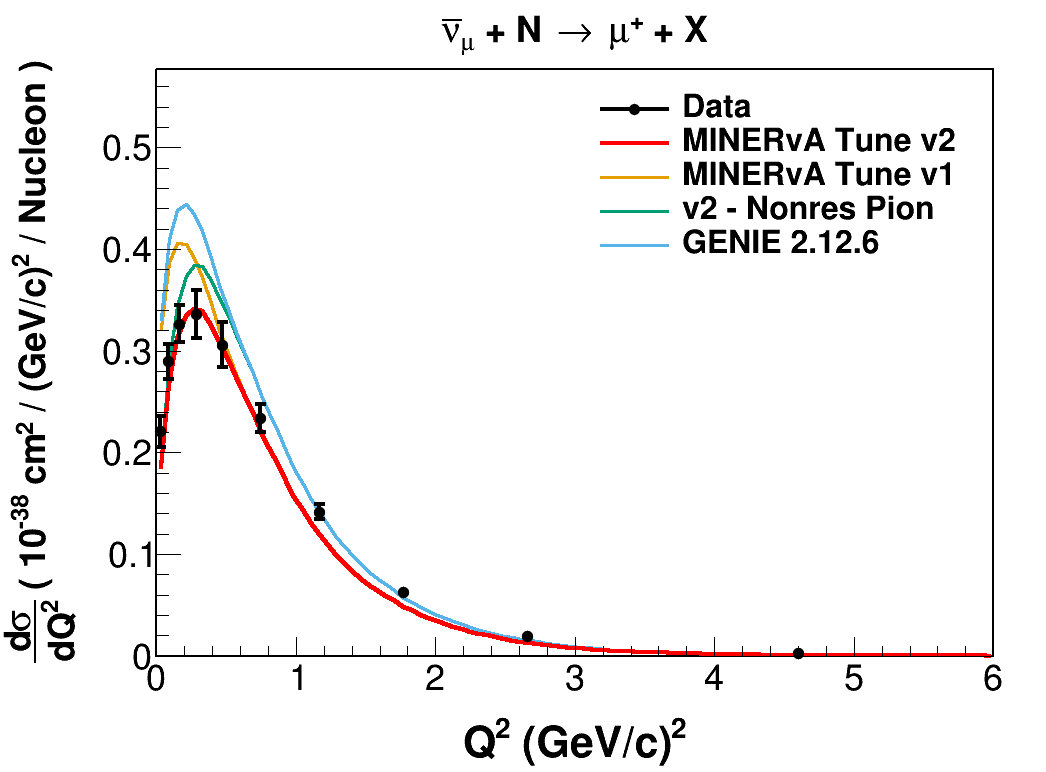}
    
    \xincludegraphics[width=0.45\textwidth,label=(c),pos=se,labelbox=false,fontsize=\large]{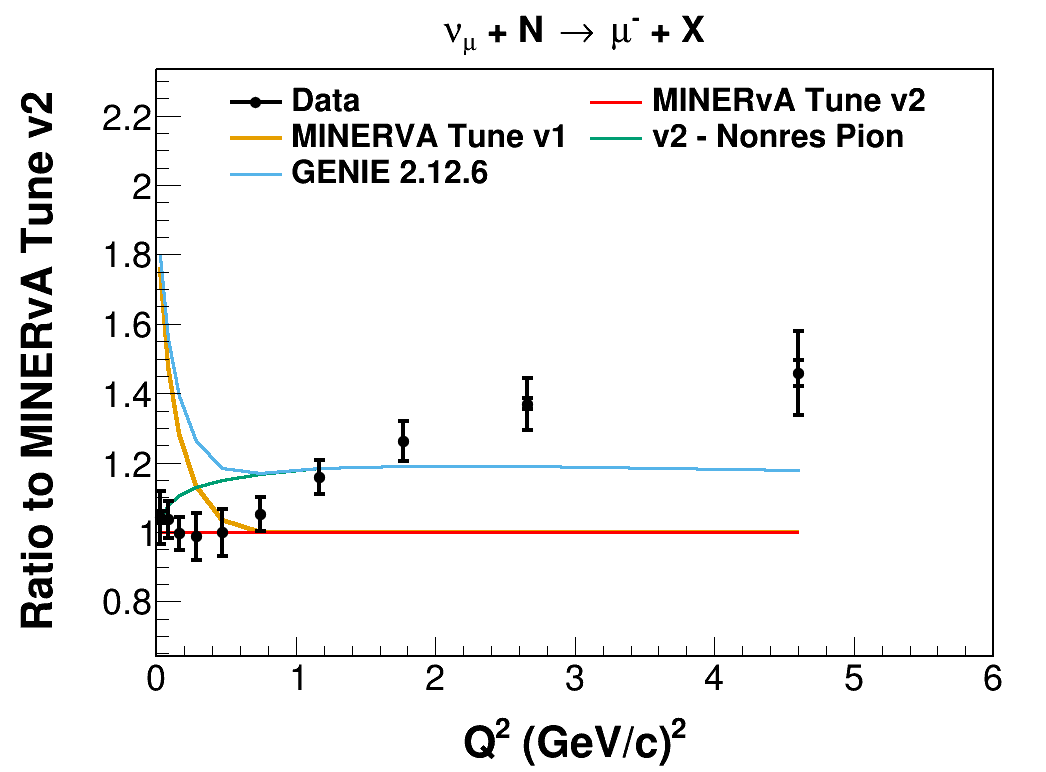}
    \xincludegraphics[width=0.45\textwidth,label=(d),pos=se,labelbox=false,fontsize=\large]{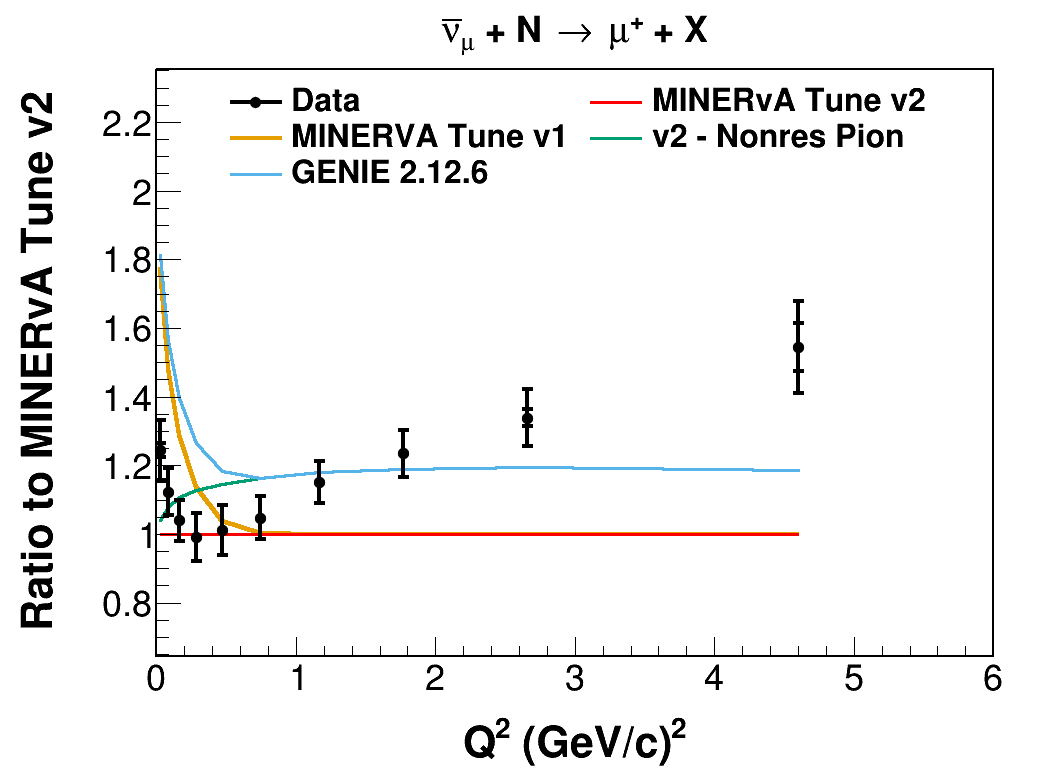}
    
    \xincludegraphics[width=0.45\textwidth,label=(e),pos=se,labelbox=false,fontsize=\large]{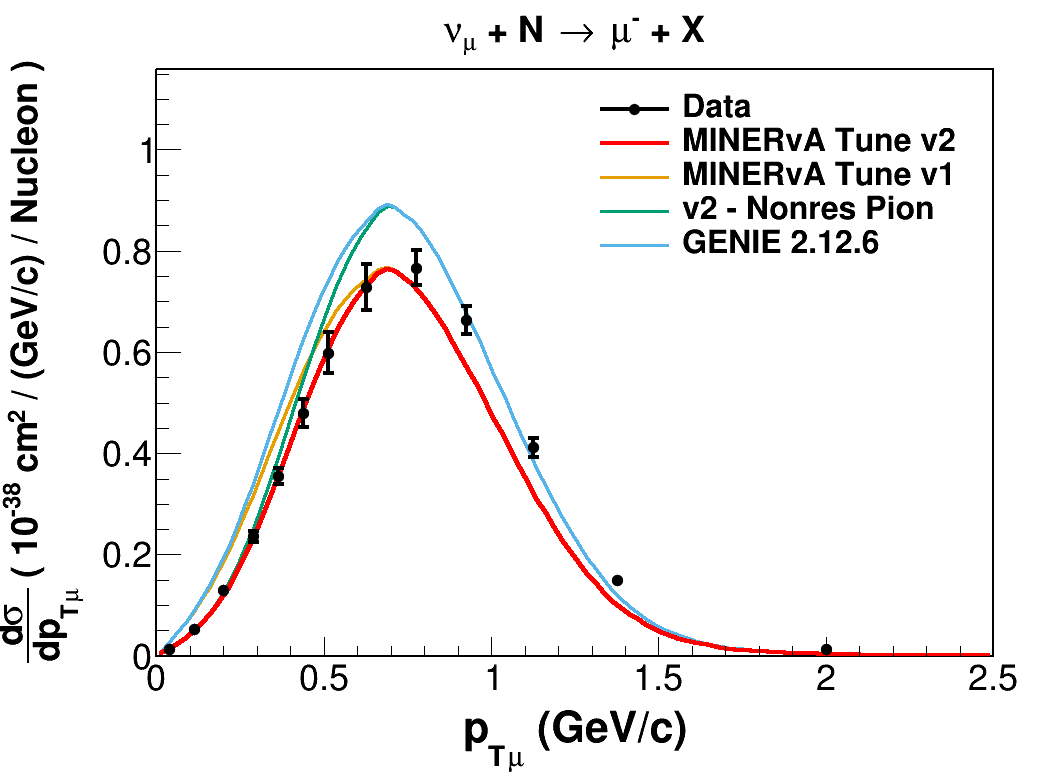}
    \xincludegraphics[width=0.45\textwidth,label=(f),pos=se,labelbox=false,fontsize=\large]{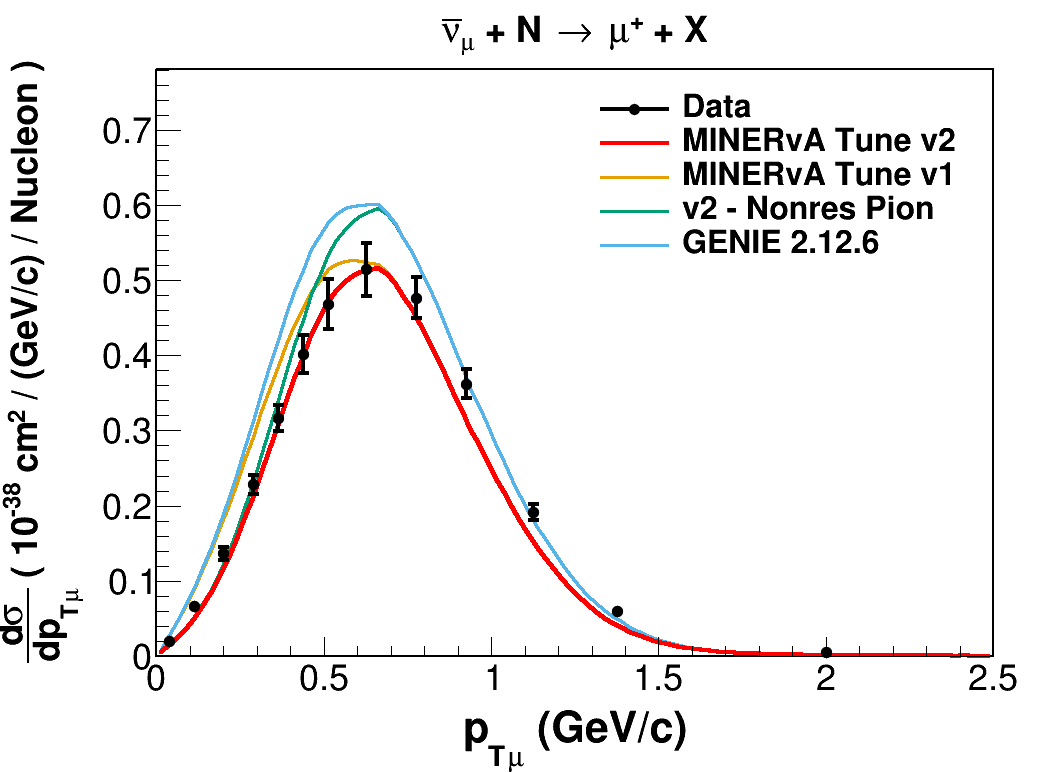}
    
    \xincludegraphics[width=0.45\textwidth,label=(g),pos=se,labelbox=false,fontsize=\large]{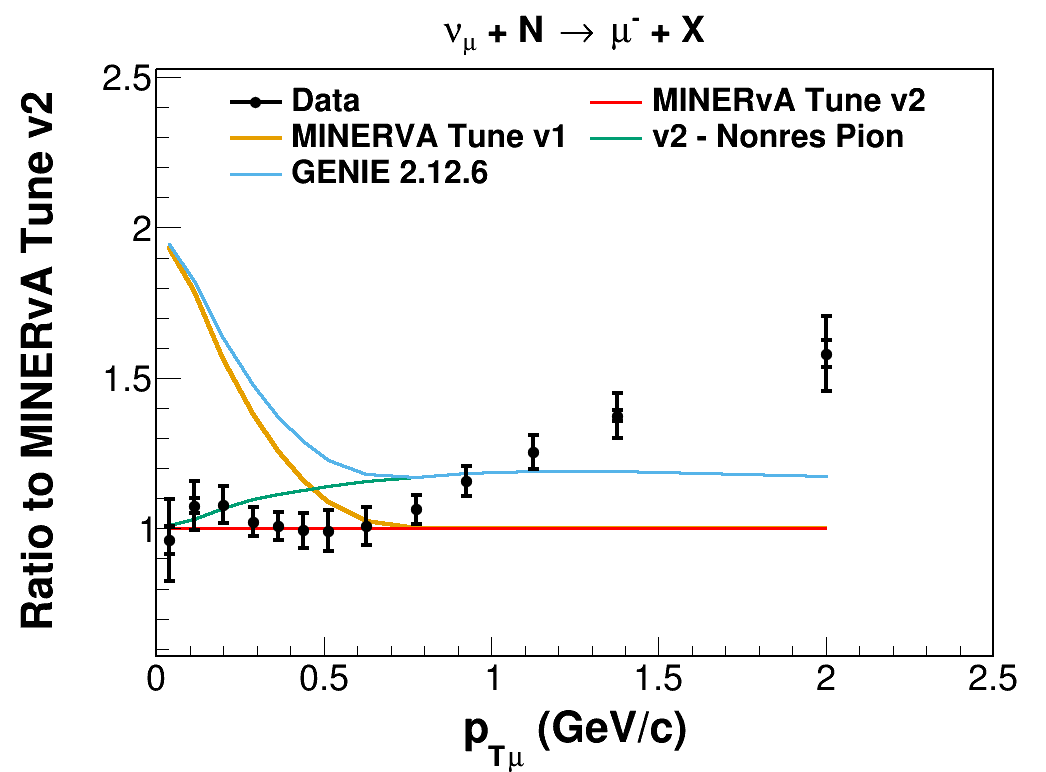}
    \xincludegraphics[width=0.45\textwidth,label=(h),pos=se,labelbox=false,fontsize=\large]{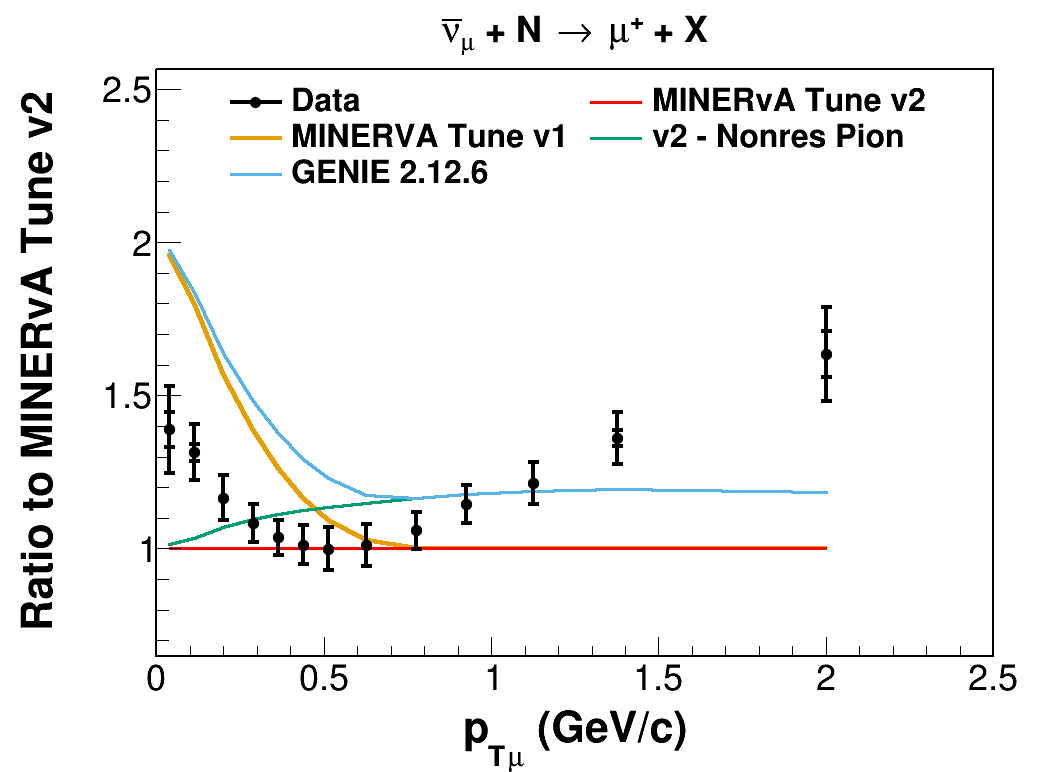}
    
    \caption{Definition I measured differential SIS cross sections (data) compared to MINERvA tunes and ratios of data and alternative MINERvA tunes to MINERvA tune v2 as a function of $Q^2$ (a-d) and $p_{T\mu}$ (e-h). Left: Neutrino. Right: Antineutrino.}
    \label{CrossSectionModelsI_Q2}
\end{figure*}

\begin{figure*}[htbp]
    \centering
    \xincludegraphics[width=0.45\textwidth,label=(a),pos=se,labelbox=false,fontsize=\large]{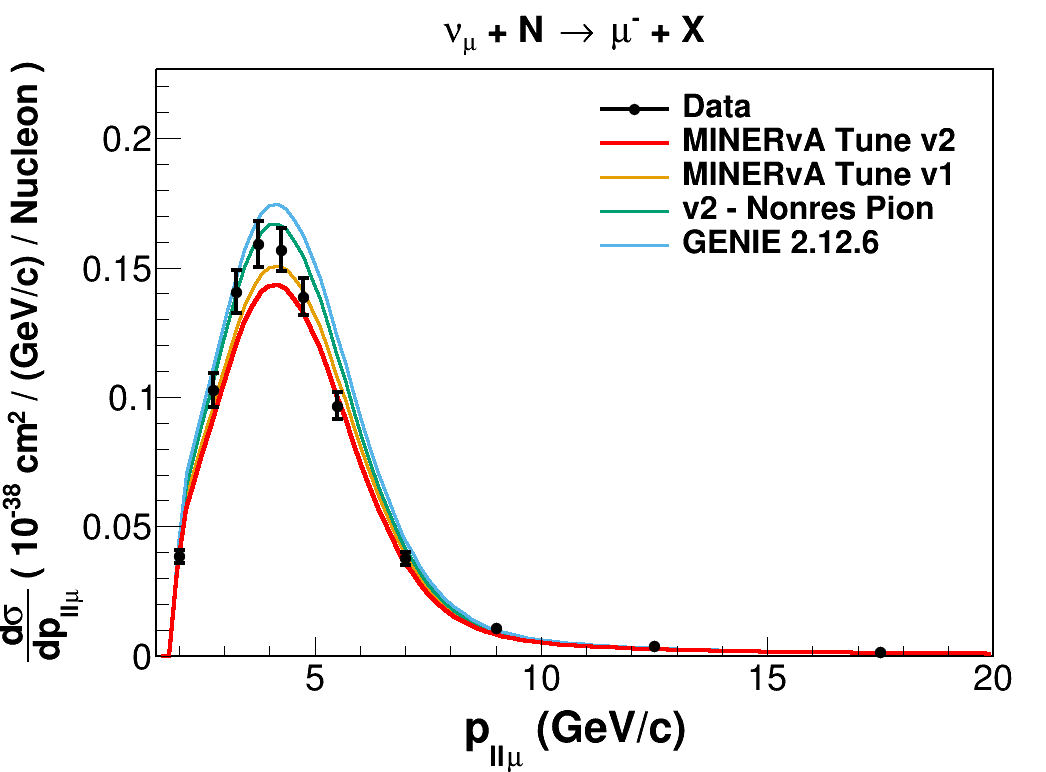}
    \xincludegraphics[width=0.45\textwidth,label=(b),pos=se,labelbox=false,fontsize=\large]{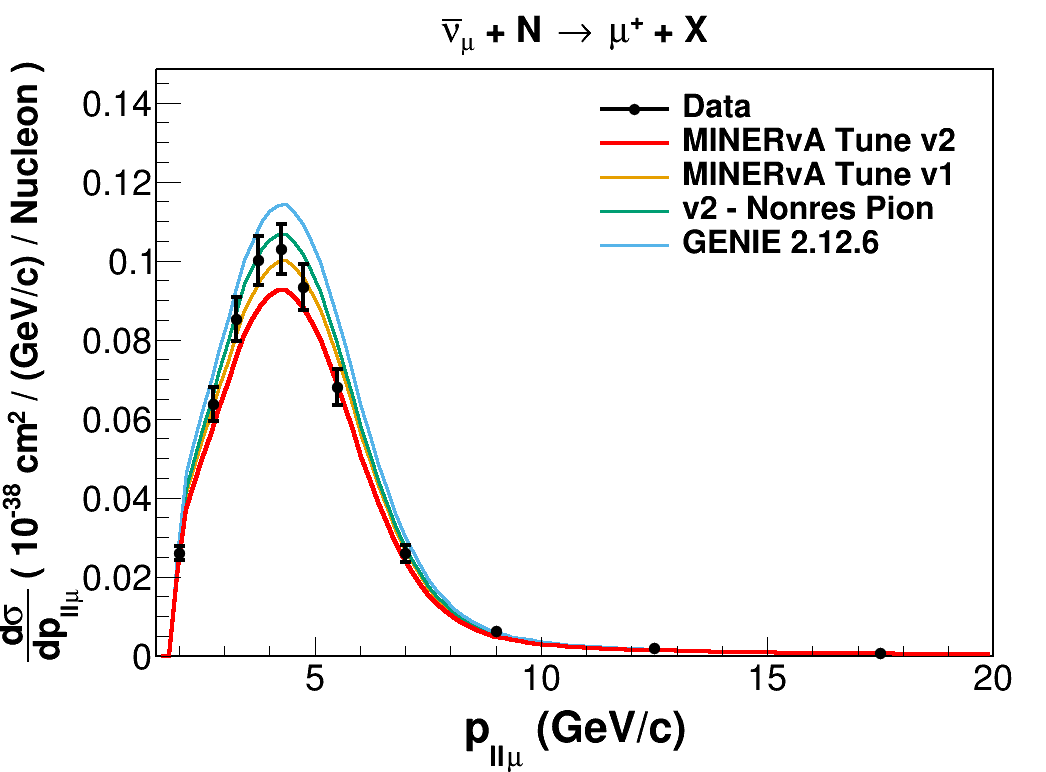}
    
    \xincludegraphics[width=0.45\textwidth,label=(c),pos=se,labelbox=false,fontsize=\large]{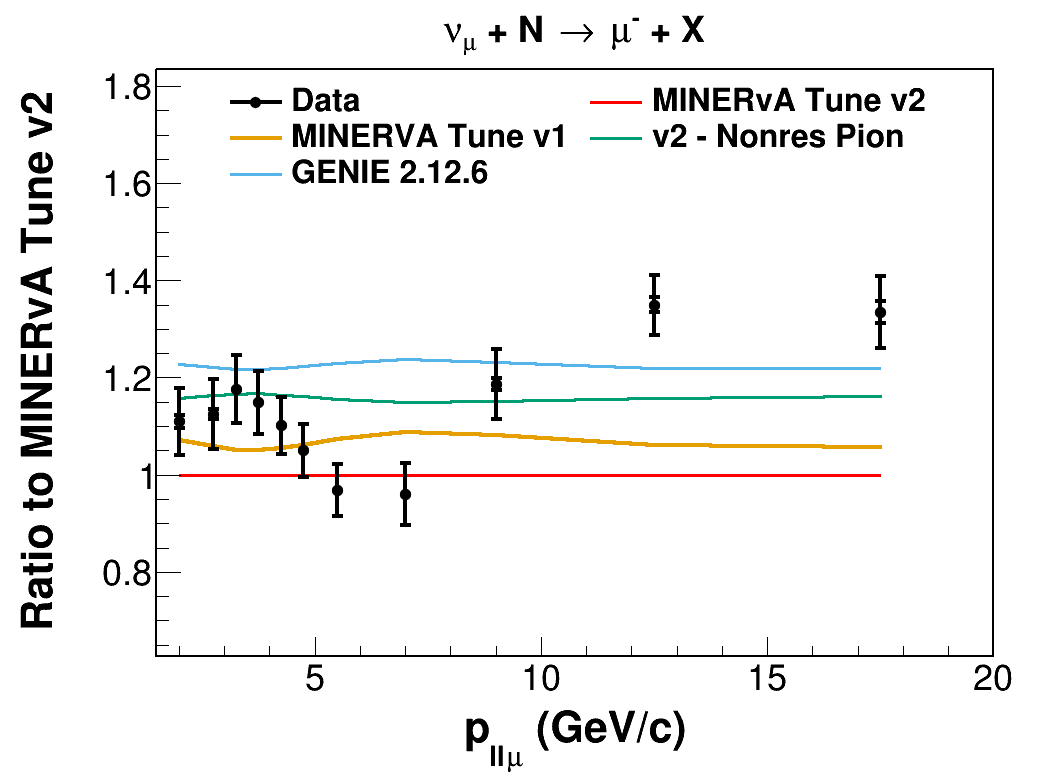}
    \xincludegraphics[width=0.45\textwidth,label=(d),pos=se,labelbox=false,fontsize=\large]{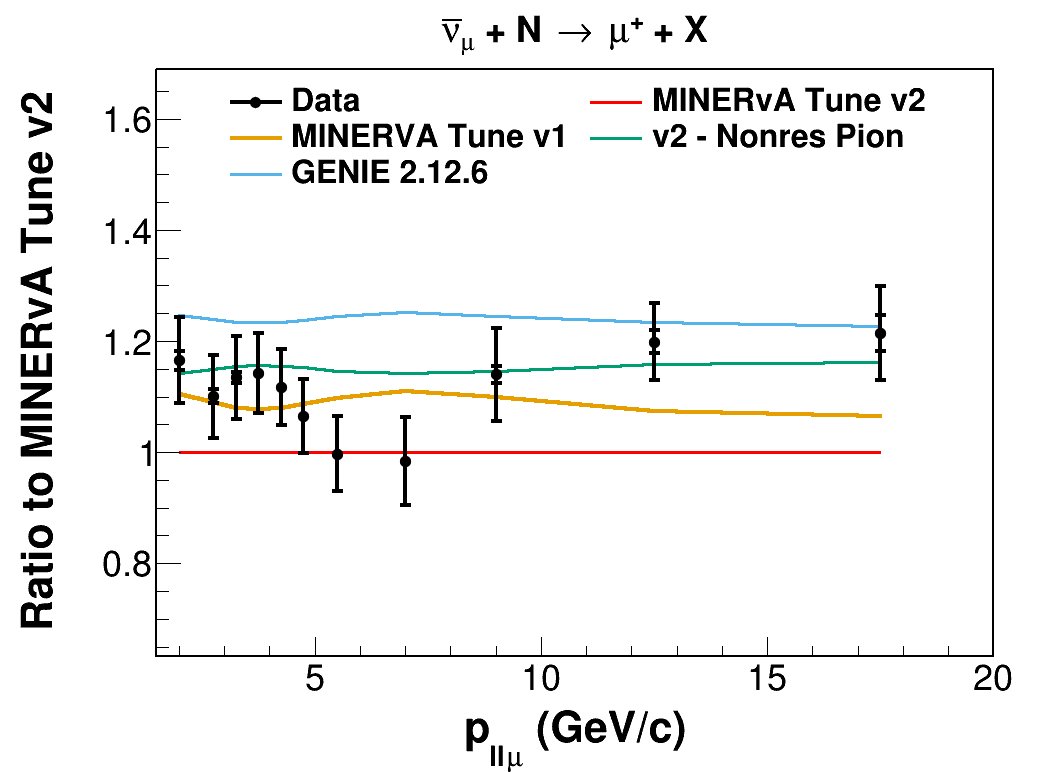}
    
    \caption{Definition I measured differential SIS cross sections (data) compared to MINERvA tunes and ratios of data and alternative MINERvA tunes to MINERvA tune v2 as a function of $p_{\parallel\mu}$. Left: Neutrino. Right: Antineutrino.}
    \label{CrossSectionModelsI_pll}
\end{figure*}

\begin{figure*}[htbp]
    \centering
    \xincludegraphics[width=0.45\textwidth,label=(a),pos=se,labelbox=false,fontsize=\large]{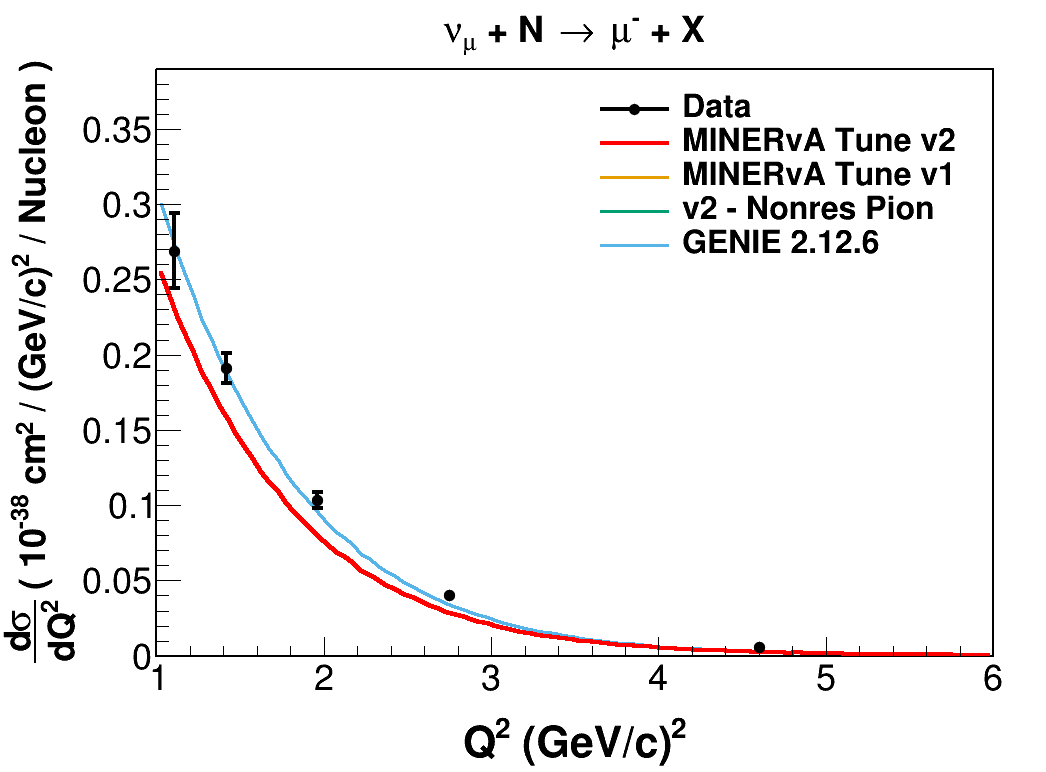}
    \xincludegraphics[width=0.45\textwidth,label=(b),pos=se,labelbox=false,fontsize=\large]{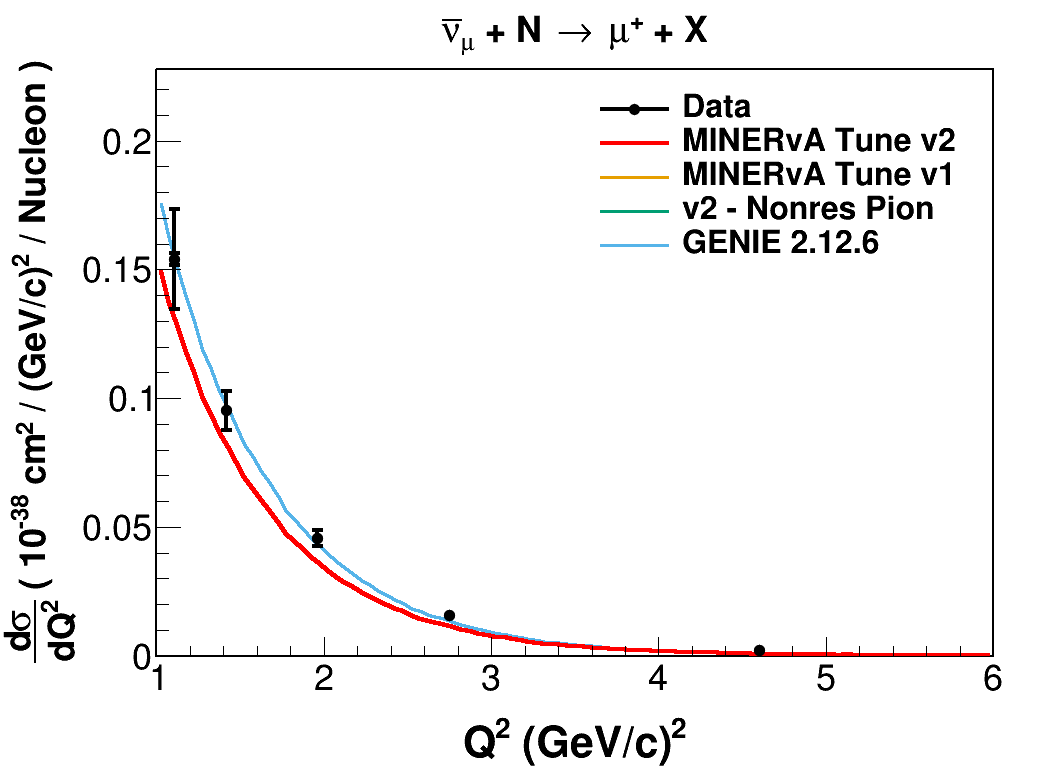}
    
    \xincludegraphics[width=0.45\textwidth,label=(c),pos=se,labelbox=false,fontsize=\large]{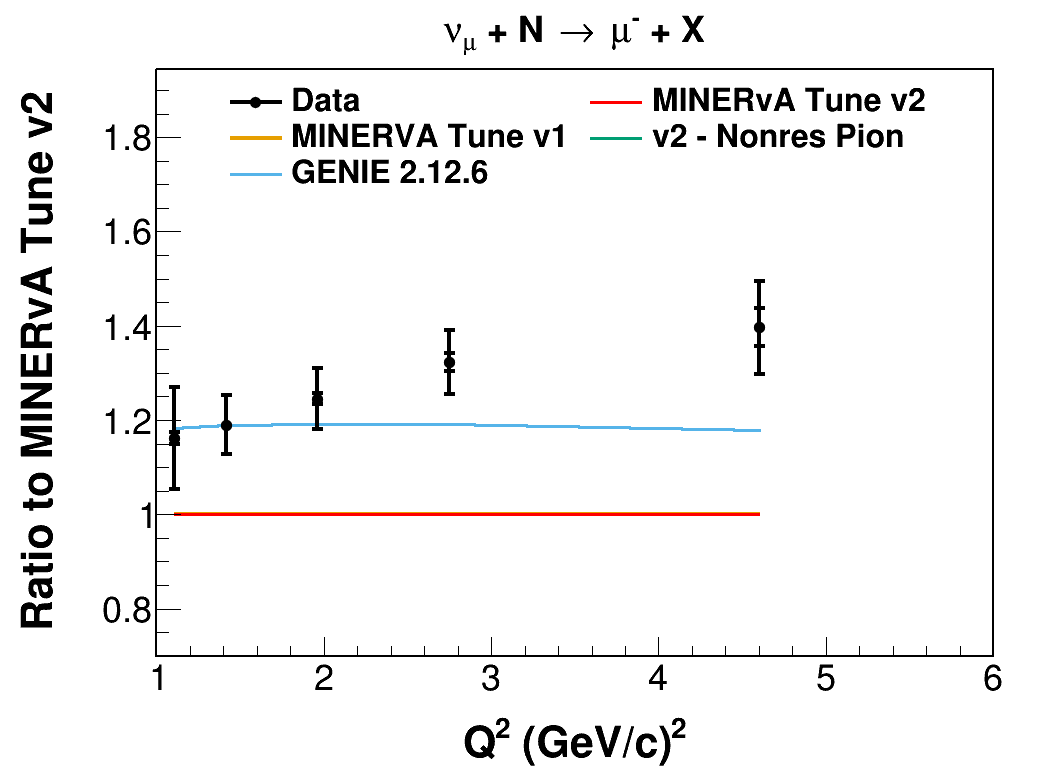}
    \xincludegraphics[width=0.45\textwidth,label=(d),pos=se,labelbox=false,fontsize=\large]{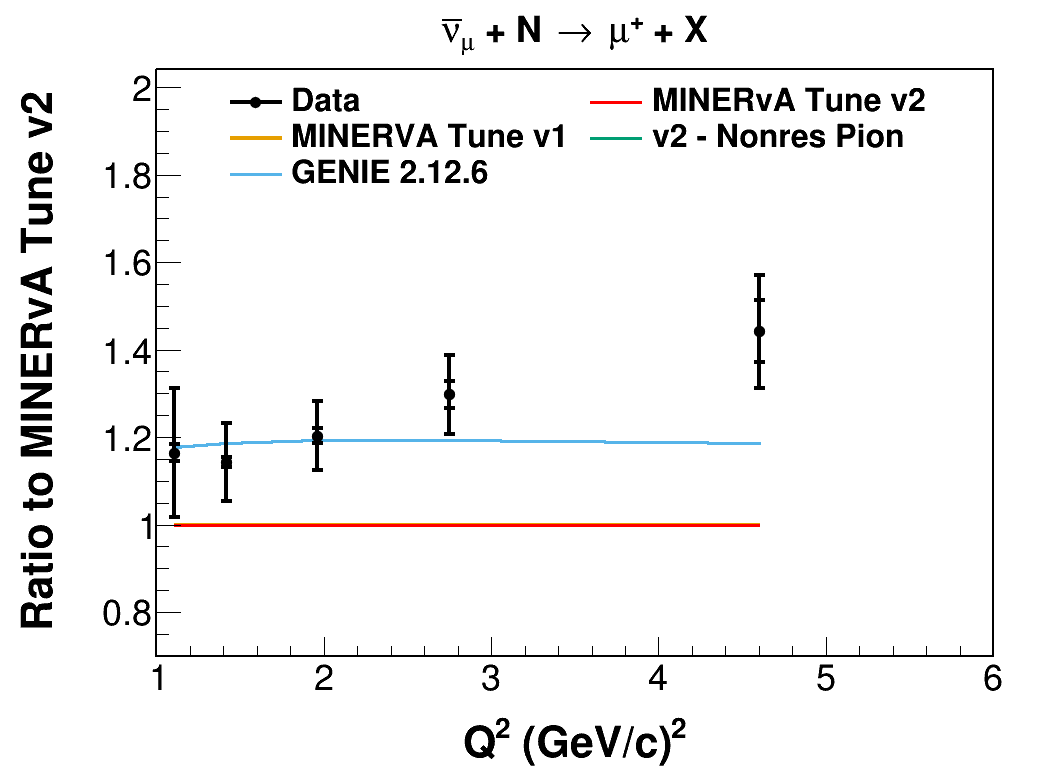}
    
    \xincludegraphics[width=0.45\textwidth,label=(e),pos=se,labelbox=false,fontsize=\large]{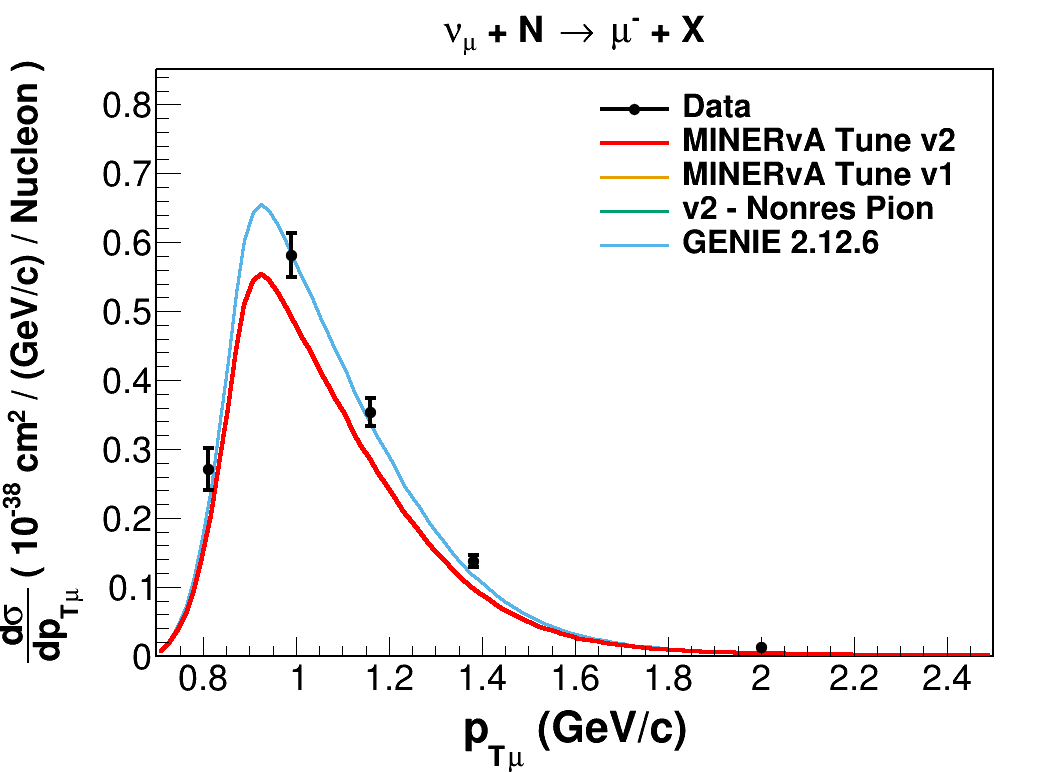}
    \xincludegraphics[width=0.45\textwidth,label=(f),pos=se,labelbox=false,fontsize=\large]{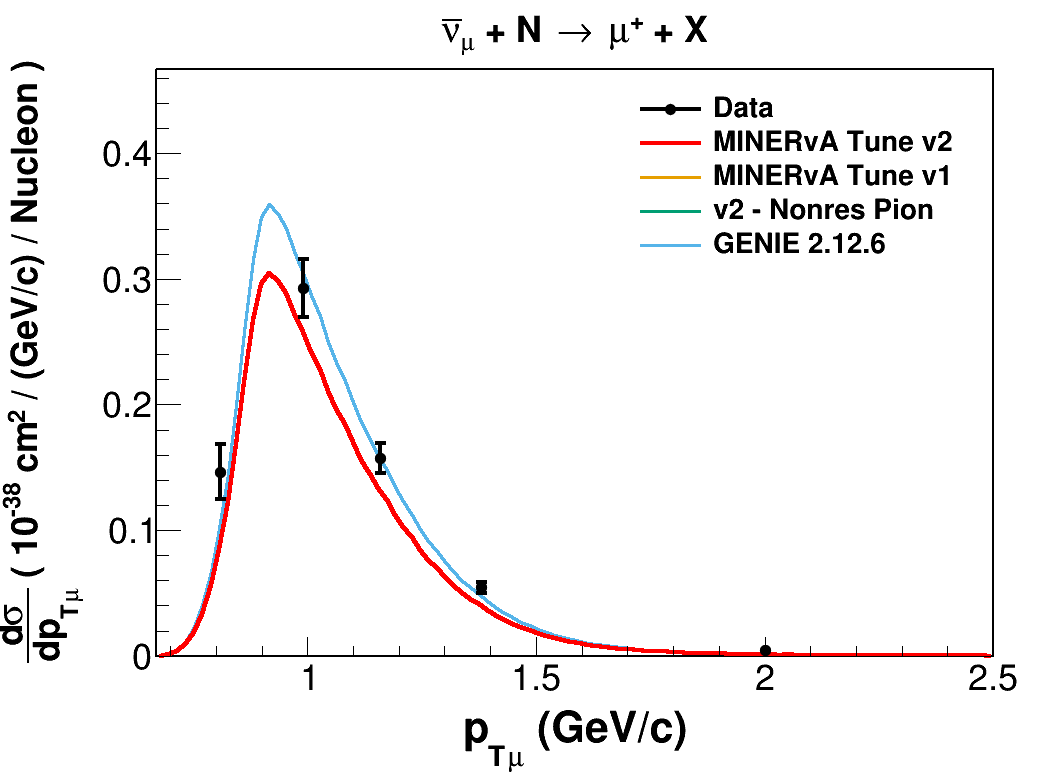}
    
    \xincludegraphics[width=0.45\textwidth,label=(g),pos=se,labelbox=false,fontsize=\large]{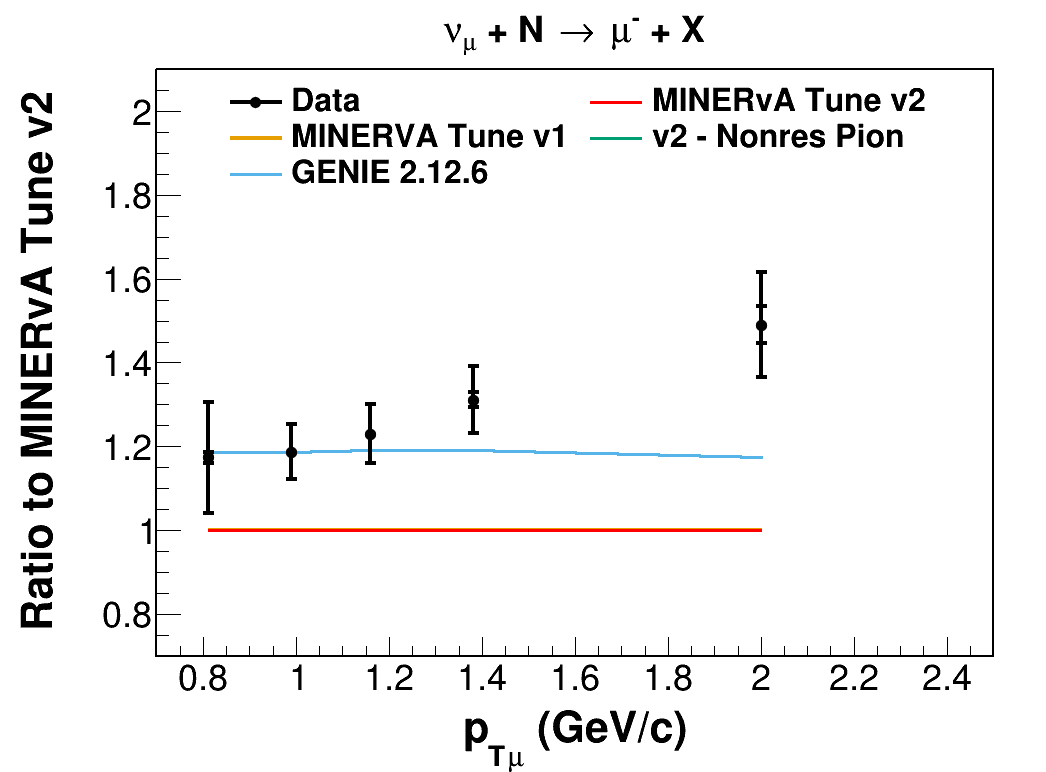}
    \xincludegraphics[width=0.45\textwidth,label=(h),pos=se,labelbox=false,fontsize=\large]{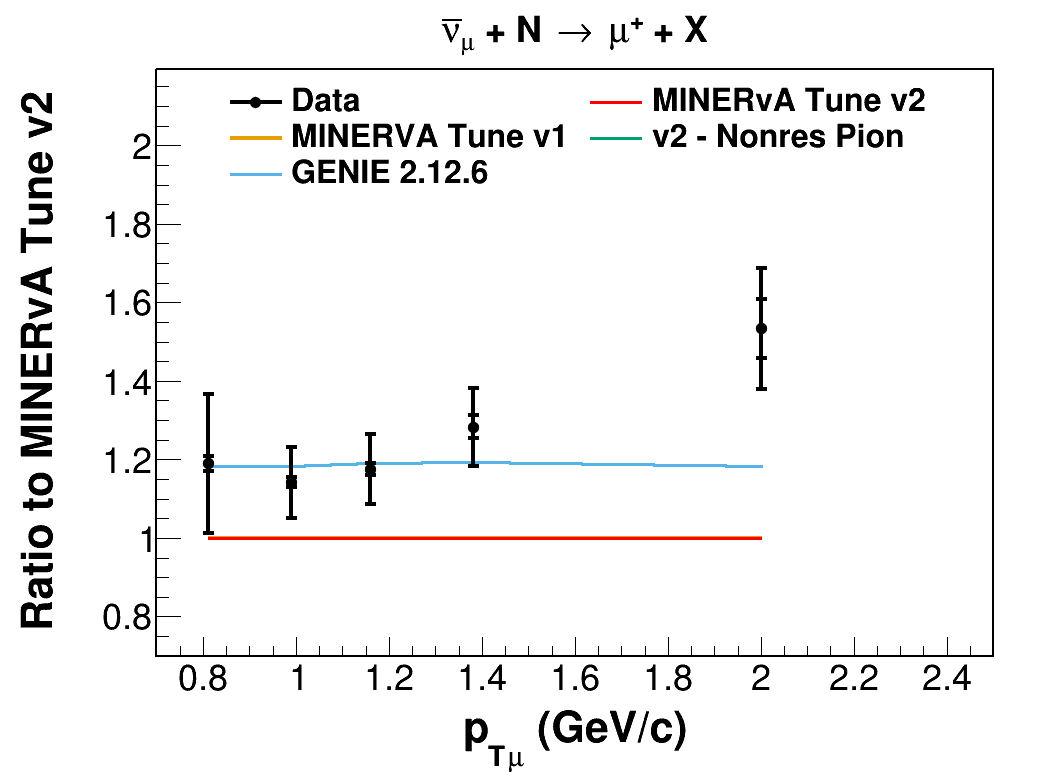}
    
    \caption{Definition II measured differential SIS cross sections (data) compared to MINERvA tunes and ratios of data and alternative MINERvA tunes to MINERvA tune v2 as a function of $Q^2$ (a-d) and $p_{T\mu}$ (e-h). Left: Neutrino. Right: Antineutrino.}
    \label{CrossSectionModelsII_Q2}
\end{figure*}

\begin{figure*}[htbp]
    \centering
    \xincludegraphics[width=0.45\textwidth,label=(a),pos=se,labelbox=false,fontsize=\large]{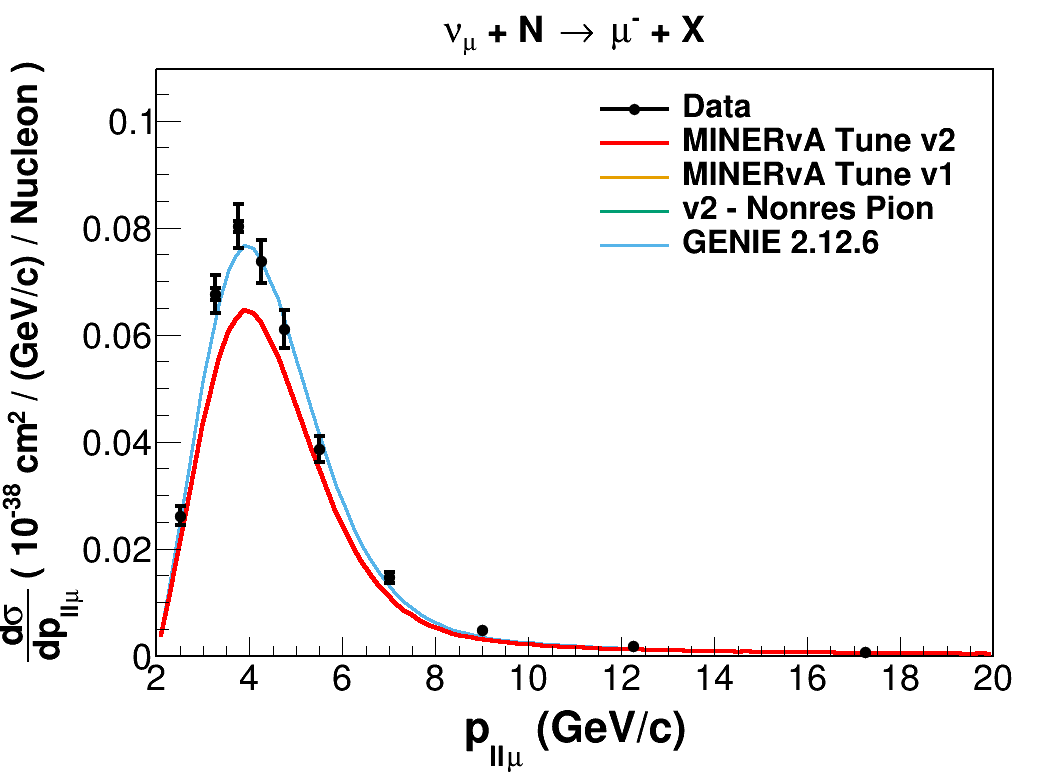}
    \xincludegraphics[width=0.45\textwidth,label=(b),pos=se,labelbox=false,fontsize=\large]{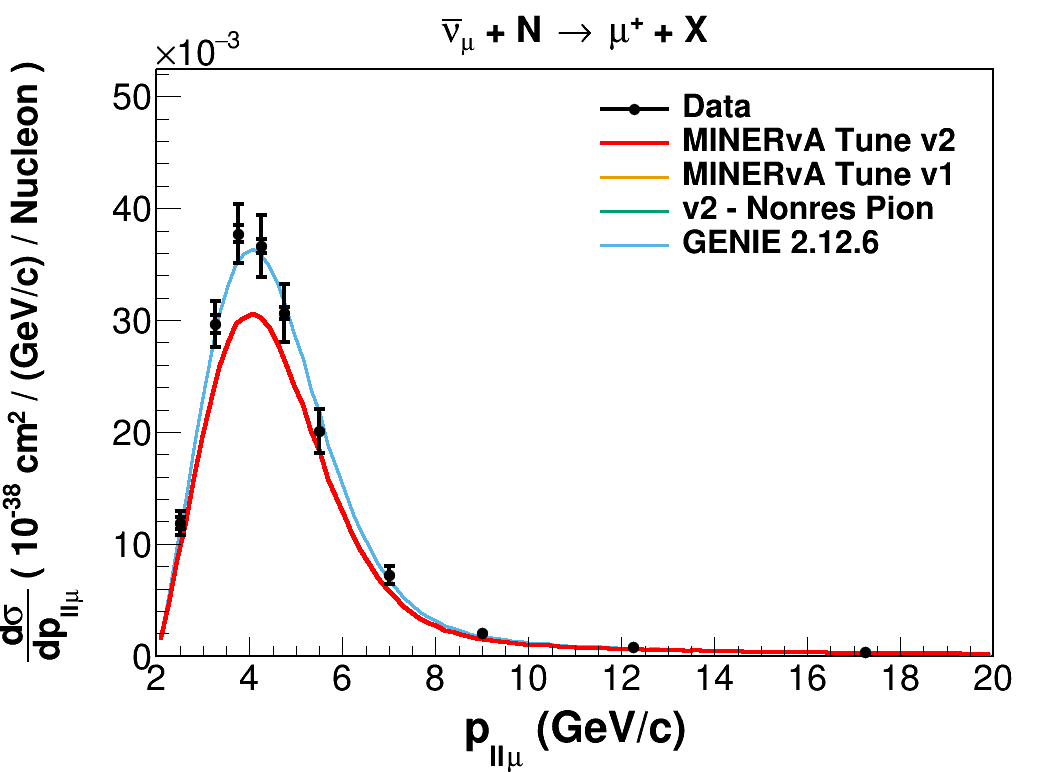}
    
    \xincludegraphics[width=0.45\textwidth,label=(c),pos=se,labelbox=false,fontsize=\large]{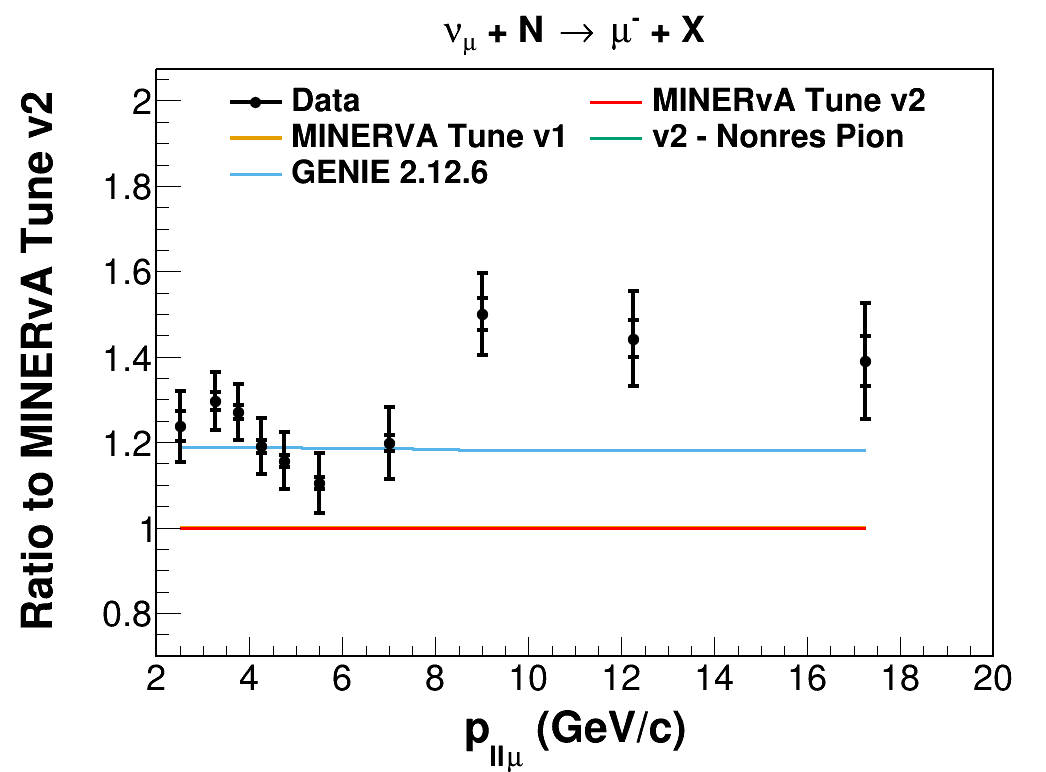}
    \xincludegraphics[width=0.45\textwidth,label=(d),pos=se,labelbox=false,fontsize=\large]{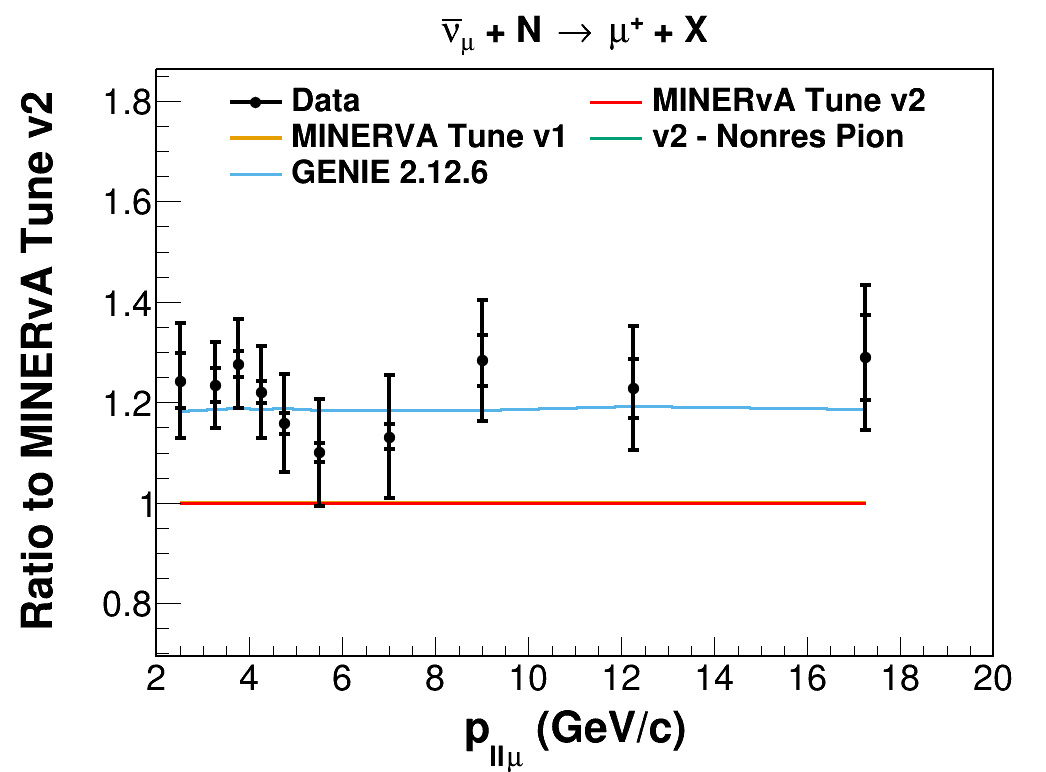}
    
    \xincludegraphics[width=0.45\textwidth,label=(e),pos=se,labelbox=false,fontsize=\large]{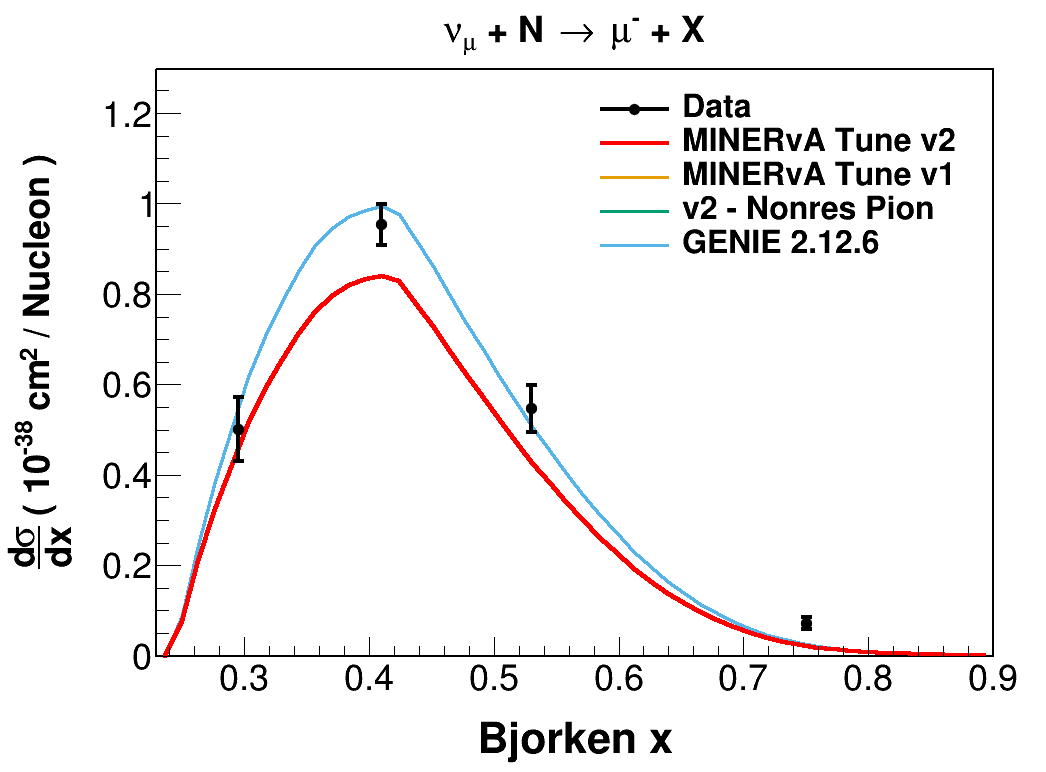}
    \xincludegraphics[width=0.45\textwidth,label=(f),pos=se,labelbox=false,fontsize=\large]{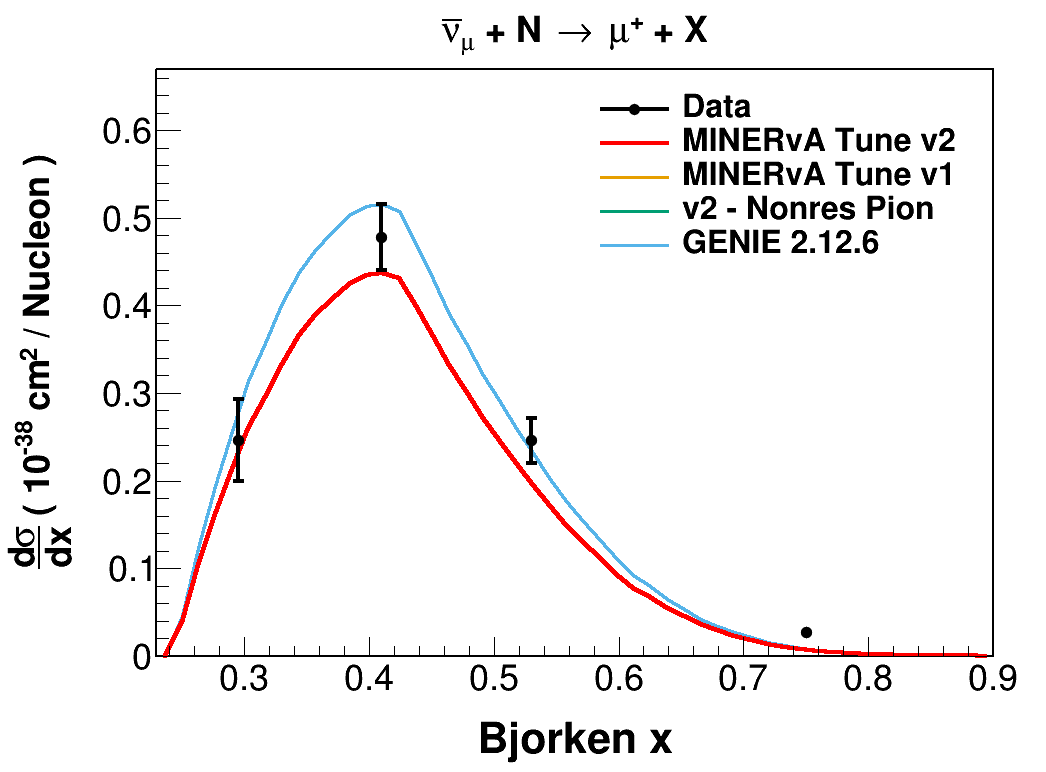}
    
    \xincludegraphics[width=0.45\textwidth,label=(g),pos=se,labelbox=false,fontsize=\large]{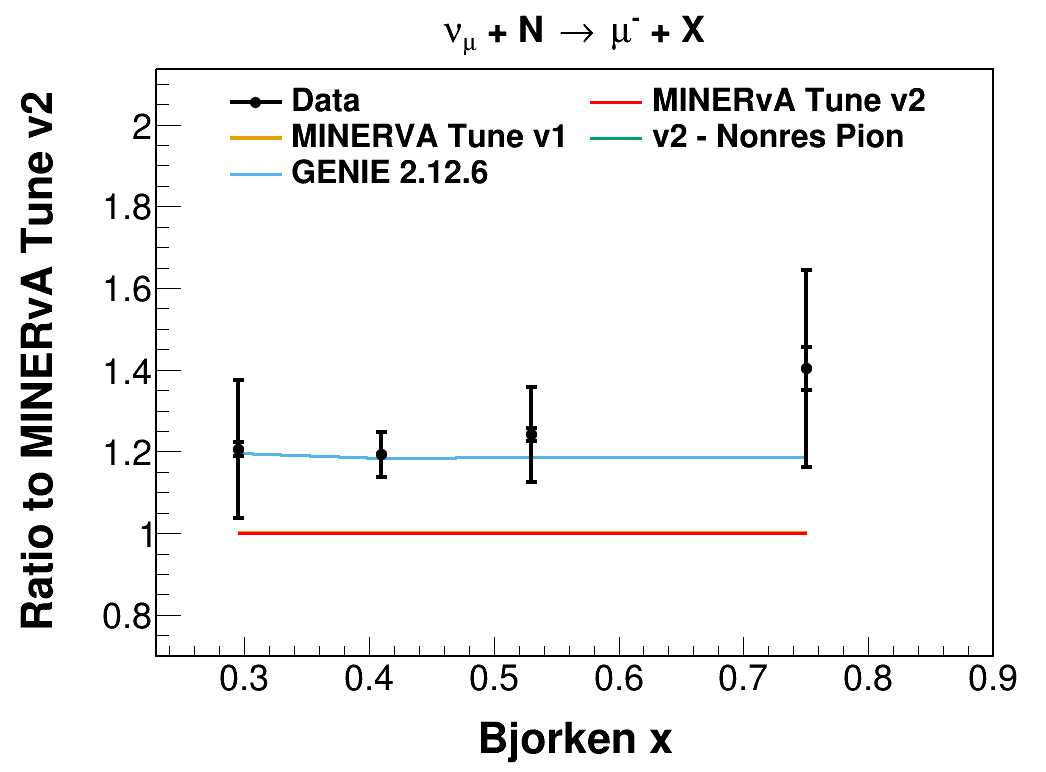}
    \xincludegraphics[width=0.45\textwidth,label=(h),pos=se,labelbox=false,fontsize=\large]{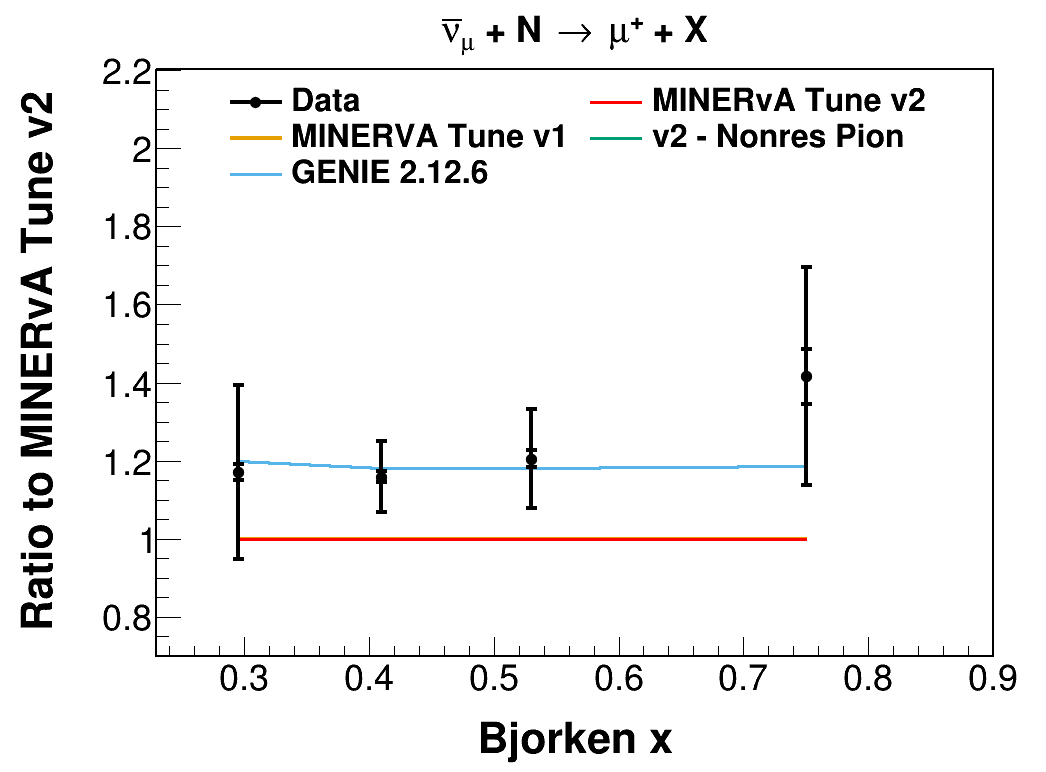}

    \caption{Definition II measured differential SIS cross sections (data) compared to MINERvA tunes and ratios of data and alternative MINERvA tunes to MINERvA tune v2 as a function of $p_{\parallel\mu}$ (a-d) and $x$ (e-h). Left: Neutrino. Right: Antineutrino.}
    \label{CrossSectionModelsII_pll}
\end{figure*}

The SIS measured and the MINERvA Tunes predicted differential cross sections of the variables $Q^2$, ${p_\mathrm{T}}_\mu$, ${p_{||}}_\mu$ are shown in Figures \ref{CrossSectionModelsI_Q2} and \ref{CrossSectionModelsI_pll} for SIS definition I and with Figures \ref{CrossSectionModelsII_Q2} and \ref{CrossSectionModelsII_pll} that also include the variable $x$ for the SIS definition II with neutrino results in the left column and antineutrino in the right column. The data points contain statistical and systematic uncertainties. The clearest representation of the proximity of SIS measurements to tuned model predictions is provided by the ratios of these measurements to predictions. Using our central value MINERvA Tune v2 as the consequent, the presented figures for each definition also display the ratio of SIS measurements and predictions of alternative MINERvA tunes to MINERvA Tune v2. 

\subsubsection{SIS definition I}
Concentrating on the ratios from Figures \ref{CrossSectionModelsI_Q2} and \ref{CrossSectionModelsI_pll} it is obvious that MINERvA Tune v2 does not correctly simulate the SIS measured data throughout the full kinematic range of the variables. This observation is supported by the top four rows of Table \ref{Chi2_ModelsData} that display the $\chi^2/ndf$ for the SIS definition I measured data relative to the predictions of the GENIE 2.12.6 - based MINERvA tunes. Although the $\chi^2/ndf$ of MINERvA Tune v2 to the measured data cannot be considered good, it is clearly a better description of the measured data than MINERvA Tune v1 that supports the introduction of the low $Q^2$ suppression of pion production of MINERvA Tune v2. 

Although the difference in $\chi^2/ndf$ of MINERvA Tune v2  with and without the nonresonant pion reduction is minimal when determined over the entire $Q^2$ $\geq$ 0 GeV$^2$/$c^2$ region, the ratios suggest that this reduction is important for neutrino $Q^2$ $<$ 1 GeV$^2$/$c^2$ and $p_{T\mu}$ $<$ 0.7 GeV/c. It is somewhat less helpful for antineutrino in these kinematic regions and contributes to the underprediction of MINERvA Tune v2 for higher values of $Q^2$ in the SIS multi-quark region for both neutrino and antineutrino. MINERvA tune v2 also underpredicts most ${p_{||}}_\mu$ measurements for neutrino and antineutrino, particularly at higher neutrino ${p_{||}}_\mu$ values. It is significant to note that for $p_{\parallel\mu}$ all antineutrino tunes predict the data better than the neutrino tunes. These $p_{\parallel\mu}$ measurements are sensitive to differences between the constrained simulation and the actual neutrino flux. 

Noting that the higher $Q^2$ region will be discussed in the SIS Definition II analysis, we specifically examine the lower $Q^2$ and the kinematically related lower ${p_\mathrm{T}}_\mu$ ratios. With the low $Q^2$ suppression of pion production, MINERvA Tune v2 neutrino correctly simulates low $Q^2$ and low ${p_\mathrm{T}}_\mu$ at and before the cross section peaks while the MINERvA Tune v2 antineutrino results show considerable underprediction of these measured low $Q^2$ and low ${p_\mathrm{T}}_\mu$ data before these peaks. This might suggest that the ad hoc low $Q^2$ suppression of pion production, still without a sound theoretical basis, is too strong when applied to very low $Q^2$ antineutrino. It is also significant to recall that a 2021/2022 modification of the axial-vector contribution of the B-Y \cite{Bodek:2021bde} model was not included in GENIE 2.12.6 or resulting tunes. The modification correctly includes the non-zero contribution of the axial vector current at $Q^2$ = 0 GeV$^2$/$c^2$ thus mainly affecting the low $Q^2$ simulations and resulting predictions.

\subsubsection{SIS definition II}
The ratio plots from Figures \ref{CrossSectionModelsII_Q2} and \ref{CrossSectionModelsII_pll} as well as the top four rows of Table~\ref{Chi2_ModelsData_SISdefinitionII} summarize the results for the SIS definition II multi-quark region. The reduction in the nonresonant single pion production is the only MINERvA tune affecting events in this region. For this reason, Tune v1 predictions become the same as the ones from Tune v2, and Tune v2 - Nonres Pion predictions become the same as the ones from GENIE 2.12.6.
With increasing values of $Q^2$ and ${p_\mathrm{T}}_\mu$ for both neutrino and antineutrino the underpredictions of Tune v2 compared to data are significant and growing.   For the measured $x$ and ${p_{||}}_\mu$ data in this SIS multi-quark region there is also significant underprediction of MINERvA Tune v2 for both neutrino and antineutrino results throughout the full range of the variables. 
Note that in this higher $Q^2$ range the measured data of all variables, including $x$, seem to be considerably better simulated by MINERvA Tune v2 without the nonresonant pion reduction. Referring to Table~\ref{Chi2_ModelsData_SISdefinitionII}, indeed $\chi^2/ndf$ decreases rather significantly for $Q^2$, ${p_\mathrm{T}}_\mu$ and $x$ when the nonresonant pion reduction is removed. Since the modifications contained in both MINERvA Tunes v2 and v1 are based on fits and observations from MINERvA and bubble chamber data primarily outside the $W$ - $Q^2$ kinematic region of this analysis, this comparison of data to predictions reveals mixed results when these modifications are extrapolated into the examined SIS kinematic region, suggesting that a $W$ dependence to the nonresonant reduction should be considered.

\subsection{Comparison of SIS Measurements to other simulation programs}
\begin{figure*}[htbp]
    \centering
    \xincludegraphics[width=0.45\textwidth,label=(a),pos=se,labelbox=false,fontsize=\large]{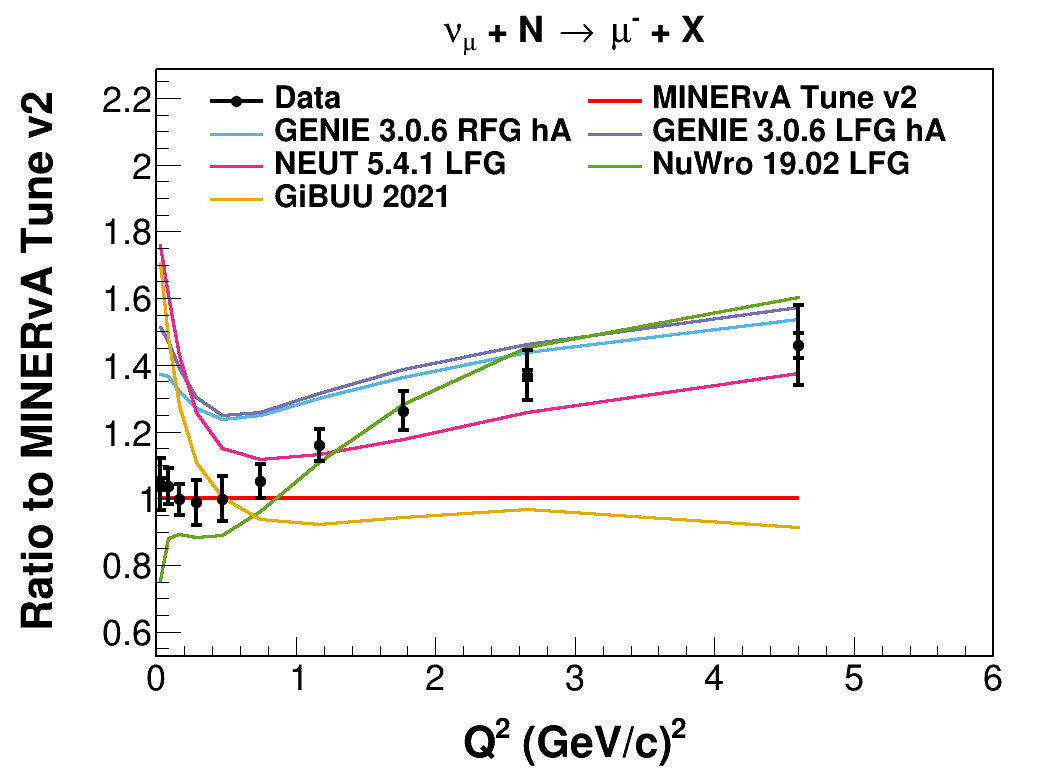}
    \xincludegraphics[width=0.45\textwidth,label=(b),pos=se,labelbox=false,fontsize=\large]{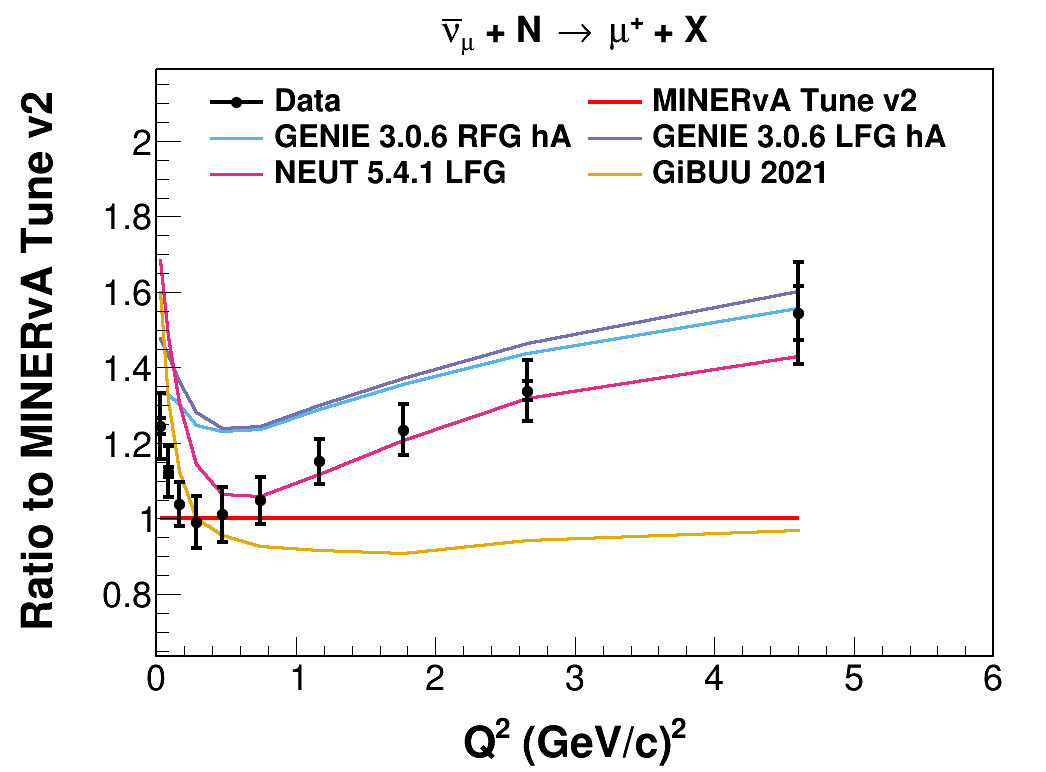}
    
    \xincludegraphics[width=0.45\textwidth,label=(c),pos=se,labelbox=false,fontsize=\large]{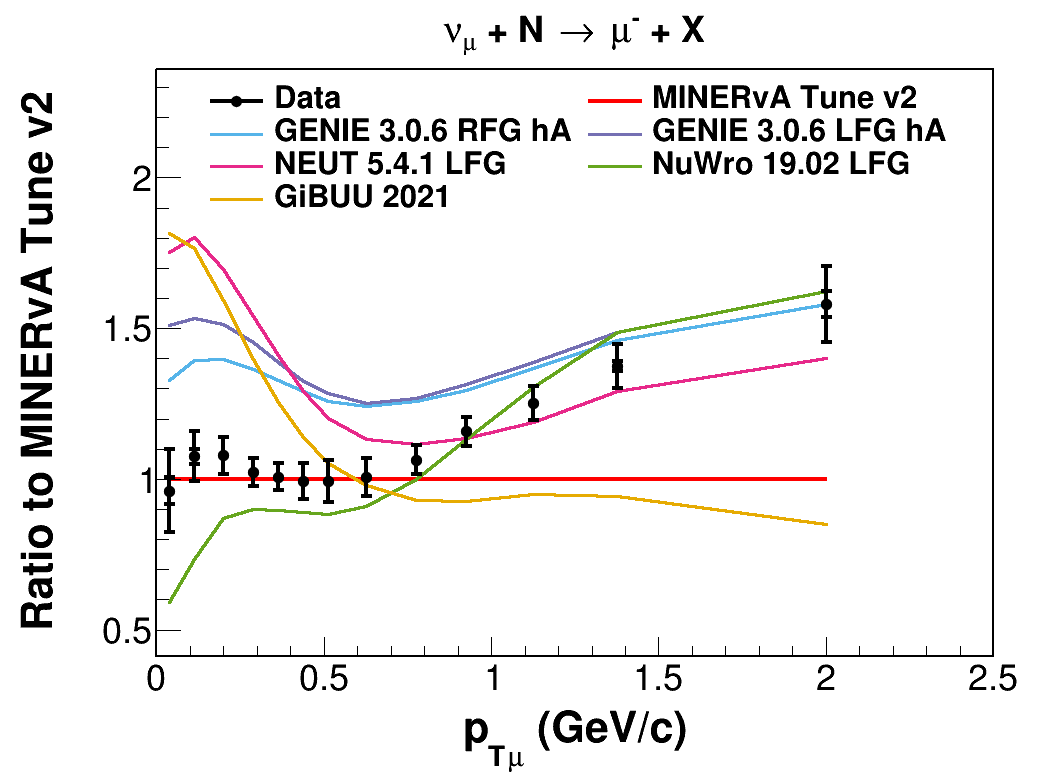}
    \xincludegraphics[width=0.45\textwidth,label=(d),pos=se,labelbox=false,fontsize=\large]{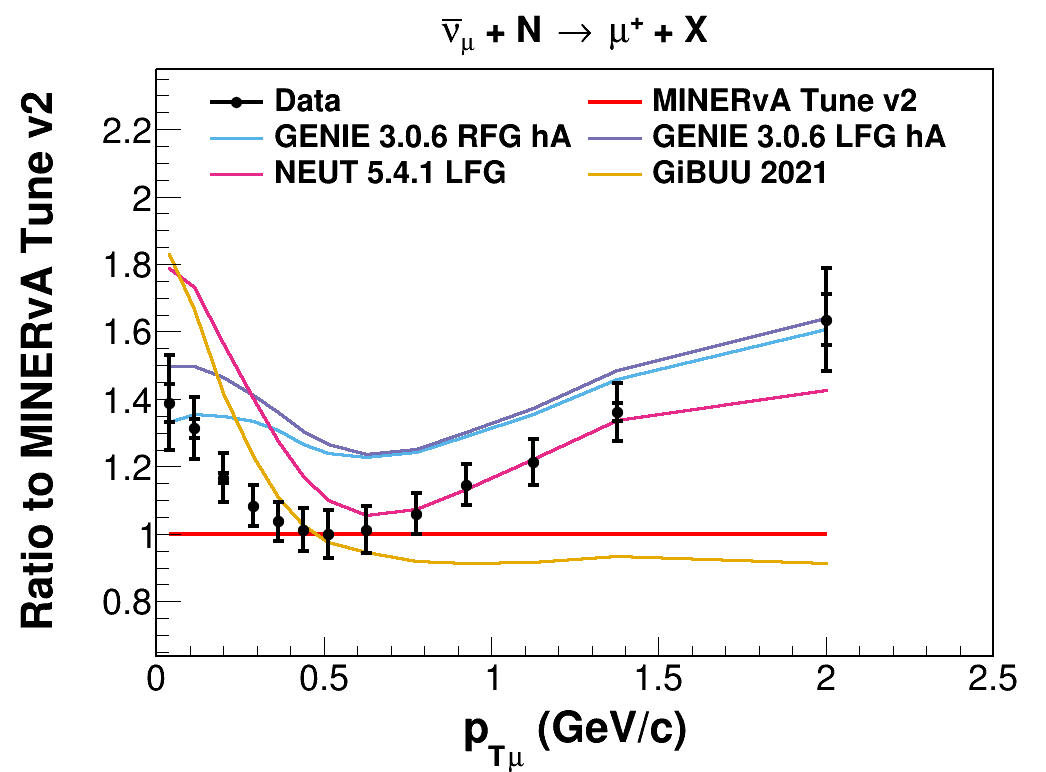}
    
    \xincludegraphics[width=0.45\textwidth,label=(e),pos=se,labelbox=false,fontsize=\large]{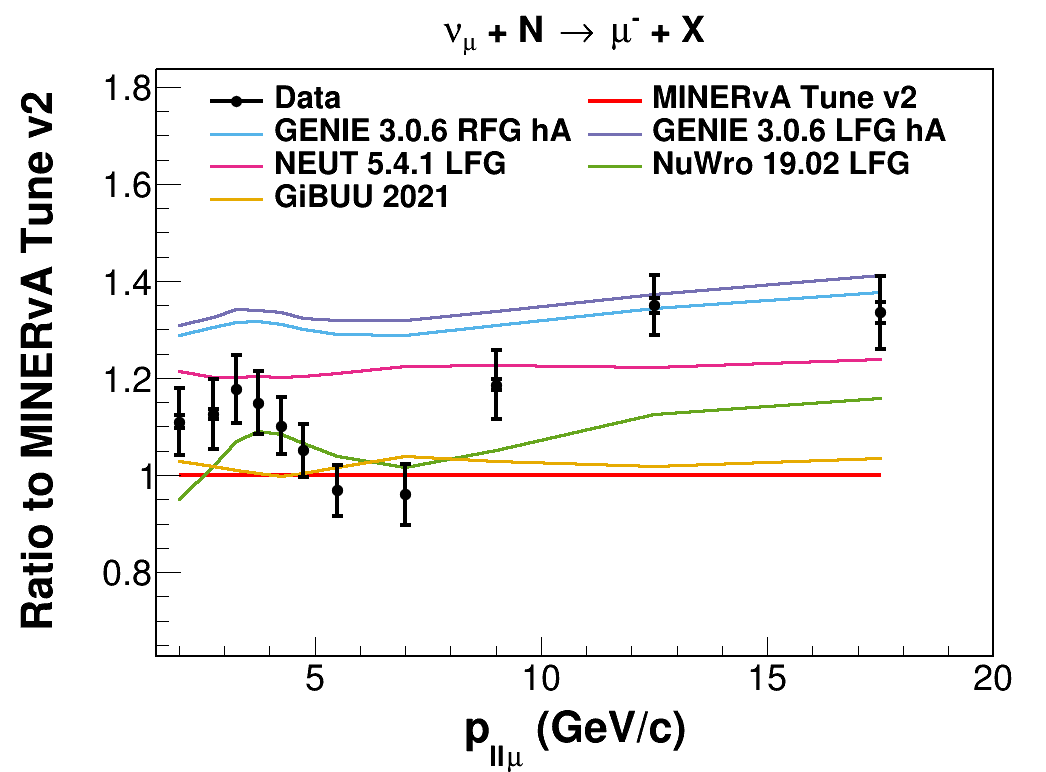}
    \xincludegraphics[width=0.45\textwidth,label=(f),pos=se,labelbox=false,fontsize=\large]{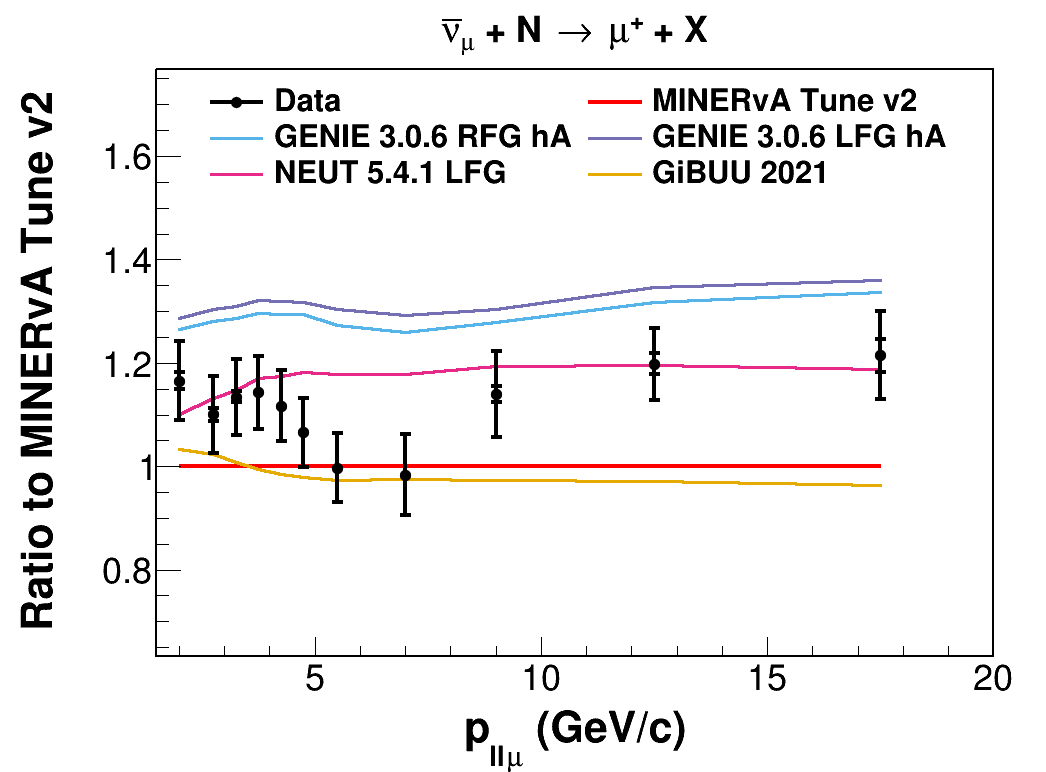}

    \caption{Definition I measured differential SIS cross section ratios compared to predictions of community neutrino generators normalized to MINERvA tune v2 as a function of $Q^2$ (a-b), $p_{T\mu}$ (c-d), and $p_{\parallel\mu}$ (e-f). Left: Neutrino. Right: Antineutrino.}
    \label{CrossSectionGenerators_Q2ge0}
\end{figure*}

\begin{figure*}[htbp]
    \centering
    \xincludegraphics[width=0.45\textwidth,label=(a),pos=se,labelbox=false,fontsize=\large]{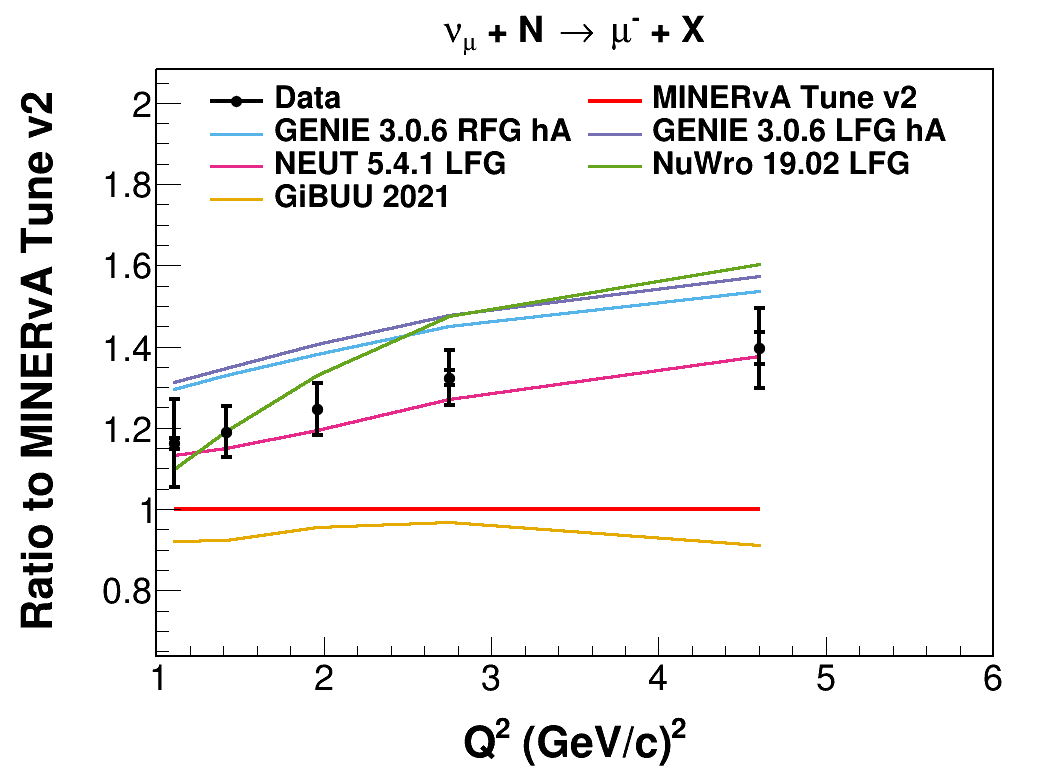}
    \xincludegraphics[width=0.45\textwidth,label=(b),pos=se,labelbox=false,fontsize=\large]{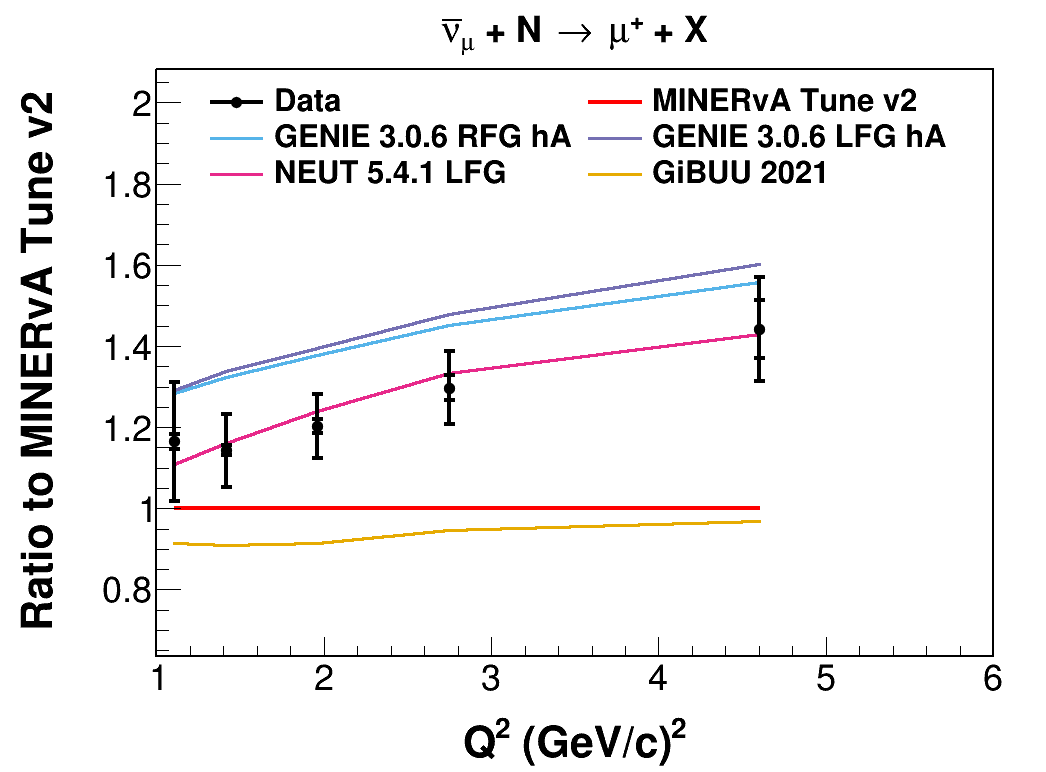}
    
    \xincludegraphics[width=0.45\textwidth,label=(c),pos=se,labelbox=false,fontsize=\large]{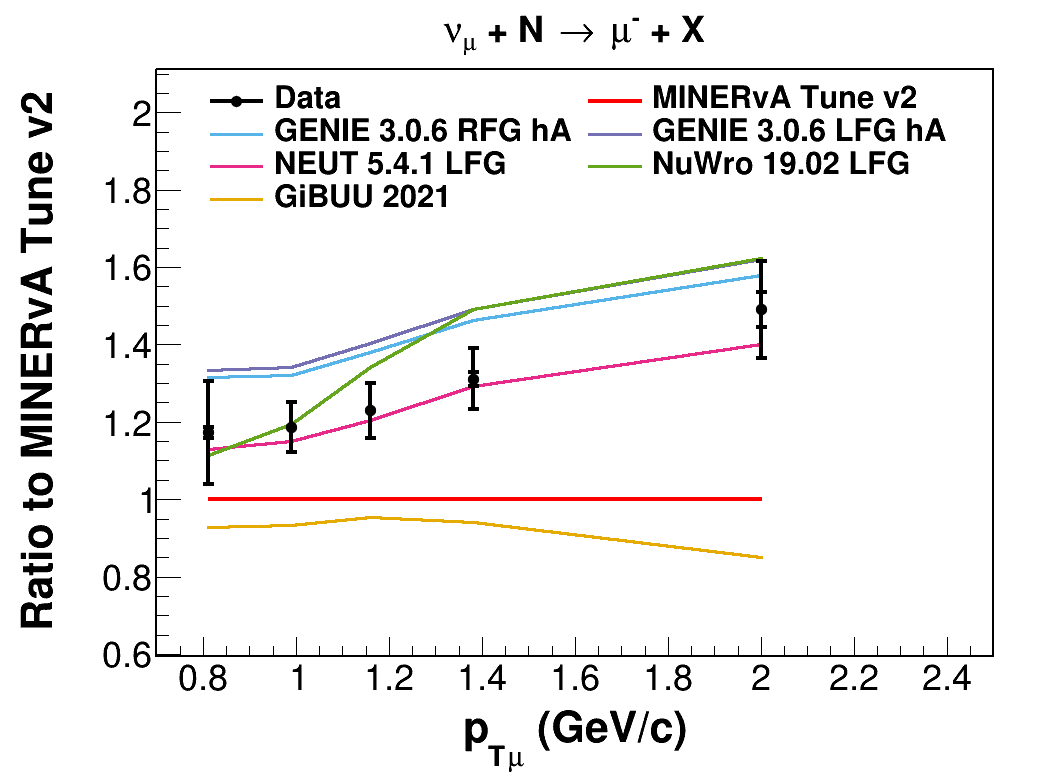}
    \xincludegraphics[width=0.45\textwidth,label=(d),pos=se,labelbox=false,fontsize=\large]{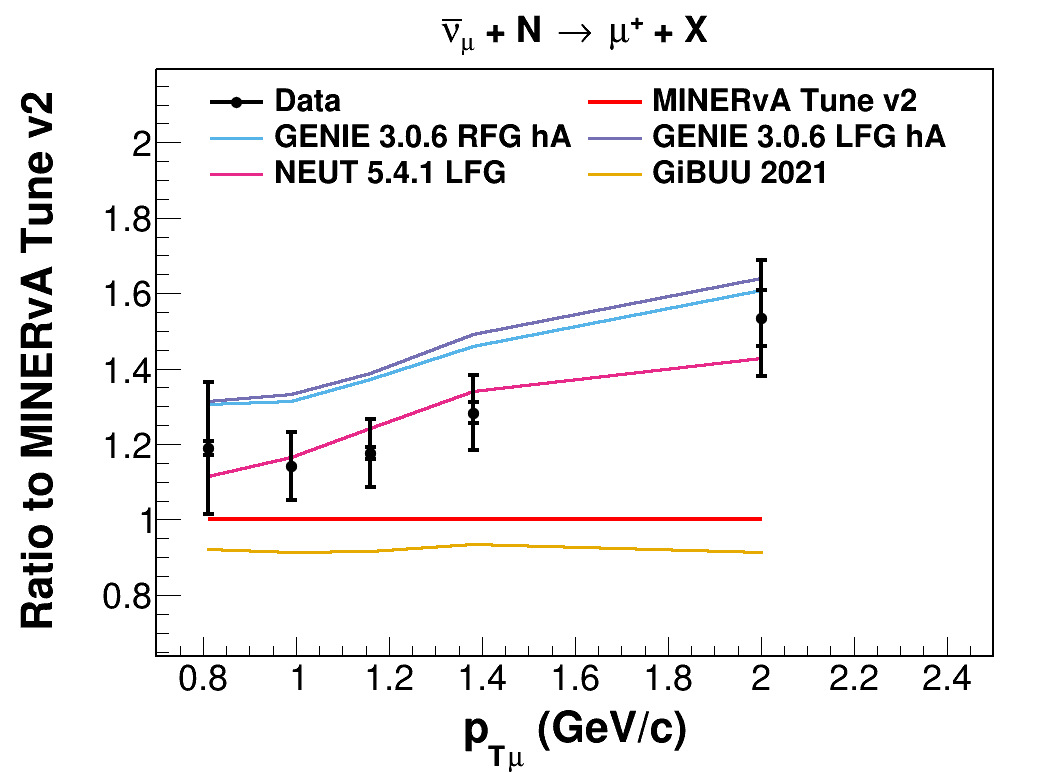}
    
    \xincludegraphics[width=0.45\textwidth,label=(e),pos=se,labelbox=false,fontsize=\large]{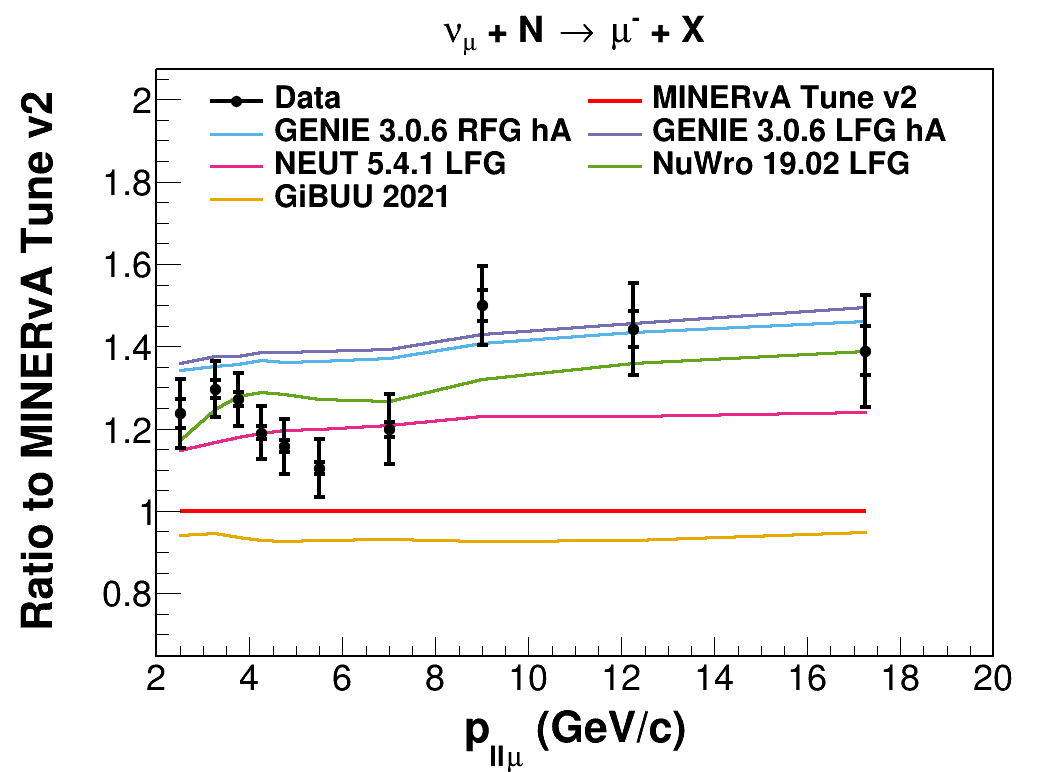}
    \xincludegraphics[width=0.45\textwidth,label=(f),pos=se,labelbox=false,fontsize=\large]{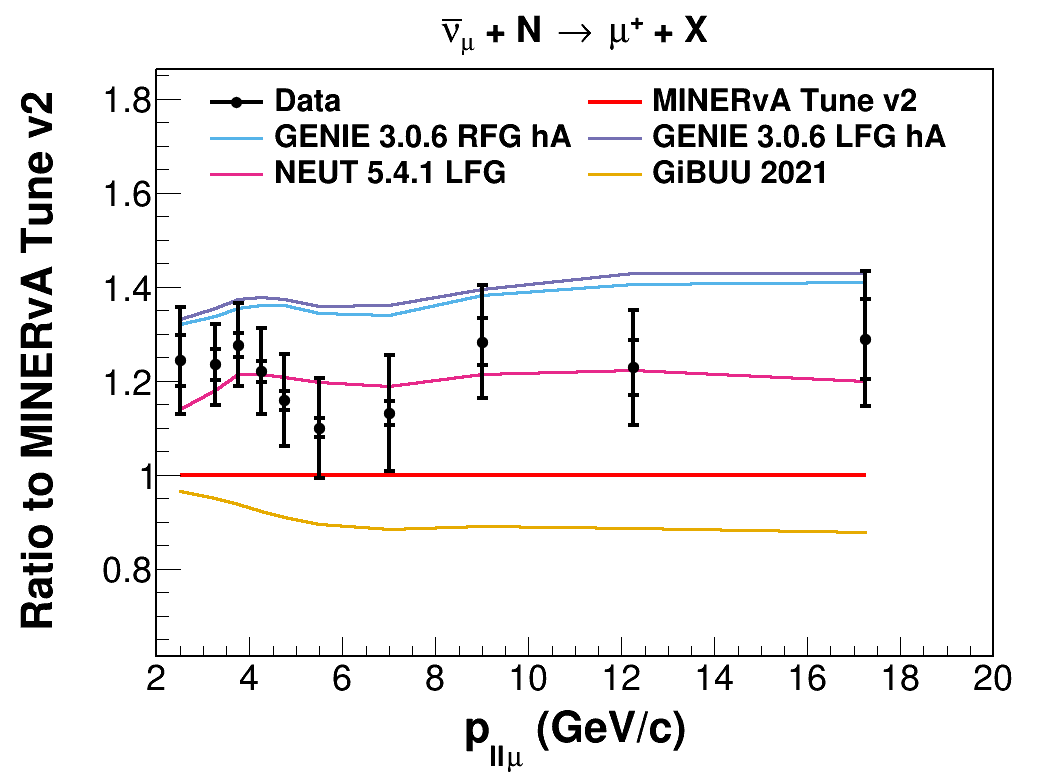}

    \xincludegraphics[width=0.45\textwidth,label=(g),pos=se,labelbox=false,fontsize=\large]{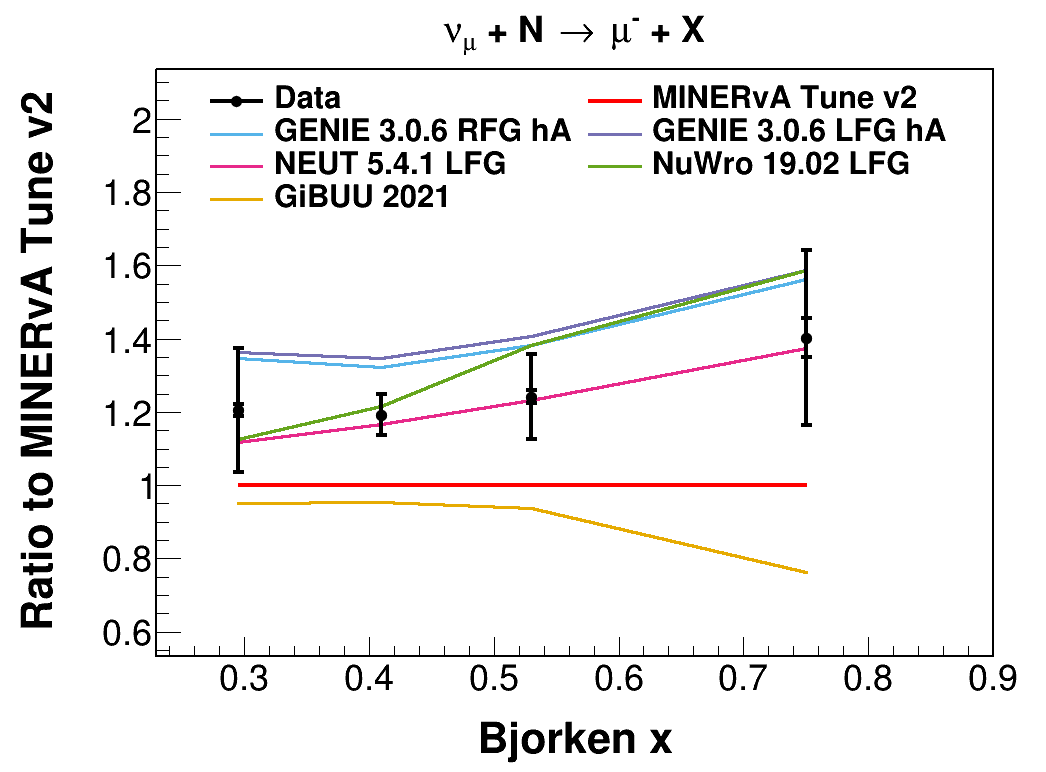}
    \xincludegraphics[width=0.45\textwidth,label=(h),pos=se,labelbox=false,fontsize=\large]{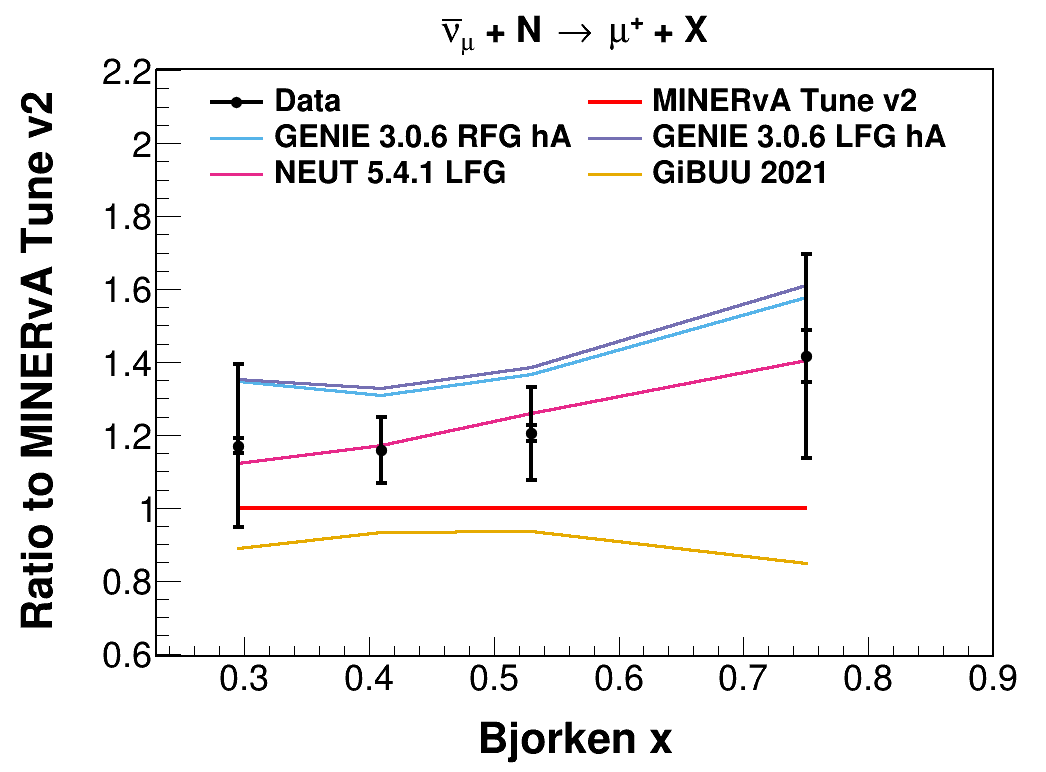}

    \caption{Definition II measured differential SIS cross section ratios compared to predictions of community neutrino generators normalized to MINERvA tune v2 as a function of $Q^2$ (a-b), $p_{T\mu}$ (c-d), $p_{\parallel\mu}$ (e-f), and $x$ (g-h). Left: Neutrino. Right: Antineutrino.}
    \label{CrossSectionGenerators_Q2ge1}
\end{figure*}

It is informative to determine if the predictions of other neutrino event simulators with alternative models are more consistent with these MINERvA SIS measurements, particularly in the higher $Q^2$ SIS multi-quark region. Selected versions of the alternative generators GENIE 3.0.6, GiBUU 2021, NEUT 5.4.1 and NuWro 19.02, have been compared with the SIS measured cross sections. The exact versions of these generators are indicated in the legends of Figure \ref{CrossSectionGenerators_Q2ge0} and Figure \ref{CrossSectionGenerators_Q2ge1} as well as Table~\ref{Chi2_ModelsData} and Table~\ref{Chi2_ModelsData_SISdefinitionII}. 

\subsubsection{SIS definition I}
The many differences between generators are a challenge to isolate, however, a few features stand out. From Table \ref{Chi2_ModelsData} and Figure \ref{CrossSectionGenerators_Q2ge0}, upon comparing simulation efforts to SIS definition I measurements throughout the full $Q^2 \geq$ 0 GeV$^2$/$c^2$ range for all variables, none of the simulation programs yielded acceptable $\chi^2/ndf$. With this stipulation, the best simulation of the SIS measured variables $Q^2$, $p_{T\mu}$ and $p_{\parallel\mu}$ was NuWro 19.02 for neutrino SIS measurements, while for antineutrino GENIE 3.0.6 RFG hA best models $Q^2$ and $p_{T\mu}$ while NEUT 5.4.1 best simulates $p_{\parallel\mu}$. However, it is the large $\chi^2/ndf$ of NEUT 5.4.1 as $Q^2$ approaches 0 that eliminates this choice as the best simulation for antineutrino $Q^2$ and $p_{T\mu}$ since it is the best for $Q^2$ $>$ 0.8 GeV$^2$/$c^2$ and $p_{T\mu}$ $>$ 0.6 GeV/c. 

Referring to Figure \ref{CrossSectionGenerators_Q2ge0} for the lower range of values of the SIS definition I variables, the overprediction of GiBUU 2021 and NEUT 5.4.1 of $Q^2$ and $p_{T\mu}$ is most extreme for neutrino and is a major contribution to the large $\chi^2/ndf$ found in Table~\ref{Chi2_ModelsData} for these simulation programs. NuWro 19.02 provides a somewhat better match to the $Q^2$ and $p_{T\mu}$ as well as $p_{\parallel\mu}$ shapes, while mildly underpredicting the low-$Q^2$ and $p_{T\mu}$ MINERvA SIS data. For this lower antineutrino kinematic range, Figure \ref{CrossSectionGenerators_Q2ge0} shows that GiBUU 2021, despite continued overprediction as $Q^2$ approaches 0, best simulates the $Q^2$ shape while none of the alternative generators simulates well the low values of $p_{T\mu}$ and $p_{\parallel\mu}$. For the lower range of $p_{\parallel\mu}$ NuWro 19.02 is best for neutrino, while NEUT 5.4.1 appears better for antineutrino.

\subsubsection{SIS definition II}
Once the considered kinematic region excludes the obviously challenging low-$Q^2$ region Table~\ref{Chi2_ModelsData_SISdefinitionII} indicates that the $\chi^2/ndf$ for all SIS variables improve significantly. For the $Q^2 \geq$ 1 GeV$^2$/$c^2$ SIS multi-quark kinematic region Figure \ref{CrossSectionGenerators_Q2ge1} as well as the $\chi^2/ndf$ of the Table indicate that GiBUU 2021 is inconsistent with both the magnitude and shape of the MINERvA data. NEUT 5.4.1 is more consistent with the shape of these higher $Q^2$ SIS measurements with Table~\ref{Chi2_ModelsData_SISdefinitionII} illustrating that NEUT 5.4.1 is the best match to MINERvA neutrino and antineutrino data and yields an acceptable $\chi^2/ndf$ in the process. Note also that in this high $Q^2$ region GENIE 3.0.6 RFG is only a somewhat less favorable alternative. The better agreement of GENIE 3.0.6 with the SIS definition II measurements comes mainly \cite{GENIE:2022qrc} from the new vector and axial form factors in the GENIE 3 resonance model, which yields a $Q^2$ dependence in this high W region more similar to the GENIE DIS model.

Since NEUT 5.4.1 so successfully predicts the MINERvA SIS cross section data in the SIS multi-quark region for both neutrino and antineutrino, a brief summary of the general differences between the cross section models of NEUT 5.4.1 and the MINERvA v2 tune of GENIE 2.12.6 could be informative. For resonance production NEUT 5.4.1 is based on the R-S model restricted to the production of a single pion with $W$ $\leq$ 2 GeV/$c^2$. MINERvA Tune v2 uses the R-S model for $W$ $\leq$ 1.7 GeV/$c^2$, dominated by one-pion production, with all resonance production for $W$ $>$ 1.7 GeV/$c^2$ included in DIS. For the DIS cross section NEUT 5.4.1 uses the B-Y model with one-pion production removed below $W$ = 2 GeV/$c^2$. MINERvA Tune v2 uses the B-Y model modulated to avoid duplication of resonances below $W$ = 1.7 GeV/$c^2$ and the complete B-Y model for $W$ $>$ 1.7 GeV/$c^2$. For the multiplicity of DIS events NEUT 5.4.1 uses several specifically developed models for pion multiplicities $\geq$ 2, one being similar but not identical to the AGKY model used by MINERvA Tune v2. Further details such as the form factors used and the method of event generation can be found in the presentations of the NuSTEC Workshop on Pion Production \cite{NuSTEC:2020nsl}.

\subsection{Generator Predicted SIS Cross Section Channel Components in the SIS Multi-quark Region}
To perhaps better understand the source of MINERvA v2 predictions consistently underpredicting the SIS measurements of all variables for both neutrino and antineutrino in the SIS multi-quark region, we examine the individual channel components that make up the $Q^2$ distribution of MINERvA Tune v2 predictions and compare them with the components of other generators $Q^2$ distributions included in this analysis. For the comparison, MINERvA Tune v2 uses GENIE classifications while for all other generators the NUISANCE \cite{Stowell:2016jfr} classification scheme is used. $Q^2$ is the chosen variable since it defines the SIS multi-quark region and provides the clearest comparison.

\begin{figure*}[htbp]
    \centering
    \xincludegraphics[width=0.45\textwidth,label=(a),pos=se,labelbox=false,fontsize=\large]{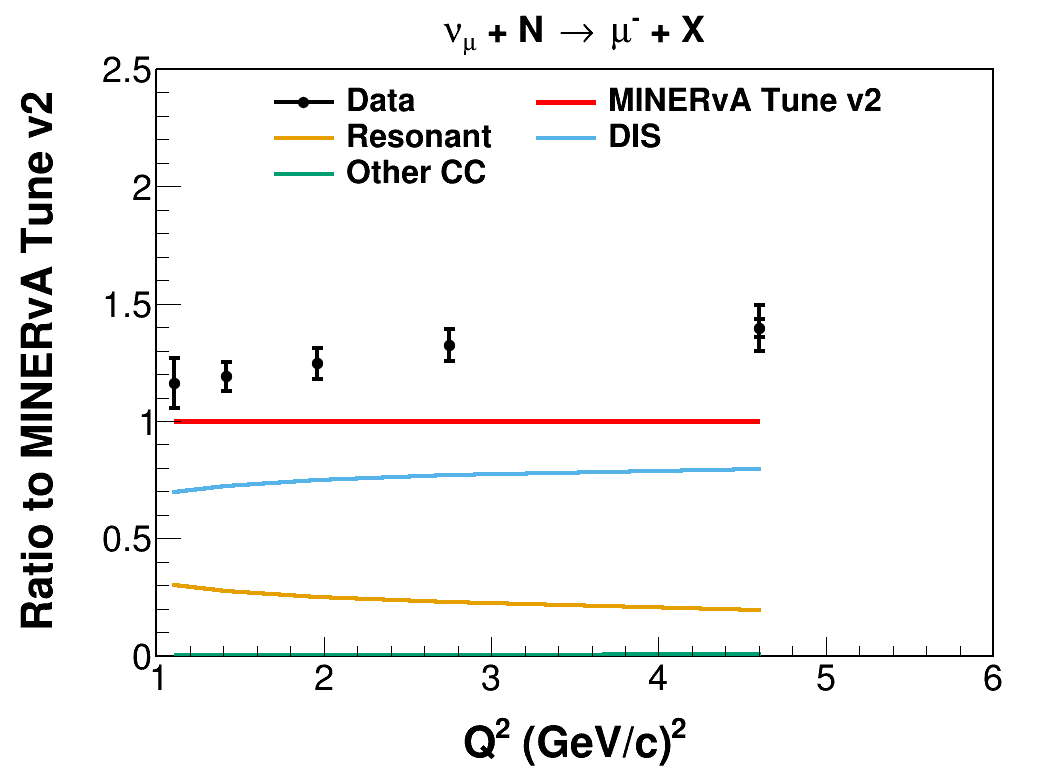}
    \xincludegraphics[width=0.45\textwidth,label=(b),pos=se,labelbox=false,fontsize=\large]{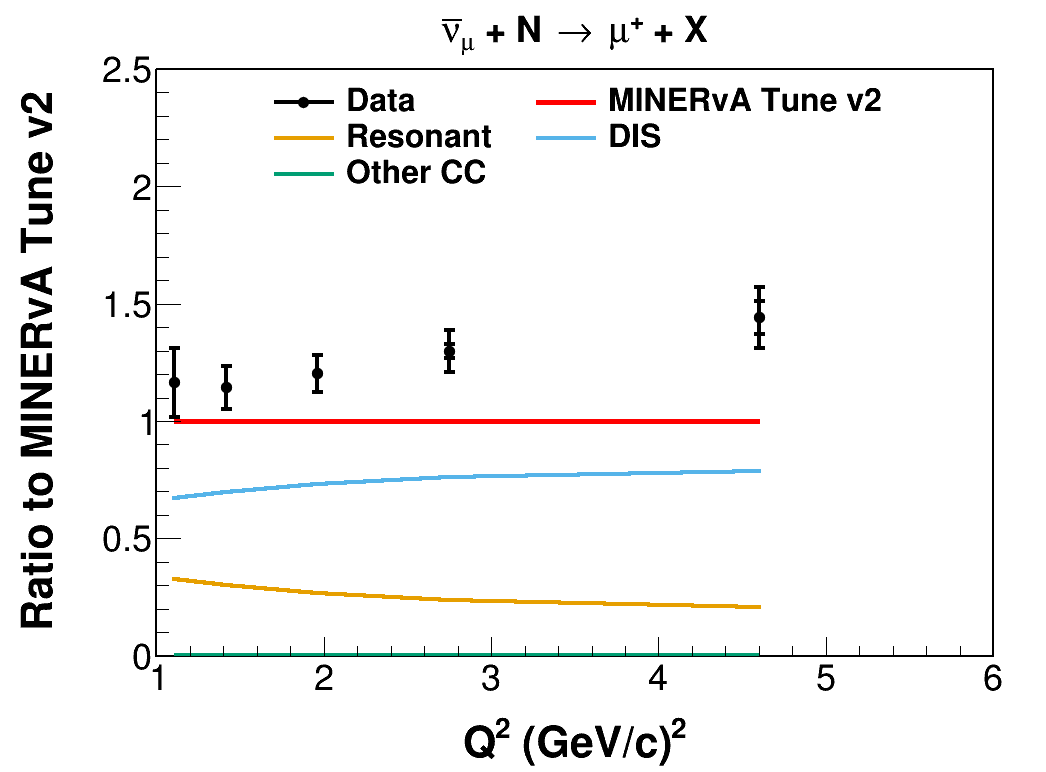}
    
    \xincludegraphics[width=0.45\textwidth,label=(c),pos=se,labelbox=false,fontsize=\large]{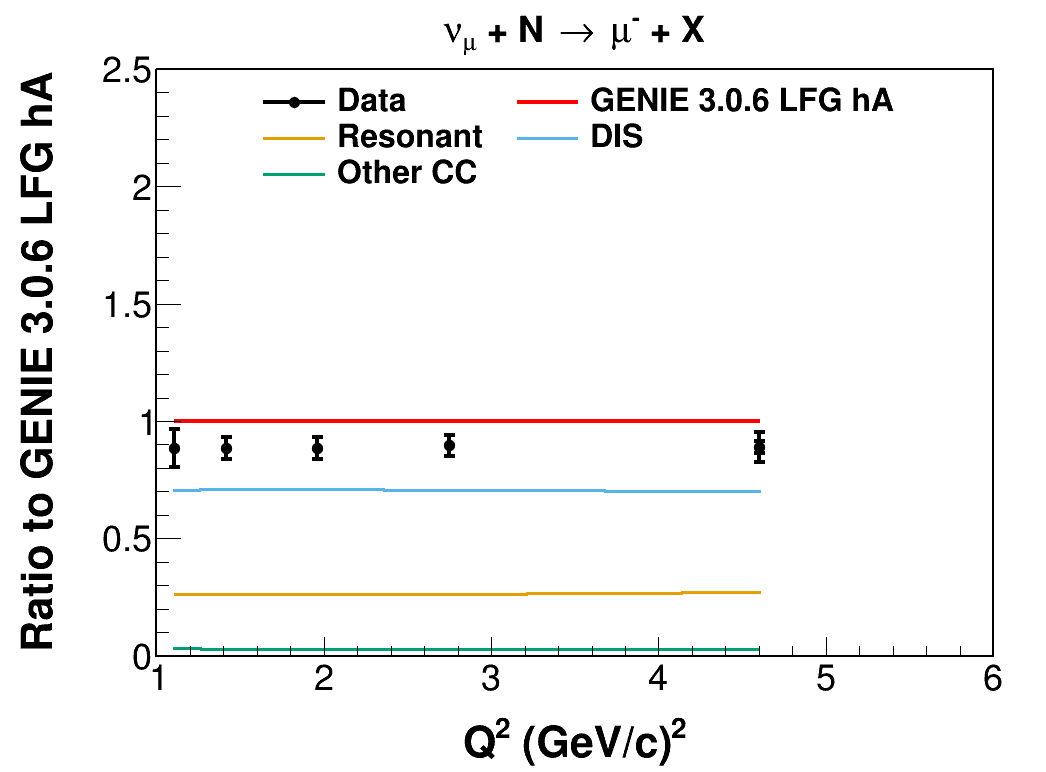}
    \xincludegraphics[width=0.45\textwidth,label=(d),pos=se,labelbox=false,fontsize=\large]{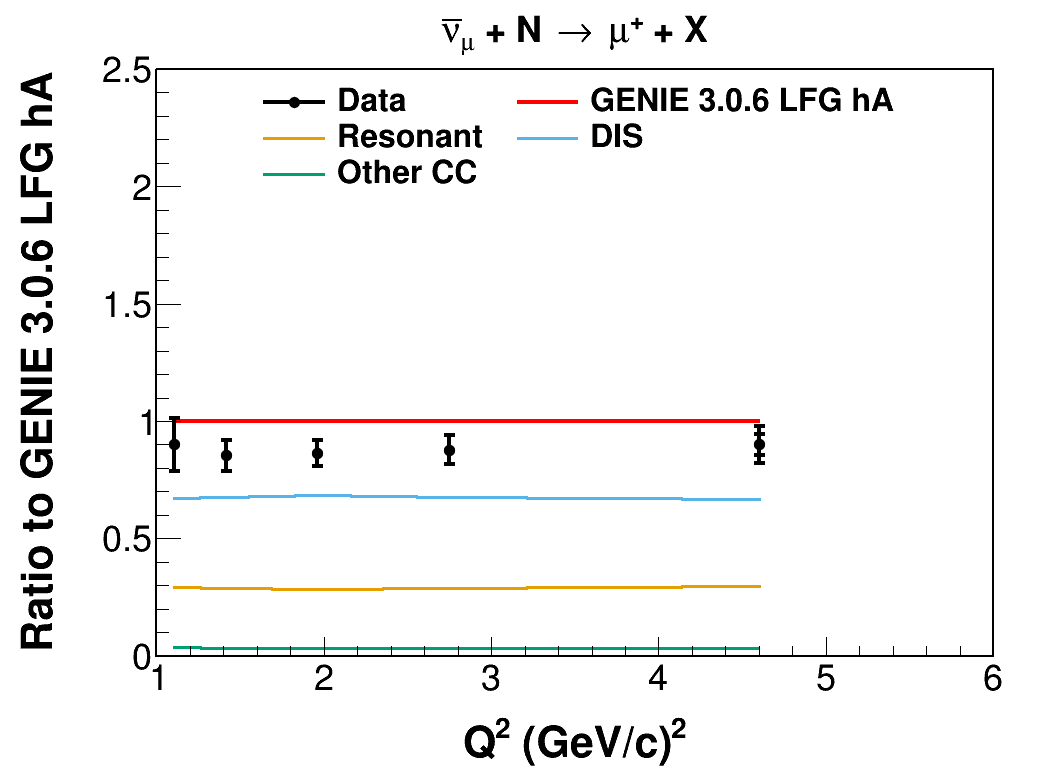}
    
    \xincludegraphics[width=0.45\textwidth,label=(e),pos=se,labelbox=false,fontsize=\large]{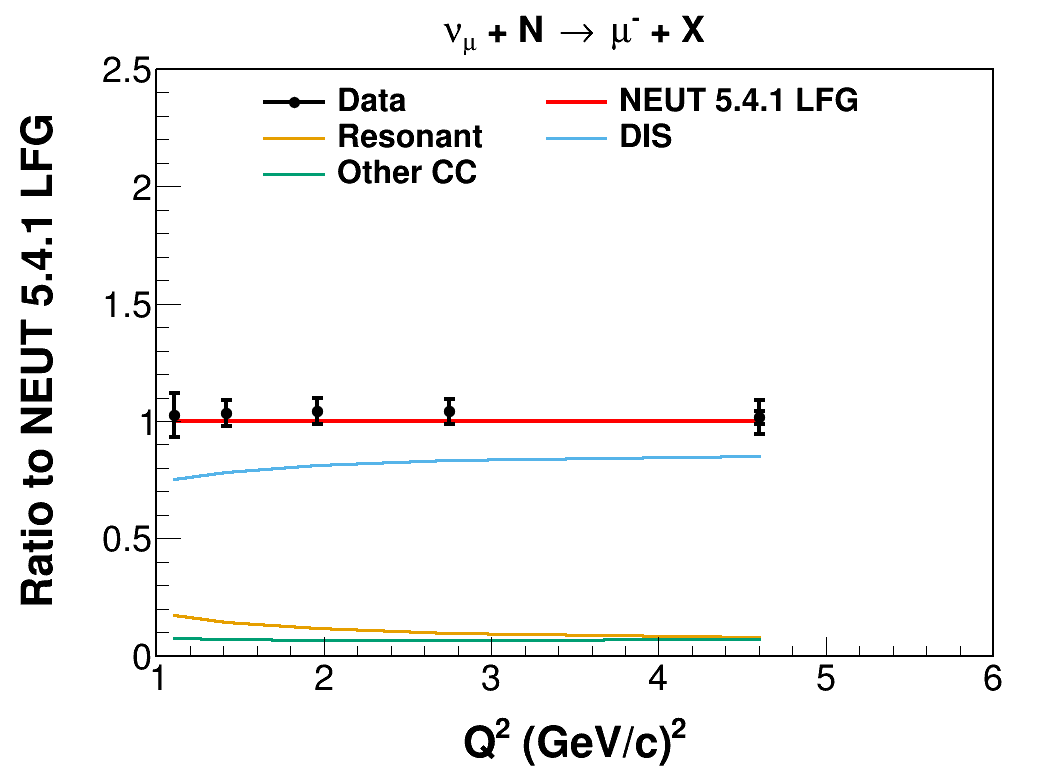}
    \xincludegraphics[width=0.45\textwidth,label=(f),pos=se,labelbox=false,fontsize=\large]{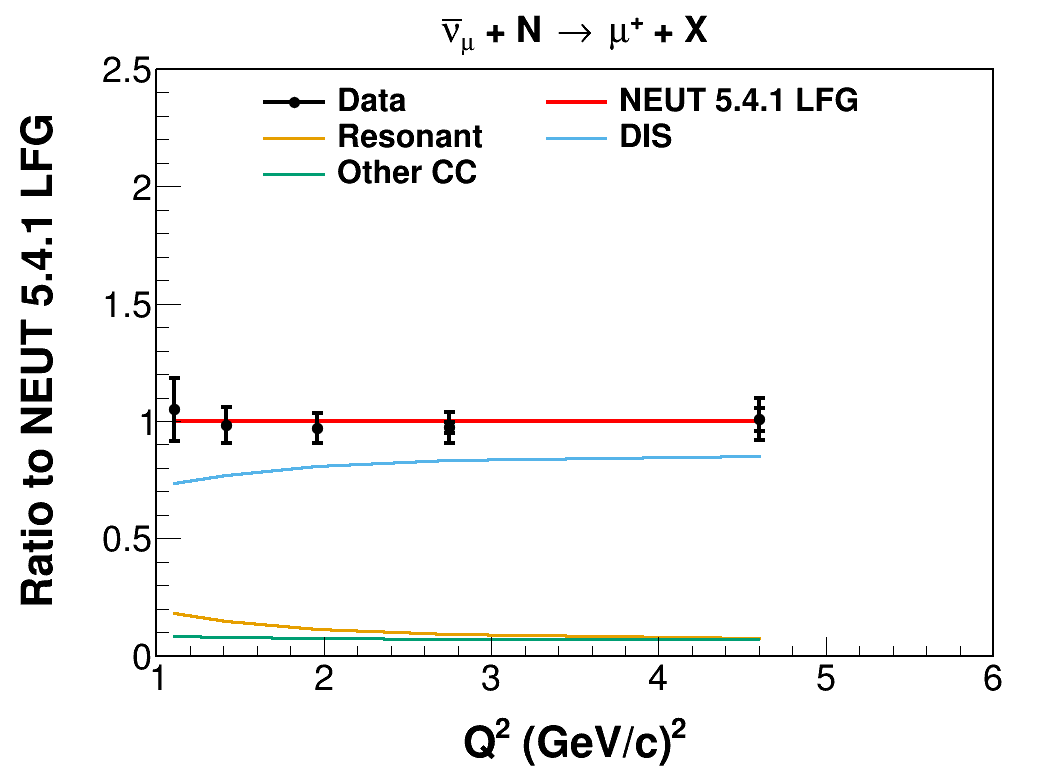}

    \xincludegraphics[width=0.45\textwidth,label=(g),pos=se,labelbox=false,fontsize=\large]{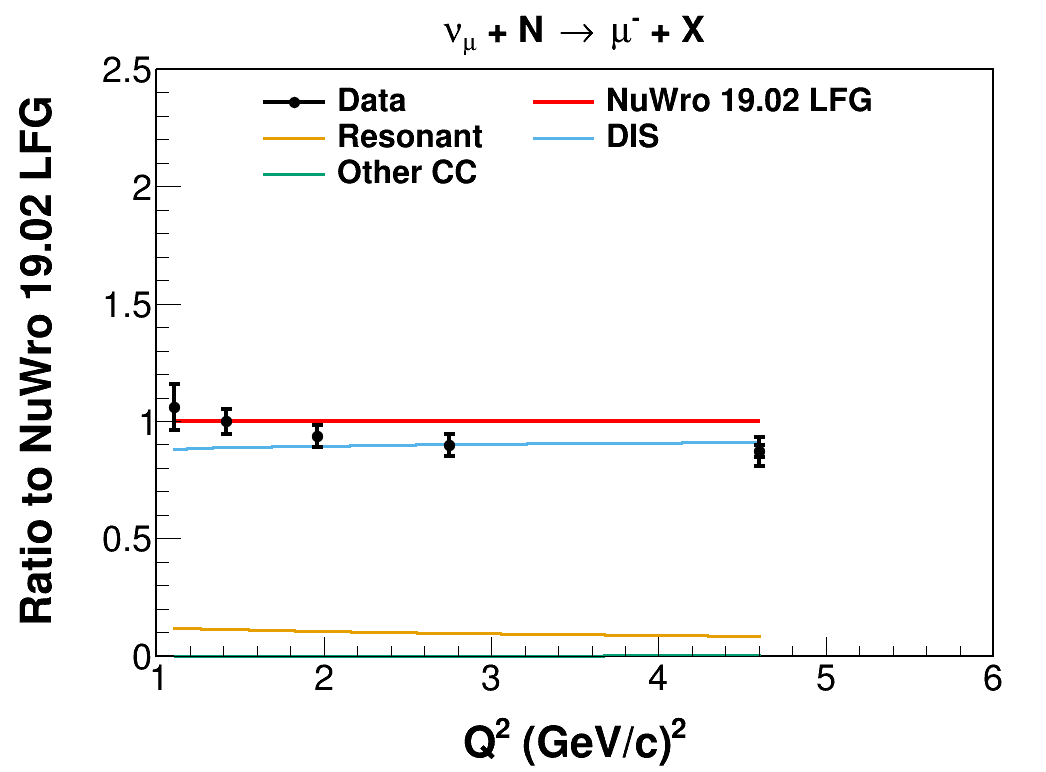}
    \xincludegraphics[width=0.45\textwidth]{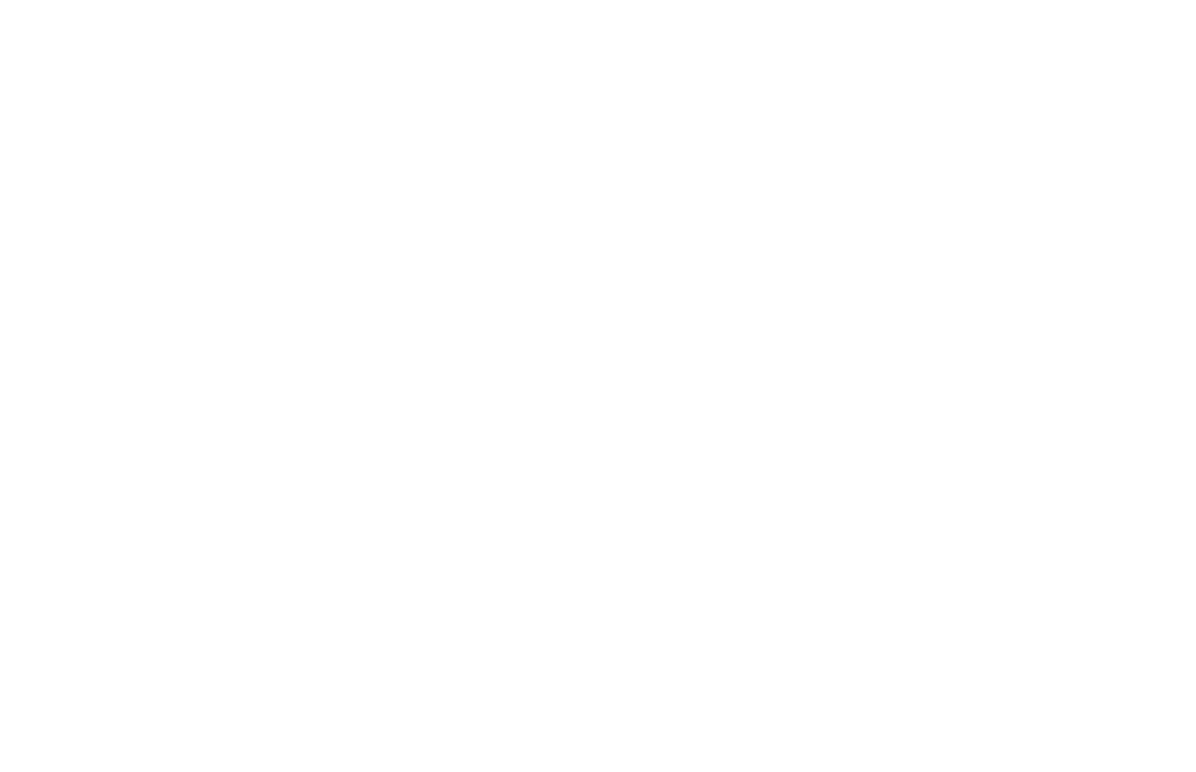}
    
    \caption{Definition II ratio of data and MC components of MINERvA Tune v2 (a-b), GENIE 3.0.6 (c-d), NEUT 5.4.1 (e-f), and NuWro 19.02 (g) as a function of $Q^2$. Left: Neutrino. Right: Antineutrino. These plots consider the QE and 2p2h events part of the Other CC category.}
    \label{CrossSection_GeneratorComponents}
\end{figure*}

Figure \ref{CrossSection_GeneratorComponents} shows the ratios of the data measured $Q^2$ and predicted individual channel components normalized to total predicted $Q^2$ of MINERvA Tune v2, GENIE 3.0.6 LFG, NEUT 5.4.1 and NuWro 19.02 neutrino simulations, as well as MINERvA Tune v2, GENIE 3.0.6 LFG and NEUT 5.4.1 antineutrino simulations. A breakdown of GiBUU 2021 predictions into channel components was also attempted, however, no combination of output resulted in similar channels as other displayed generators.

The major contribution comes from the respective DIS models, with a smaller contribution from the resonance models, which decrease as $Q^2$ increases. Each of the generators has their own resonance and nonesonance model and their own transition mix of resonance and DIS before reaching a region of pure DIS the start of which is also model-dependent. The actual magnitude of the DIS contribution most often then reflects the difference between the B-Y predicted total inelastic cross section and the generator's modeled resonance production. In the SIS multi-quark region it is obvious that GENIE 3.0.6 has a much larger resonance contribution to the cross section (about 30\%) than either NEUT 5.4.1 or NuWro 19.02, which have around 10\% contribution. This larger resonance contribution of GENIE 3.0.6 most likely reflects the inclusion of multipion decays, compared to NEUT 5.4.1 that includes only one $\pi$ production, and the much higher $W_{cut}$ value of 1.93 GeV/$c^2$. This higher value $W_{cut}$ includes a larger fraction of resonance production than NuWro 19.02, which is purely DIS after $W$ = 1.6 GeV/$c^2$ or MINERvA Tune v2 with a $W_{cut}$ value of 1.7 GeV/$c^2$.

For MINERvA Tune v2, the dominance of GENIE DIS is 75\% of the predicted total $Q^2$ cross section with resonance contributing the remaining 25\%. Since in general the resonance contribution decreases with growing $Q^2$, this suggests that even with the nonresonant subtraction removed in this high $W$ and high $Q^2$ region the growing underprediction of MINERvA Tune v2 with $Q^2$ is probably due to the current $Q^2$ dependence of the GENIE DIS contribution that needs consideration.  

The comparison of generators components suggests a general trend that the larger the contribution of DIS to the prediction, the better the fit to SIS measured cross sections. It is noteworthy that even though GENIE 3.0.6 has included a much broader range of $W$ and multi-pion decays in its resonances classification, and with new resonance form factors with a $Q^2$ shape more like the DIS $Q^2$ shape, resonances still only account for about 30\% of the total cross section in this higher $Q^2$ region.

\section{Conclusions}
The results presented here represent the first analysis of neutrino and antineutrino cross sections of $Q^2$, $p_{T\mu}$, $p_{\parallel\mu}$, and $x$ in the 1.5 $\leq$ $W$ $\leq$ 2 GeV/$c^2$ region. Since the nonresonant meson and multi-quark scattering components of SIS are dependent on $Q^2$ within any $W$ region, two $Q^2$ regions have been defined. In addition to the inclusive differential cross sections, to better enhance any multi-quark component of SIS a subsample with $Q^2$ $\geq$ 1 ~GeV$^2$/$c^2$ has been studied.

This has provided an opportunity to test various simulations of neutrino and antineutrino scattering in these kinematic regions. The measured data were compared primarily with the predictions of tuned variations from the GENIE 2.12.6 neutrino generator. To test how other simulation models predict these measured results, GENIE 3.0.6, NEUT 5.4.1, NuWro 19.02 neutrino simulation generators, and GiBUU 2021 nuclear transport code were employed. Comparisons were made through $\chi^2/ndf$ with covariance matrix considerations included. 

In general, the main MINERvA tune of the GENIE 2.12.6 simulation considered here, labeled MINERvA Tune v2, does not describe the data across the full kinematic region for all variables considered. However, in distinct regions that led to a specific tune, such as low $Q^2$ (and consequently $\approx$ low $p_{T\mu}$) for neutrinos, it is an accurate prediction. The same tune was found to be a much less accurate prediction for antineutrino scattering, and we suggest a less strong suppression of pion production at low $Q^2$ values for antineutrinos. For the higher $Q^2$ SIS multi-quark region the tune of GENIE 2.12.6 without the large reduction in nonresonant pion production was considerably superior for all measured variables. It is recommended that the large reduction in nonresonant pion production in MINERvA tunes be modified with a $W$ and perhaps $Q^2$ dependence that better reflects the kinematic region in which the reduction was measured.

The measured data was compared with the predictions of other simulation programs used in the community; GENIE 3.0.6, NEUT 5.4.1, NuWro 19.02 and GiBUU 2021 that employed different nuclear effects as well as different resonance and DIS models. In the $Q^2$ $\geq$ 0  range, heavily weighted by $Q^2$ $<$ 1 GeV$^2$/$c^2$ data, it was found that NuWro 19.02 is the best simulation of our data for neutrinos while MINERvA Tune v2 is best for antineutrinos. In the higher $Q^2$ $\geq$ 1 GeV$^2$/$c^2$ SIS multi-quark range, NEUT 5.4.1 is the best overall simulation of the higher value regions for all considered SIS variables in both neutrino and antineutrino. 

Finally, consider that in the SIS multi-quark region Figure \ref{CrossSection_GeneratorComponents} shows that for all generators considered, the resonance contribution is a relatively reduced fraction of the total cross section. In particular for GENIE 3.0.6, with its expanded multi-$\pi$ resonance model covering nearly the full $W$ region of this study, resonances still only contribute less than 1/3 of the total inelastic cross section. This could suggest that in what is designated as the DIS region of all considered generators, besides any still unmodeled associated resonance production, there could be and most likely is a significant contribution of SIS multi-quark interactions. As noted, the best overall simulation of this possibly significant SIS multi-quark region is NEUT 5.4.1. with its specific multipion model. 

These first measured total inelastic cross sections for neutrino and antineutrino in Shallow Inelastic Scatting regions clearly indicate that the generators need improvement for specific ranges of all measured variables. The many differences between the predictions of neutrino generators as well as the observation that there could be a significant contribution of SIS multi-quark interactions emphasize that there is an extreme lack of both the necessary detailed experimental investigations and the theoretical studies of interactions in this region that could enable a better understanding of the SIS contributions. This would lead to better modeling of neutrino-nucleus scattering physics in this SIS region, which would help to further reduce the systematic uncertainties for neutrino oscillation experiments.

\section{Acknowledgments}
We thankfully acknowledge the important contributions of L. Alvarez Ruso, C. Bronner, N. Jachowicz and C. Keppel to this study. 
This document was prepared by members of the MINERvA Collaboration using the resources of the Fermi National Accelerator Laboratory (Fermilab), a U.S. Department of Energy, Office of Science, Office of High Energy Physics HEP User Facility. Fermilab is managed by Fermi Forward Discovery Group, LLC, acting under Contract No. 89243024CSC000002. These resources included support for the MINERvA construction project, and support for construction also was granted by the United States National Science Foundation under Award No. PHY-0619727 and by the University of Rochester. Support for participating scientists was provided by NSF and DOE (USA); by CAPES and CNPq (Brazil); by CoNaCyT (Mexico); by ANID PIA / APOYO AFB180002, CONICYT PIA ACT1413, and Fondecyt 3170845 and 11130133 (Chile); by CONCYTEC (Consejo Nacional de Ciencia, Tecnolog\'ia e Innovaci\'on Tecnol\'ogica), DGI-PUCP (Direcci\'on de Gesti\'on de la Investigaci\'on - Pontificia Universidad Cat\'olica del Peru), and VRI-UNI (Vice-Rectorate for Research of National University of Engineering) (Peru); NCN Opus Grant No. 2016/21/B/ST2/01092 (Poland); by Science and Technology Facilities Council (UK); by EU Horizon 2020 Marie Skłodowska-Curie Action; by a Cottrell Postdoctoral Fellowship from the Research Corporation for Scientific Advancement; by an Imperial College London President's PhD Scholarship. We thank the MINOS Collaboration for use of its near detector data. Finally, we thank the staff of Fermilab for support of the beam line, the detector, and computing infrastructure.

\bibliographystyle{apsrev4-2}
\bibliography{main}

\end{document}